\documentclass[11pt]{article}

\usepackage{amsfonts,amssymb,chet,dsfont,graphicx,mathrsfs,pgfplots,tikz}
\allowdisplaybreaks

\newcommand{\1}{\mathds{1}}

\newcommand{\Op}[2]{\mathcal{O}_{#1}(\eta_{#2})}

\newcommand{\ee}[3]{(\eta_{#1}\cdot\eta_{#2})^{#3}}

\newcommand{\D}{\mathcal{D}}

\newcommand{\cOPE}[4]{{}_{#1}c_{#2#3}^{\phantom{#2#3}#4}}

\title{Seven-Point Conformal Blocks in the\\Extended Snowflake Channel and Beyond}

\author{Jean-Fran\c{c}ois Fortin$^{\ast,}$\email{jean-francois.fortin@phy.ulaval.ca}, Wen-Jie Ma$^{\ast,}$\email{wenjie.ma.1@ulaval.ca} and Witold Skiba$^{\dagger,}$\email{witold.skiba@yale.edu}}

\affiliation{
$^\ast$D\'epartement de Physique, de G\'enie Physique et d'Optique\\Universit\'e Laval, Qu\'ebec, QC G1V 0A6, Canada\\
$^\dagger$Department of Physics, Yale University, New Haven, CT 06520, USA
}%Choices for affiliations $^{\ast,\dagger,\$,\S,\ddag,}$

\abstract{Seven-point functions have two inequivalent topologies or channels.  The comb channel has been computed previously and here we compute scalar conformal blocks in the extended snowflake channel in $d$ dimensions.  Our computation relies on the known action of the differential operator that sets up the operator product expansion in embedding space.  The scalar conformal blocks in the extended snowflake channel are obtained as a power series expansion in the conformal cross-ratios whose coefficients are a triple sum of the hypergeometric type.  This triple sum factorizes into a single sum and a double sum.  The single sum can be seen as originating from the comb channel and is given in terms of a ${}_3F_2$-hypergeometric function, while the double sum originates from the snowflake channel which corresponds to a Kamp\'e de F\'eriet function.  We verify that our results satisfy the symmetry properties of the extended snowflake topology.  Moreover, we check that the behavior of the extended snowflake conformal blocks under several limits is consistent with known results.  Finally, we conjecture rules leading to a partial construction of scalar $M$-point conformal blocks in arbitrary topologies.}

\date{June 2020} %Uncomment this line for month to be fixed

\begin{document}

\maketitle

%\toc

%%%%%%%%%%%%%%%%%%%%%%%%%%%%%%%%%%%%%%%%%%%%%%%%%%
%%%%%%%%%%%%%%%%%%%%%%%%%%%%%%%%%%%%%%%%%%%%%%%%%%

\section{Introduction}\label{SecIntro}

With their extended spacetime symmetry group, conformal field theories (CFTs) are possibly amenable to exact non-perturbative solutions.  Indeed, from the operator product expansion (OPE), which expresses the product of two quasi-primary operators in terms of an infinite sum of quasi-primary operators and their descendants, conformal correlation functions can be expanded in terms of the CFT data (the OPE coefficients and operator dimensions, which encode in particular the spectrum of quasi-primary operators of the CFT and their couplings) and conformal blocks.  The blocks are functions of the conformal cross-ratios (scalar quantities built from the spacetime coordinates that are invariant under the conformal group) and are completely determined by conformal covariance, although it is notoriously difficult to compute them in all generality.  Once the four-point conformal blocks are known, the CFT data is in principle constrained from associativity of the four-point correlation functions, the so-called conformal bootstrap introduced in \cite{Ferrara:1973yt,Polyakov:1974gs}.  For this reason, a lot of effort has been put towards computing four-point conformal blocks, see for example the Casimir equations \cite{Dolan:2003hv,Dolan:2011dv,Kravchuk:2017dzd}, the shadow formalism \cite{Ferrara:1972xe,Ferrara:1972uq,SimmonsDuffin:2012uy}, the weight-shifting formalism \cite{Karateev:2017jgd,Costa:2018mcg}, integrability \cite{Isachenkov:2016gim,Schomerus:2016epl,Schomerus:2017eny,Isachenkov:2017qgn,Buric:2019dfk}, AdS/CFT \cite{Hijano:2015zsa,Nishida:2016vds,Castro:2017hpx,Dyer:2017zef,Chen:2017yia,Sleight:2017fpc}, and the OPE \cite{Ferrara:1971vh,Ferrara:1971zy,Ferrara:1972cq,Ferrara:1973eg,Ferrara:1973vz,Ferrara:1974nf,Dolan:2000ut,Fortin:2016lmf,Fortin:2016dlj,Comeau:2019xco,Fortin:2019fvx,Fortin:2019dnq,Fortin:2019xyr,Fortin:2019pep,Fortin:2019gck,Fortin:2020ncr}.

The study of higher-point conformal blocks is a relatively recent field of research in CFT.  For example, the scalar $M$-point blocks in the comb channel in one and two spacetime dimensions as well as the scalar five-point blocks in any spacetime dimensions were first computed in \cite{Alkalaev:2015fbw,Rosenhaus:2018zqn,Goncalves:2019znr,Parikh:2019ygo,Jepsen:2019svc}.  In any number of dimensions, the scalar $M$-point conformal blocks in the comb channel were introduced in \cite{Parikh:2019dvm,Fortin:2019zkm}, and the scalar six-point conformal blocks in the snowflake channel were obtained in \cite{Fortin:2020yjz}.\footnote{See also \cite{Anous:2020vtw} for specific snowflake conformal blocks in two-dimensional CFTs.}

The method used in \cite{Fortin:2019zkm,Fortin:2020yjz} to compute the higher-point conformal blocks is based on the embedding space OPE formalism introduced in \cite{Fortin:2019fvx,Fortin:2019dnq}.  Since the embedding space OPE formalism is general, it can in principle be used for any correlation functions in any topology, including internal and external quasi-primary operators in arbitrary irreducible representations of the Lorentz group.  In practice, we start with known $(M-1)$-point correlation functions, express them in terms of the natural conformal cross-ratios for the OPE differential operator, trivially apply the latter on the known $(M-1)$-point correlation functions,\footnote{The action of the OPE differential operator has been worked out explicitly in \cite{Fortin:2019dnq}.} express the resulting $M$-point correlation functions in terms of the appropriate conformal cross-ratios, and finally evaluate the extra sums using simple hypergeometric identities.  All the steps are straightforward, although the final re-summations can be somewhat tedious.

One interesting property of higher-point conformal blocks is the existence of several inequivalent topologies, as exemplified for example by the two topologies appearing at six points, namely the comb channel and the snowflake channel.  In \cite{Fortin:2020yjz}, we argued that the symmetry group of a given topology, denoted by $H_{M|\text{channel}}$, plays an interesting role in determining the number of equivalent ways of writing the conformal blocks, or in other words the number of identities the conformal blocks must satisfy.  This observation also led to an understanding of the number of $M$-point conformal bootstrap channels one should expect.  Moreover, we conjectured an identity between the number of topologies and the symmetry groups, given by
\eqn{\sum_{1\leq i\leq T_0(M)}\frac{1}{|H_{M|\text{channel $i$}}|}=\frac{T(M)}{|S_M|}=\frac{(2M-5)!!}{M!}.}[EqT]
In \eqref{EqT}, $T_0(M)$ is the number of inequivalent $M$-point topologies, which corresponds also to the number of unrooted binary trees with $M$ unlabeled leaves;\footnote{Unfortunately, there does not exist an analytic expression for $T_0(M)$.  Starting at $M=2$, the first few numbers in the sequence are $(1,1,1,1,2,2,4,6,11,\ldots)$.  See \textit{The On-line Encyclopedia of Integer Sequences} at \href{https://oeis.org/A000672}{https://oeis.org/A000672} and \href{https://oeis.org/A129860}{https://oeis.org/A129860} for more details.} $T(M)$ is the number of different ways of expressing the same full $M$-point correlation functions, which is given by the number of unrooted binary trees with $M$ labeled leaves; and $S_M$ is the symmetry group of the full $M$-point correlation functions, which is simply the symmetry group of $M$ elements.

Another interesting feature of higher-point conformal blocks is the appearance of extra sums, denoted by the $F$-function [see \eqref{EqG}], in the conformal cross-ratio power series.  We argued in \cite{Fortin:2019zkm,Fortin:2020yjz} that the minimal number of extra sums for $M$-point conformal blocks in an arbitrary channel is $M-4$ and that it is always possible to express $F_M$ as $M-4$ sums (extra sums can be evaluated).  This was proven by direct computation in the comb channel and for six-point conformal blocks in the snowflake channel.  We also noticed that the $F$-function factorizes in the comb channel, but not in the six-point snowflake channel.

In this paper, we continue the analysis of higher-point correlation functions by computing scalar seven-point conformal blocks in all topologies, \textit{i.e.} in the comb channel and the so-called extended snowflake channel.  From the topologies, it is clear that the symmetry groups are
\eqn{H_{7|\text{comb}}=(\mathbb{Z}_2)^2\rtimes\mathbb{Z}_2\qquad\text{and}\qquad H_{7|\substack{\text{extended}\\\text{snowflake}}}=\mathbb{Z}_2\times\left((\mathbb{Z}_2)^2\rtimes\mathbb{Z}_2\right),}
in agreement with \eqref{EqT}.  Focusing on the coset cardinalities of the two topologies $|S_7/H_{7|\text{comb}}|=630$ and $S_7/H_{7|\substack{\text{extended}\\\text{snowflake}}}|=315$, we conclude that there are $945$ different ways of writing the full seven-point correlation functions.

Moreover, we prove that $F_{7|\substack{\text{extended}\\\text{snowflake}}}$ has interesting factorization properties.  Although this function does not factorize into three independent sums as in the comb channel, we show that it factorizes into one sum and an object with two intertwined sums.  In fact, these two factors correspond to $F_{5|\text{comb}}$ and $F_{6|\text{snowflake}}$, respectively.  From this observation, we conjecture rules to partially write down scalar $M$-point conformal blocks in any topology, up to the knowledge of the conformal cross-ratios.

This paper is organized as follows: Section \ref{SecCB} summarizes the embedding space OPE formalism.  We first review the OPE differential operator in the scalar case and then we give its action on products of conformal cross-ratios.  After, we describe our notation for contributions to scalar $M$-point correlation functions from exchanged quasi-primary operators in trivial irreducible representations, and we discuss the recurrence relation transforming scalar $(M-1)$-point correlation functions to scalar $M$-point correlation functions.  We finally review the scalar $M$-point conformal blocks in the comb channel as well as the scalar six-point conformal blocks in the snowflake channel before introducing the scalar seven-point conformal blocks in the extended snowflake channel.  In Section \ref{SecChecks}, we perform several consistency checks for the scalar seven-point conformal blocks in the extended snowflake channel.  We start by verifying that the extended snowflake conformal blocks transform appropriately under the three generators of its topology symmetry group $H_{7|\substack{\text{extended}\\\text{snowflake}}}$.  We then study the OPE limit, where two embedding space coordinates are taken to coincide, and the limit of unit operator, where one external quasi-primary operator is set to the identity operator.  Finally, we conclude in Section \ref{SecConc} with a lengthy discussion of the implications of our results by conjecturing rules to construct scalar conformal blocks in arbitrary topologies from partially fixed building blocks.  Appendix \ref{SAppCB7} presents the computations necessary to obtain the scalar seven-point conformal blocks in the extended snowflake channel from the scalar six-point conformal blocks in the snowflake channel.  Appendix \ref{SAppSym} gives the proofs for the identities of the extended snowflake conformal blocks under the symmetry generators of $H_{7|\substack{\text{extended}\\\text{snowflake}}}$.  Lastly, Appendix \ref{SAppLim} contains the remaining proofs related to the OPE limit and the limit of unit operator.

%%%%%%%%%%%%%%%%%%%%%%%%%%%%%%%%%%%%%%%%%%%%%%%%%%
%%%%%%%%%%%%%%%%%%%%%%%%%%%%%%%%%%%%%%%%%%%%%%%%%%

\section{Higher-Point Conformal Blocks}\label{SecCB}

This section presents the scalar seven-point conformal blocks in the extended snowflake channel.  Following \cite{Fortin:2019dnq,Fortin:2019zkm,Fortin:2020yjz}, we first give a quick review of the OPE formalism and the scalar conformal blocks in all channels for four-, five-, and six-point correlation functions, as well as for the seven-point correlation functions in the comb channel only.  We then give results for the remaining scalar seven-point conformal blocks in the extended snowflake channel, exhausting all the seven-point topologies.  The relevant proofs are left for the appendixes.

%%%%%%%%%%%%%%%%%%%%%%%%%%%%%%%%%%%%%%%%%%%%%%%%%%

\subsection{\texorpdfstring{$M$}{M}-Point Correlation Functions from the OPE}

The form of the OPE as introduced in \cite{Fortin:2019fvx,Fortin:2019dnq} is applicable to operators with any spins in any dimension $d$.  Using the explicit action obtained in \cite{Fortin:2016dlj,Comeau:2019xco,Fortin:2019dnq}, it is straightforward to compute $M$-point correlation functions from $(M-1)$-point correlation functions.  By recurrence, we can thus determine any conformal blocks by simply applying the OPE recursively in the proper order.  For example, the OPE has been used to compute two-, three-, and four-point correlation functions in \cite{Fortin:2019xyr,Fortin:2019pep,Fortin:2019gck} and general rules for four-point conformal blocks in arbitrary irreducible representations, relevant for the bootstrap, were presented in \cite{Fortin:2020ncr}.  Recently, higher-point scalar conformal blocks in the comb channel were obtained in \cite{Fortin:2019zkm} and scalar six-point conformal blocks in the snowflake channel were evaluated in \cite{Fortin:2020yjz} using this method.

The OPE in embedding space simplifies considerably for scalar operators because there are no free Lorentz indices
\eqn{
\begin{gathered}
\Op{i}{1}\Op{j}{2}=\cOPE{}{i}{j}{k}\frac{1}{\ee{1}{2}{p_{ijk}}}(\D_{12}^2)^{h_{ijk}}\Op{k}{2}+\ldots,\\
p_{ijk}=\frac{1}{2}(\Delta_i+\Delta_j-\Delta_k),\qquad h_{ijk}=-\frac{1}{2}(\Delta_i-\Delta_j+\Delta_k),			   
\end{gathered}
}[EqOPE]
where $\Delta$'s are the dimensions of operators $\Op{}{}$, and $\eta_{1,2}$ are the embedding space coordinates.  We omitted an infinite sum over quasi-primary operators and kept only the scalar primary in the OPE.  The differential operator $(\D_{12}^2)^{h_{ijk}}$ generates the appropriate infinite towers of descendants of $\Op{k}{2}$, while $\cOPE{}{i}{j}{k}$ is the OPE coefficient.  All quantities defined in \eqref{EqOPE} are introduced and explained in detail in \cite{Fortin:2019dnq} as is the more complicated case of operators with spin.

The scalar differential operator $\D_{12}^2$ in \eqref{EqOPE} can be defined for any pairs of coordinates as
\eqn{ \D_{ij}^2= \ee{i}{j}{}\partial_j^2-(d+2\eta_j\cdot\partial_j)\eta_i\cdot\partial_j,}
where $\partial_j=\frac{\partial}{\partial\eta_j}$ and $d$ is the dimension of spacetime.  $\D_{ij}^2$ is homogeneous of degree $1$ with respect to $\eta_i$ and of degree $-1$ with respect to $\eta_j$.  The scalar OPE differential operator can be made homogeneous of degree 0 with respect to all coordinates, where for brevity $\eta_{ij}\equiv\eta_i\cdot\eta_j$,
\eqn{\bar{\D}_{ij;kl;m}^2=\frac{\eta_{ij}\eta_{kl}}{\eta_{ik}\eta_{il}}\D_{ij}^2,}
for some $k$, $l$, and $m$.  While $\eta_m$ does not appear above, it is singled out in the choice of the cross-ratios below.  The action of $\bar{\D}_{ij;kl;m}^2$ on the conformal cross-ratios
\eqn{
\begin{gathered}
x_m=\frac{\eta_{ij}\eta_{kl}\eta_{im}}{\eta_{ik}\eta_{il}\eta_{jm}},\\
y_a=1-\frac{\eta_{im}\eta_{ja}}{\eta_{ia}\eta_{jm}},\qquad1\leq a\leq M,\qquad a\neq i,j,m,\\
z_{ab}=\frac{\eta_{ik}\eta_{il}\eta_{ab}}{\eta_{kl}\eta_{ia}\eta_{ib}},\qquad1\leq a<b\leq M,\qquad a,b\neq i,j,
\end{gathered}
}[EqCROPE]
can be written as an infinite series of the hypergeometric type as
\eqna{
&\bar{\D}_{ij;kl;m}^{2h}x_m^{\bar{q}}\prod_{\substack{1\leq a\leq M\\a\neq i,j,m}}(1-y_a)^{-q_a}\\
&\qquad=x_m^{\bar{q}+h}\sum_{\{n_a,n_{am},n_{ab}\}\geq0}\frac{(-h)_{\bar{n}_m+\bar{\bar{n}}}(q_m)_{\bar{n}_m}(\bar{q}+h)_{\bar{n}-\bar{\bar{n}}}}{(\bar{q})_{\bar{n}+\bar{n}_m}(\bar{q}+1-d/2)_{\bar{n}_m+\bar{\bar{n}}}}\\
&\qquad\phantom{=}\qquad\times\prod_{\substack{1\leq a\leq M\\a\neq i,j,m}}\frac{(q_a)_{n_a}}{n_{am}!(n_a-n_{am}-\bar{n}_a)!}y_a^{n_a}\left(\frac{x_mz_{am}}{y_a}\right)^{n_{am}}\prod_{\substack{1\leq a<b\leq M\\a,b\neq i,j,m}}\frac{1}{n_{ab}!}\left(\frac{x_mz_{ab}}{y_ay_b}\right)^{n_{ab}},
}[EqD]
for any real $h$.  We note here that the OPE differential operator \eqref{EqD} differs by a numerical scaling factor from the one defined in \cite{Fortin:2019dnq}.  Moreover, in \eqref{EqD} we introduced $\bar{q}=\sum_{\substack{1\leq a\leq M\\a\neq i,j}}q_a$ as well as
\eqn{
\begin{gathered}
\bar{n}=\sum_{\substack{1\leq a\leq M\\a\neq i,j,m}}n_a,\qquad\qquad\bar{n}_m=\sum_{\substack{1\leq a\leq M\\a\neq i,j,m}}n_{am},\\
\bar{n}_a=\sum_{\substack{1\leq b\leq M\\b\neq i,j,m,a}}n_{ab},\qquad\qquad\bar{\bar{n}}=\sum_{\substack{1\leq a<b\leq M\\a,b\neq i,j,m}}n_{ab}.
\end{gathered}
}

On the one hand, the action of the OPE is most elementary with the choice of conformal cross-ratios shown above in \eqref{EqCROPE}.  On the other hand, the expressions for conformal blocks may be simpler when written in terms of \textit{a priori} unknown set of cross-ratios.  Once a convenient set of cross-ratios for the conformal blocks has been found one needs to change variables between the different choices for the cross-ratios.  First, it is necessary to transform an $(M-1)$-point correlation function to be given in terms of the cross-ratios that work well with the OPE differential operator.  Then, one acts with with the OPE.  Finally, the result is re-expressed using the cross-ratios that simplify the $M$-point conformal blocks.  This last step implies that we explicitly evaluate series of the hypergeometric type to reach the simplest possible form for the conformal blocks.

%%%%%%%%%%%%%%%%%%%%%%%%%%%%%%%%%%%%%%%%%%%%%%%%%%

\subsection{Scalar \texorpdfstring{$M$}{M}-Point Correlation Functions}

We now focus on the contributions from exchanged scalar quasi-primary operators with conformal dimensions $\Delta_{k_a}$ to $M$-point correlation functions of external scalar quasi-primary operators with conformal dimensions $\Delta_{i_a}$, denoted by
\eqna{
\left.I_{M(\Delta_{k_1},\ldots,\Delta_{k_{M-3}})}^{(\Delta_{i_2},\ldots,\Delta_{i_M},\Delta_{i_1})}\right|_{\text{channel}}&=L_{M|\text{channel}}^{(\Delta_{i_2},\ldots,\Delta_{i_M},\Delta_{i_1})}\left[\prod_{1\leq a\leq M-3}(u_a^M)^{\frac{\Delta_{k_a}}{2}}\right]G_{M|\text{channel}}^{(d,\boldsymbol{h};\boldsymbol{p})}(\boldsymbol{u}^M,\textbf{v}^M).
}[EqI]
Obviously, the full scalar correlation functions are sums of $I_M$ for scalar exchanges as in \eqref{EqI} (the focus of this paper) and $I_M$ for exchanges of quasi-primary operators in non-trivial irreducible representations of the Lorentz group (not discussed here).

In \eqref{EqI}, the quantity $L_M$ corresponds to the external operators, or the ``legs,'' and is a product of $M$ factors of embedding space coordinates responsible for the proper scaling behavior of the correlation functions.  Meanwhile, $G_M$ are the conformal blocks in the channel (or topology) of interest, given by
\eqna{
G_{M|\text{channel}}^{(d,\boldsymbol{h};\boldsymbol{p})}(\boldsymbol{u}^M,\textbf{v}^M)&=\sum_{\{m_a,m_{ab}\}\geq0}C_{M|\text{channel}}^{(d,\boldsymbol{h};\boldsymbol{p})}(\boldsymbol{m},\textbf{m})F_{M|\text{channel}}^{(d,\boldsymbol{h};\boldsymbol{p})}(\boldsymbol{m},\textbf{m})\\
&\phantom{=}\qquad\times\prod_{1\leq a\leq M-3}\frac{(u_a^M)^{m_a}}{m_a!}\prod_{1\leq a\leq b\leq M-3}\frac{(1-v_{ab}^M)^{m_{ab}}}{m_{ab}!}.
}[EqG]

Equations \eqref{EqI} and \eqref{EqG} are functions of a set of conformal cross-ratios denoted by the vector $\boldsymbol{u}^M$ of $u_a^M$ and the matrix $\textbf{v}^M$ of $v_{ab}^M$.  We stress here that the best sets of conformal cross-ratios are not known \textit{a priori} (there are usually more than one) and that they are dependent on the topology.  The scalar conformal blocks \eqref{EqG} are thus expressed as series of the hypergeometric type in powers of the conformal cross-ratios, with extra sums encoded in the functions $F_M$.  Although there is freedom in the division between $C_M$ and $F_M$, we conjectured in \cite{Fortin:2020yjz} that the latter is a function of the vector $\boldsymbol{m}$ of powers of the cross-ratios $\boldsymbol{u}^M$ only, implying it would not depend on the matrix $\textbf{m}$ of powers of the cross-ratios $1-v^M$.  With that in mind, we also construct $F_M$ such that it satisfies interesting symmetry properties related to the symmetries of the associated topology.

Finally, we mention that the vectors of parameters $\boldsymbol{h}$ and $\boldsymbol{p}$ originate from the action of the OPE \eqref{EqOPE}.  They are thus simple linear combinations of the conformal dimensions for the external and internal quasi-primary operators.  This statement can be seen directly by recursion of the OPE, from which it is straightforward to obtain the following identity for some choice of $k$, $l$ and $m$,
\eqn{\left.I_{M(\Delta_{k_1},\ldots,\Delta_{k_{M-3}})}^{(\Delta_{i_2},\ldots,\Delta_{i_M},\Delta_{i_1})}\right|_{\substack{\text{channel}\\\text{$M$ points}}}=\frac{1}{\eta_{1M}^{p_M}}\left(\frac{\eta_{kM}\eta_{lM}}{\eta_{1M}\eta_{kl}}\right)^{h_{M-1}}\bar{\D}_{M1;kl;m}^{2h_{M-1}}\left.I_{M-1(\Delta_{k_1},\ldots,\Delta_{k_{M-4}})}^{(\Delta_{i_2},\ldots,\Delta_{i_{M-1}},\Delta_{k_{M-3}})}\right|_{\substack{\text{channel}\\\text{$M-1$ points}}}.}[EqIfromOPE]
It is important to point out that to reach the desired channel on the LHS of \eqref{EqIfromOPE}, it is necessary to start from some appropriate channels on the RHS of \eqref{EqIfromOPE}.

Before proceeding with the scalar seven-point conformal blocks in the extended snowflake channel, we now state the results for the scalar $M$-point correlation functions in the comb channel as well as the scalar six-point correlation functions in the snowflake channel that were obtained previously by using \eqref{EqIfromOPE}.  While these channels were computed elsewhere, the results will be important for the factorization properties of the extended snowflake channel and for certain general observations made in Section \ref{SecConc}.

%%%%%%%%%%%%%%%%%%%%%%%%%%%%%%%%%%%%%%%%%%%%%%%%%%

\subsection{Scalar \texorpdfstring{$M$}{M}-Point Correlation Functions in the Comb Channel}

The comb channel has the topology depicted in Figure \ref{FigComb}.
\begin{figure}[t]
\centering
\resizebox{13cm}{!}{%
\begin{tikzpicture}[thick]
\begin{scope}
\node at (-2.2,0) {$\left.I_{M(\Delta_{k_1},\ldots,\Delta_{k_{M-3}})}^{(\Delta_{i_2},\ldots,\Delta_{i_M},\Delta_{i_1})}\right|_{\text{comb}}$};
\node at (0,0) {$=$};
\node at (1,0) {$\mathcal{O}_{i_2}$};
\draw[-] (1.5,0)--(7.5,0);
\node at (8.1,0) {$\mathcal{O}_{i_1}$};
\draw[-] (2.5,0)--(2.5,1) node[above] {$\mathcal{O}_{i_3}$};
\draw[-] (3.5,0)--(3.5,1) node[above] {$\mathcal{O}_{i_4}$};
\node at (5,0.5) {$\ldots$};
\draw[-] (6.5,0)--(6.5,1) node[above] {$\mathcal{O}_{i_M}$};
\node at (3,-0.5) {$\mathcal{O}_{k_1}$};
\end{scope}
\end{tikzpicture}
}
\caption{Topology of scalar $M$-point conformal blocks in the comb channel.}
\label{FigComb}
\end{figure}
Starting from the scalar four-point correlation functions and using \eqref{EqIfromOPE}, we found in \cite{Fortin:2019zkm} that the $M$-point correlation functions in the comb channel can be expressed as
\eqn{
\begin{gathered}
L_{M|\text{comb}}^{(\Delta_{i_2},\ldots,\Delta_{i_M},\Delta_{i_1})}=\left(\frac{\eta_{34}}{\eta_{23}\eta_{24}}\right)^{\frac{\Delta_{i_2}}{2}}\left[\prod_{1\leq a\leq M-2}\left(\frac{\eta_{a+1,a+3}}{\eta_{a+1,a+2}\eta_{a+2,a+3}}\right)^{\frac{\Delta_{i_{a+2}}}{2}}\right]\left(\frac{\eta_{M-1,M}}{\eta_{1,M-1}\eta_{1M}}\right)^{\frac{\Delta_{i_1}}{2}},\\
u_a^M=\frac{\eta_{1+a,2+a}\eta_{3+a,4+a}}{\eta_{1+a,3+a}\eta_{2+a,4+a}},\qquad v_{ab}^M=\frac{\eta_{2-a+b,4+b}}{\eta_{2+b,4+b}}\prod_{1\leq c\leq a}\frac{\eta_{3+b-c,4+b-c}}{\eta_{2+b-c,4+b-c}},
\end{gathered}
}[EqCBCRcomb]
with
\eqn{
\begin{gathered}
C_{M|\text{comb}}^{(d,\boldsymbol{h};\boldsymbol{p})}(\boldsymbol{m},\textbf{m})=\frac{(p_3)_{m_1+\text{tr}_0\textbf{m}}(p_2+h_2)_{m_1+\text{tr}_1\textbf{m}}}{(p_3)_{m_1+\text{tr}_1\textbf{m}}}\left[\prod_{1\leq a\leq M-3}\frac{(\bar{p}_{a+2}+\bar{h}_{a+2})_{m_a+m_{a+1}+\bar{m}_a+\bar{\bar{m}}_a}}{(\bar{p}_{a+2}+\bar{h}_{a+1})_{2m_a+\bar{m}_{a-1}+\bar{m}_a+\bar{\bar{m}}_a}}\right.\\
\phantom{=}\qquad\qquad\qquad\left.\times(p_{a+2}-m_{a-1})_{m_a+\text{tr}_a\textbf{m}}\frac{(-h_{a+2})_{m_a}(-h_{a+2}+m_a-m_{a+1})_{\bar{m}_{a-1}}}{(\bar{p}_{a+2}+\bar{h}_{a+1}+1-d/2)_{m_a}}\right],\\
\text{tr}_a\textbf{m}=\sum_bm_{b,a+b},\qquad\qquad\bar{m}_a=\sum_{b\leq a}m_{ba},\qquad\qquad\bar{\bar{m}}_a=\sum_{b>a}(\bar{m}_b-\text{tr}_b\textbf{m}),
\end{gathered}
}[EqCBCcomb]
and
\eqn{F_{M|\text{comb}}^{(d,\boldsymbol{h};\boldsymbol{p})}(\boldsymbol{m})=\prod_{1\leq a\leq M-4}{}_3F_2\left[\begin{array}{c}-m_a,-m_{a+1},-\bar{p}_{a+2}-\bar{h}_{a+1}+d/2-m_a\\p_{a+3}-m_a,h_{a+2}+1-m_a\end{array};1\right].}[EqCBFcomb]
We define $\bar{p}_a=\sum_{b=2}^ap_b$ and $\bar{h}_a=\sum_{b=2}^ah_b$, while the explicit expressions for the vectors $\boldsymbol{h}$ and $\boldsymbol{p}$ in the comb channel are given by
\eqn{
\begin{gathered}
2h_2=\Delta_{k_1}-\Delta_{i_2}-\Delta_{i_3},\qquad2h_a=\Delta_{k_{a-1}}-\Delta_{k_{a-2}}-\Delta_{i_{a+1}},\\
p_2=\Delta_{i_3},\qquad2p_3=\Delta_{i_2}+\Delta_{k_1}-\Delta_{i_3},\qquad2p_a=\Delta_{i_a}+\Delta_{k_{a-2}}-\Delta_{k_{a-3}},
\end{gathered}
}[EqCBhpcomb]
where $k_{M-2}\equiv i_1$.

As mentioned earlier, the set of conformal cross-ratios is not unique, and \eqref{EqCBCRcomb} was chosen based on the OPE limit while other choice are possible.  Moreover, we point out that $F_M$ in the comb channel is a product of $M-4$ hypergeometric functions.  The factorization property of $F_{M|\text{comb}}$ will play an important role later.

%%%%%%%%%%%%%%%%%%%%%%%%%%%%%%%%%%%%%%%%%%%%%%%%%%

\subsection{Scalar Six-Point Correlation Functions in the Snowflake Channel}

Starting at six points, there are more topologies than the comb channel.
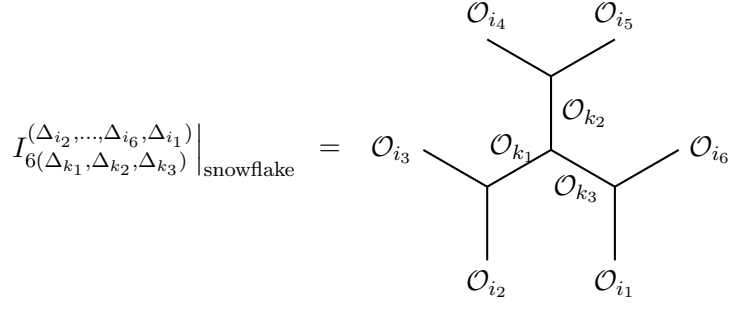
\begin{figure}[t]
\centering
\resizebox{10cm}{!}{%
\begin{tikzpicture}[thick]
\begin{scope}
\node at (-1.4,0) {$\left.I_{6(\Delta_{k_1},\Delta_{k_2},\Delta_{k_3})}^{(\Delta_{i_2},\ldots,\Delta_{i_6},\Delta_{i_1})}\right|_{\text{snowflake}}$};
\node at (1,0) {$=$};
\draw[-] (4,0)--+(-150:1) node[pos=0.6,above]{$\mathcal{O}_{k_1}$};
\draw[-] (4,0)++(-150:1)--+(-90:1) node[below]{$\mathcal{O}_{i_2}$};
\draw[-] (4,0)++(-150:1)--+(150:1) node[left]{$\mathcal{O}_{i_3}$};
\draw[-] (4,0)--+(90:1) node[pos=0.5,right]{$\mathcal{O}_{k_2}$};
\draw[-] (4,0)++(90:1)--+(30:1) node[above]{$\mathcal{O}_{i_5}$};
\draw[-] (4,0)++(90:1)--+(150:1) node[above]{$\mathcal{O}_{i_4}$};
\draw[-] (4,0)--+(-30:1) node[pos=0.4,below]{$\mathcal{O}_{k_3}$};
\draw[-] (4,0)++(-30:1)--+(30:1) node[right]{$\mathcal{O}_{i_6}$};
\draw[-] (4,0)++(-30:1)--+(-90:1) node[below]{$\mathcal{O}_{i_1}$};
\end{scope}
\end{tikzpicture}
}
\caption{Topology of the six-point conformal blocks in the snowflake channel.}
\label{Fig6pt}
\end{figure}
For six-point correlation functions, the remaining topology is the snowflake channel, depicted in Figure \ref{Fig6pt} and studied in \cite{Fortin:2020yjz}.

Using the known five-point correlation functions in the comb channel and applying the OPE as in \eqref{EqIfromOPE} on the appropriate quasi-primary operator, we generate the scalar six-point correlation functions in the snowflake channel for which their decomposition as in \eqref{EqI} and \eqref{EqG} is
\eqn{
\begin{gathered}
L_{6|\text{snowflake}}^{(\Delta_{i_2},\ldots,\Delta_{i_6},\Delta_{i_1})}=\left(\frac{\eta_{13}}{\eta_{12}\eta_{23}}\right)^{\frac{\Delta_{i_2}}{2}}\left(\frac{\eta_{12}}{\eta_{13}\eta_{23}}\right)^{\frac{\Delta_{i_3}}{2}}\left(\frac{\eta_{35}}{\eta_{34}\eta_{45}}\right)^{\frac{\Delta_{i_4}}{2}}\\
\qquad\qquad\qquad\qquad\qquad\qquad\times\left(\frac{\eta_{34}}{\eta_{35}\eta_{45}}\right)^{\frac{\Delta_{i_5}}{2}}\left(\frac{\eta_{15}}{\eta_{16}\eta_{56}}\right)^{\frac{\Delta_{i_6}}{2}}\left(\frac{\eta_{56}}{\eta_{15}\eta_{16}}\right)^{\frac{\Delta_{i_1}}{2}},\\
u_1^6=\frac{\eta_{15}\eta_{23}}{\eta_{12}\eta_{35}},\qquad u_2^6=\frac{\eta_{13}\eta_{45}}{\eta_{15}\eta_{34}},\qquad u_3^6=\frac{\eta_{16}\eta_{35}}{\eta_{13}\eta_{56}},\\
v_{11}^6=\frac{\eta_{13}\eta_{25}}{\eta_{12}\eta_{35}},\qquad v_{12}^6=\frac{\eta_{14}\eta_{35}}{\eta_{15}\eta_{34}},\qquad v_{22}^6=\frac{\eta_{13}\eta_{24}}{\eta_{12}\eta_{34}},\\
v_{13}^6=\frac{\eta_{15}\eta_{26}}{\eta_{12}\eta_{56}},\qquad v_{23}^6=\frac{\eta_{15}\eta_{36}}{\eta_{13}\eta_{56}},\qquad v_{33}^6=\frac{\eta_{35}\eta_{46}}{\eta_{34}\eta_{56}},
\end{gathered}
}[EqCB6CR]
with
\eqna{
&C_{6|\text{snowflake}}^{(d,\boldsymbol{h};\boldsymbol{p})}(\boldsymbol{m},\textbf{m})\\
&\qquad=\frac{(p_2+h_3)_{m_1+m_{23}}(p_3)_{-m_1+m_2+m_3+m_{12}+m_{33}}(-h_3)_{m_1+m_{11}+m_{22}+m_{13}}}{(p_2)_{2m_1+m_{11}+m_{13}+m_{22}+m_{23}}(p_2+1-d/2)_{m_1}}\\
&\qquad\phantom{=}\times\frac{(p_3-h_2+h_4)_{m_2+m_{11}}(p_2+h_2)_{m_1-m_2+m_3+m_{13}+m_{23}}(-h_4)_{m_2+m_{12}+m_{22}+m_{33}}}{(p_3-h_2)_{2m_2+m_{11}+m_{12}+m_{22}+m_{33}}(p_3-h_2+1-d/2)_{m_2}}\\
&\qquad\phantom{=}\times\frac{(\bar{p}_3+h_2+h_5)_{m_3+m_{12}}(-h_2)_{m_1+m_2-m_3+m_{11}+m_{22}}(-h_5)_{m_3+m_{13}+m_{23}+m_{33}}}{(\bar{p}_3+h_2)_{2m_3+m_{12}+m_{13}+m_{23}+m_{33}}(\bar{p}_3+h_2+1-d/2)_{m_3}},
}[EqCB6C]
and
\eqna{
F_{6|\text{snowflake}}^{(d,\boldsymbol{h};\boldsymbol{p})}(\boldsymbol{m})&=\frac{1}{(p_3)_{-m_1}(-p_3+h_2+d/2)_{-m_2}(-h_2)_{-m_3}}\\
&\phantom{=}\qquad\times F_{2,1,0}^{1,3,2}\left[\left.\begin{array}{c}\bar{p}_3-d/2;-m_2,-h_2,p_3;-m_1,-m_3\\-h_2-m_3,p_3-m_1;p_3-h_2+1-d/2;-\end{array}\right|1,1\right]\\
&=\frac{1}{(p_3)_{-m_1}(-p_3+h_2+d/2)_{-m_2}(-\bar{p}_3-h_2+d/2)_{-m_3}}\\
&\phantom{=}\qquad\times F_{1,1,1}^{2,1,1}\left[\left.\begin{array}{c}\bar{p}_3-d/2,p_3;-m_2;-m_3\\p_3-m_1;p_3-h_2+1-d/2;\bar{p}_3+h_2+1-d/2\end{array}\right|1,1\right].
}[EqCB6F]
where the vectors $\boldsymbol{h}$ and $\boldsymbol{p}$ are given by
\eqn{
\begin{gathered}
2h_2=\Delta_{k_3}-\Delta_{k_2}-\Delta_{k_1},\qquad2h_3=\Delta_{i_3}-\Delta_{i_2}-\Delta_{k_1},\\
2h_4=\Delta_{i_5}-\Delta_{i_4}-\Delta_{k_2},\qquad2h_5=\Delta_{i_1}-\Delta_{i_6}-\Delta_{k_3},\\
p_2=\Delta_{k_1},\qquad2p_3=\Delta_{k_2}+\Delta_{k_3}-\Delta_{k_1},\qquad2p_4=\Delta_{i_2}+\Delta_{i_3}-\Delta_{k_1},\\
2p_5=\Delta_{i_4}+\Delta_{i_5}-\Delta_{k_2},\qquad2p_6=\Delta_{i_6}+\Delta_{i_1}-\Delta_{k_3}.
\end{gathered}
}[EqCB6hp]

Contrary to the comb channel where the extra sums in $F_{M|\text{comb}}$ factorize, the two extra sums appearing in $F_{6|\text{snowflake}}$ do not factorize.  In fact, \eqref{EqCB6F} is written in terms of Kamp\'e de F\'eriet functions \cite{exton1976multiple,srivastava1985multiple}, which are defined as
\eqn{F_{q,s,v}^{p,r,u}\left[\left.\begin{array}{c}\boldsymbol{a};\boldsymbol{c};\boldsymbol{f}\\\boldsymbol{b};\boldsymbol{d};\boldsymbol{g}\end{array}\right|x,y\right]=\sum_{m,n\geq0}\frac{(\boldsymbol{a})_{m+n}(\boldsymbol{c})_m(\boldsymbol{f})_n}{(\boldsymbol{b})_{m+n}(\boldsymbol{d})_m(\boldsymbol{g})_n}\frac{x^my^n}{m!n!},}[EqKdF]
for
\eqn{
\begin{gathered}
(\boldsymbol{a})_{m+n}=(a_1)_{m+n}\cdots(a_p)_{m+n},\qquad(\boldsymbol{b})_{m+n}=(b_1)_{m+n}\cdots(b_q)_{m+n},\\
(\boldsymbol{c})_m=(c_1)_m\cdots(c_r)_m,\qquad(\boldsymbol{d})_m=(d_1)_m\cdots(d_s)_m,\\
(\boldsymbol{f})_n=(f_1)_n\cdots(f_u)_n,\qquad(\boldsymbol{g})_n=(g_1)_n\cdots(g_v)_n.
\end{gathered}
}

We now turn to the scalar seven-point correlation functions in the extended snowflake channel.

%%%%%%%%%%%%%%%%%%%%%%%%%%%%%%%%%%%%%%%%%%%%%%%%%%

\subsection{Scalar Seven-Point Correlation Functions in the Extended Snowflake Channel}

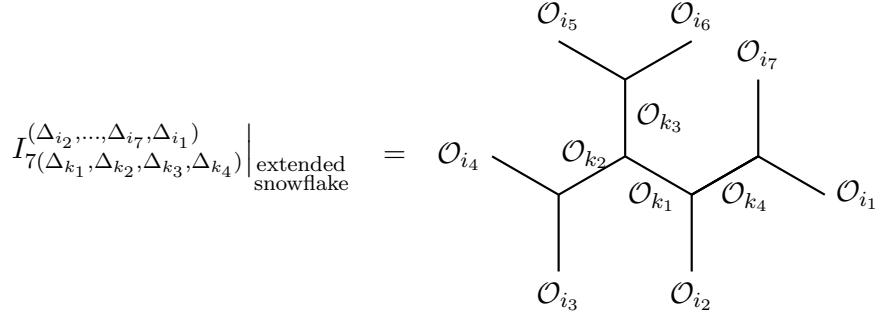
\begin{figure}[t]
\centering
\resizebox{12cm}{!}{%
\begin{tikzpicture}[thick]
\begin{scope}
\node at (-1.8,0) {$\left.I_{7(\Delta_{k_1},\Delta_{k_2},\Delta_{k_3},\Delta_{k_4})}^{(\Delta_{i_2},\ldots,\Delta_{i_7},\Delta_{i_1})}\right|_{\substack{\text{extended}\\\text{snowflake}}}$};
\node at (1,0) {$=$};
\draw[-] (4,0)--+(-150:1) node[pos=0.6,above]{$\mathcal{O}_{k_2}$};
\draw[-] (4,0)++(-150:1)--+(-90:1) node[below]{$\mathcal{O}_{i_3}$};
\draw[-] (4,0)++(-150:1)--+(150:1) node[left]{$\mathcal{O}_{i_4}$};
\draw[-] (4,0)--+(90:1) node[pos=0.5,right]{$\mathcal{O}_{k_3}$};
\draw[-] (4,0)++(90:1)--+(30:1) node[above]{$\mathcal{O}_{i_6}$};
\draw[-] (4,0)++(90:1)--+(150:1) node[above]{$\mathcal{O}_{i_5}$};
\draw[-] (4,0)--+(-30:1) node[pos=0.4,below]{$\mathcal{O}_{k_1}$};
\draw[-] (4,0)++(-30:1)--+(-90:1) node[below]{$\mathcal{O}_{i_2}$};
\draw[-] (4,0)++(-30:1)--+(30:1) node[pos=0.8,yshift=-0.1cm,below]{$\mathcal{O}_{k_4}$};
\draw[-] (4,0)++(-30:1)--++(30:1)--+(90:1) node[above]{$\mathcal{O}_{i_7}$};
\draw[-] (4,0)++(-30:1)--++(30:1)--+(-30:1) node[right]{$\mathcal{O}_{i_1}$};
\end{scope}
\end{tikzpicture}
}
\caption{Topology of the seven-point conformal blocks in the extended snowflake channel.}
\label{Fig7pt}
\end{figure}

As for six-point correlation functions, there are two different topologies for seven-point correlation functions: the comb channel discussed before and the extended snowflake channel illustrated in Figure \ref{Fig7pt}.
The scalar seven-point correlation functions in the extended snowflake channel can be computed from the scalar six-point correlation functions in both channels.

Starting from the comb channel, it is necessary to act with the OPE on one of the two quasi-primary operators inside the comb, \textit{i.e.} on $\mathcal{O}_{i_4}$ or $\mathcal{O}_{i_5}$ in Figure \ref{FigComb} (acting on any one of the other quasi-primary operators leads to the seven-point comb channel).  On the contrary, due to the symmetry properties of the snowflake channel seen in Figure \ref{Fig6pt}, the OPE can be applied anywhere on the snowflake to reach the extended snowflake.  In Appendix \ref{SAppCB7} we follow the second route to compute the scalar seven-point conformal blocks in the extended snowflake channel.

To summarize, we start from the scalar six-point correlation functions in the snowflake channel \eqref{EqCB6C}, redefine the quasi-primary operators such that $\mathcal{O}_{i_a}(\eta_a)\to\mathcal{O}_{i_{a-1}}(\eta_{a-1})$ with the knowledge that $\mathcal{O}_{i_0}(\eta_0)\equiv\mathcal{O}_{i_6}(\eta_6)$, and finally apply \eqref{EqIfromOPE} with $k=5$, $l=6$ and $m=6$, and we obtain
\eqn{
\begin{gathered}
L_{7|\substack{\text{extended}\\\text{snowflake}}}^{(\Delta_{i_2},\ldots,\Delta_{i_7},\Delta_{i_1})}=\left(\frac{\eta_{16}}{\eta_{12}\eta_{26}}\right)^{\frac{\Delta_{i_2}}{2}}\left(\frac{\eta_{24}}{\eta_{23}\eta_{34}}\right)^{\frac{\Delta_{i_3}}{2}}\left(\frac{\eta_{23}}{\eta_{24}\eta_{34}}\right)^{\frac{\Delta_{i_4}}{2}}\qquad\qquad\qquad\qquad\\
\qquad\qquad\qquad\qquad\qquad\times\left(\frac{\eta_{46}}{\eta_{45}\eta_{56}}\right)^{\frac{\Delta_{i_5}}{2}}\left(\frac{\eta_{45}}{\eta_{46}\eta_{56}}\right)^{\frac{\Delta_{i_6}}{2}}\left(\frac{\eta_{16}}{\eta_{17}\eta_{67}}\right)^{\frac{\Delta_{i_7}}{2}}\left(\frac{\eta_{67}}{\eta_{16}\eta_{17}}\right)^{\frac{\Delta_{i_1}}{2}},\\
u_1^7=\frac{\eta_{12}\eta_{46}}{\eta_{16}\eta_{24}},\qquad u_2^7=\frac{\eta_{26}\eta_{34}}{\eta_{23}\eta_{46}},\qquad u_3^7=\frac{\eta_{24}\eta_{56}}{\eta_{26}\eta_{45}},\qquad u_4^7=\frac{\eta_{17}\eta_{26}}{\eta_{12}\eta_{67}},\\
v_{11}^7=\frac{\eta_{14}\eta_{26}}{\eta_{16}\eta_{24}},\qquad v_{12}^7=\frac{\eta_{24}\eta_{36}}{\eta_{23}\eta_{46}},\qquad v_{13}^7=\frac{\eta_{15}\eta_{46}}{\eta_{16}\eta_{45}},\qquad v_{14}^7=\frac{\eta_{16}\eta_{27}}{\eta_{12}\eta_{67}},\\
v_{22}^7=\frac{\eta_{13}\eta_{26}}{\eta_{16}\eta_{23}},\qquad v_{23}^7=\frac{\eta_{25}\eta_{46}}{\eta_{26}\eta_{45}},\qquad v_{24}^7=\frac{\eta_{26}\eta_{37}}{\eta_{23}\eta_{67}},\qquad v_{33}^7=\frac{\eta_{24}\eta_{35}}{\eta_{23}\eta_{45}},\\
v_{34}^7=\frac{\eta_{26}\eta_{47}}{\eta_{24}\eta_{67}},\qquad v_{44}^7=\frac{\eta_{46}\eta_{57}}{\eta_{45}\eta_{67}},
\end{gathered}
}[EqCB7CR]
with
\eqna{
&C_{7|\substack{\text{extended}\\\text{snowflake}}}^{(d,\boldsymbol{h};\boldsymbol{p})}(\boldsymbol{m},\textbf{m})\\
&\qquad=\frac{(p_2+h_3)_{m_1-m_4+m_{23}}(p_3)_{-m_1+m_2+m_3+m_{12}+m_{33}}(-h_3)_{m_1+m_4+m_{11}+m_{13}+m_{22}+m_{24}+m_{34}+m_{44}}}{(p_2)_{2m_1+m_{11}+m_{13}+m_{22}+m_{23}+m_{24}+m_{34}+m_{44}}(p_2+1-d/2)_{m_1}}\\
&\qquad\phantom{=}\times\frac{(p_3-h_2+h_4)_{m_{2}+m_{11}+m_{34}}(p_2+h_2)_{m_1-m_2+m_3+m_{13}+m_{23}+m_{44}}(-h_4)_{m_2+m_{12}+m_{22}+m_{24}+m_{33}}}{(p_3-h_2)_{2m_2+m_{11}+m_{12}+m_{22}+m_{24}+m_{33}+m_{34}}(p_3-h_2+1-d/2)_{m_2}}\\
&\qquad\phantom{=}\times\frac{(\bar{p}_3+h_2+h_5)_{m_3+m_{12}}(-h_2)_{m_1+m_2-m_3+m_{11}+m_{22}+m_{24}+m_{34}}(-h_5)_{m_3+m_{13}+m_{23}+m_{33}+m_{44}}}{(\bar{p}_3+h_2)_{2m_3+m_{12}+m_{13}+m_{23}+m_{33}+m_{44}}(\bar{p}_3+h_2+1-d/2)_{m_3}}\\
&\qquad\phantom{=}\times\frac{(p_4-m_1)_{m_4+m_{14}}(-h_6)_{m_4+m_{14}+m_{24}+m_{34}+m_{44}}(p_4-h_3+h_6)_{m_4+m_{11}+m_{13}+m_{22}}}{(p_2+h_3+m_1)_{-m_4}(p_4-h_3)_{2m_4+m_{11}+m_{13}+m_{14}+m_{22}+m_{24}+m_{34}+m_{44}}(p_4-h_3+1-d/2)_{m_4}},
}[EqCB7C]
as well as
\eqna{
F_{7|\substack{\text{extended}\\\text{snowflake}}}^{(d,\boldsymbol{h};\boldsymbol{p})}(\boldsymbol{m})&=\frac{1}{(p_3)_{-m_1}(-p_3+h_2+d/2)_{-m_2}(-\bar{p}_3-h_2+d/2)_{-m_3}}\\
&\phantom{=}\qquad\times F_{1,1,1}^{2,1,1}\left[\left.\begin{array}{c}\bar{p}_3-d/2,p_3;-m_2;-m_3\\p_3-m_1;p_3-h_2+1-d/2;\bar{p}_3+h_2+1-d/2\end{array}\right|1,1\right]\\
&\phantom{=}\qquad\times{}_3F_2\left[\begin{array}{c}-m_1,-m_4,-p_2+d/2-m_1\\p_4-m_1,1-p_2-h_3-m_1\end{array};1\right],
}[EqCB7F]
and finally
\eqn{
\begin{gathered}
2h_2=\Delta_{k_3}-\Delta_{k_2}-\Delta_{k_1},\qquad2h_3=\Delta_{i_2}-\Delta_{k_4}-\Delta_{k_1},\qquad2h_4=\Delta_{i_4}-\Delta_{i_3}-\Delta_{k_2},\\
2h_5=\Delta_{i_6}-\Delta_{i_5}-\Delta_{k_3},\qquad2h_6=\Delta_{i_1}-\Delta_{i_7}-\Delta_{k_4},\\
p_2=\Delta_{k_1},\qquad2p_3=\Delta_{k_2}+\Delta_{k_3}-\Delta_{k_1},\qquad2p_4=\Delta_{k_4}+\Delta_{i_2}-\Delta_{k_1},\\
2p_5=\Delta_{i_3}+\Delta_{i_4}-\Delta_{k_2},\qquad2p_6=\Delta_{i_5}+\Delta_{i_6}-\Delta_{k_3},\qquad2p_7=\Delta_{i_7}+\Delta_{i_1}-\Delta_{k_4}.
\end{gathered}
}[EqCB7hp]

As conjectured and argued for based on consistency under several limits in \cite{Fortin:2019zkm,Fortin:2020yjz}, the number of extra sums for scalar seven-point conformal blocks [see \eqref{EqCBFcomb} with $M=7$ and \eqref{EqCB7F}] is three.  Moreover, a comparison between \eqref{EqCB7F} on one side and $\eqref{EqCBFcomb}$ and $\eqref{EqCB6F}$ on the other side shows that $F_{7|\substack{\text{extended}\\\text{snowflake}}}=F_{6|\text{snowflake}}F_{5|\text{comb}}$, a factorization which is reminiscent of the factorization seen in the comb channel.  We will come back to this interesting observation in Section \ref{SecConc}.

The following sections analyse the extended snowflake results \eqref{EqCB7CR}, \eqref{EqCB7C}, \eqref{EqCB7F}, and \eqref{EqCB7hp}.  We first verify that the symmetry properties of the extended snowflake topology extend to the scalar conformal blocks $G_7$ \eqref{EqCB7C} and \eqref{EqCB7F}.  We then check that the scalar conformal blocks in the extended snowflake channel exhibit the proper behavior under the OPE limit and the limit of unit operator.  All proofs are left for Appendixes \ref{SAppSym} and \ref{SAppLim}.

%%%%%%%%%%%%%%%%%%%%%%%%%%%%%%%%%%%%%%%%%%%%%%%%%%
%%%%%%%%%%%%%%%%%%%%%%%%%%%%%%%%%%%%%%%%%%%%%%%%%%

\section{Sanity Checks}\label{SecChecks}

Conformal blocks must satisfy several interesting properties.  In this section, we verify that the symmetries of the scalar seven-point conformal blocks in the extended snowflake channel coincide with the symmetry group $H_{7|\substack{\text{extended}\\\text{snowflake}}}=\mathbb{Z}_2\times\left((\mathbb{Z}_2)^2\rtimes\mathbb{Z}_2\right)$ of the extended snowflake topology.  We also check that the scalar seven-point correlation functions reduce to the corresponding scalar six-point correlation functions under the OPE limit and the limit of unit operator.

%%%%%%%%%%%%%%%%%%%%%%%%%%%%%%%%%%%%%%%%%%%%%%%%%%

\subsection{Symmetry Properties}

For a given topology, there exists an associated symmetry group which we denote by $H_{M|\text{channel}}$.  Each element of the symmetry group is a symmetry transformation of the topology, \textit{i.e.} acting with a symmetry element transforms the topology back to itself.  As a consequence, the scalar conformal blocks must verify several identities related to the symmetry group $H_{M|\text{channel}}$ where the conformal cross-ratios and the vectors $\boldsymbol{h}$ and $\boldsymbol{p}$ are transformed according to the symmetries.  All the identities can be generated from a smaller subset of identities corresponding to the subset of generators of the symmetry group of the associated topology.

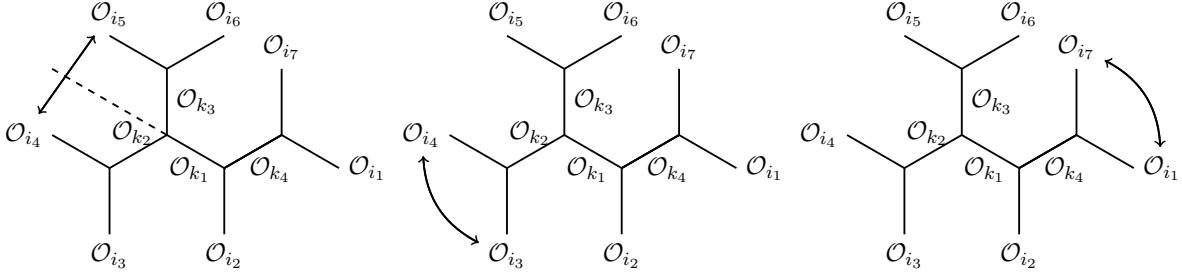
\begin{figure}[t]
\centering
\resizebox{16cm}{!}{%
\begin{tikzpicture}[thick]
\begin{scope}[xshift=0cm]
\draw[-] (4,0)--+(-150:1) node[pos=0.6,above]{$\mathcal{O}_{k_2}$};
\draw[-] (4,0)++(-150:1)--+(-90:1) node[below] {$\mathcal{O}_{i_3}$};
\draw[-] (4,0)++(-150:1)--+(150:1) node[left] (1) {$\mathcal{O}_{i_4}$};
\draw[-] (4,0)--+(90:1) node[pos=0.5,right]{$\mathcal{O}_{k_3}$};
\draw[-] (4,0)++(90:1)--+(30:1) node[above]{$\mathcal{O}_{i_6}$};
\draw[-] (4,0)++(90:1)--+(150:1) node[above] (2) {$\mathcal{O}_{i_5}$};
\draw[-] (4,0)--+(-30:1) node[pos=0.4,below]{$\mathcal{O}_{k_1}$};
\draw[-] (4,0)++(-30:1)--+(-90:1) node[below]{$\mathcal{O}_{i_2}$};
\draw[-] (4,0)++(-30:1)--+(30:1) node[pos=0.8,yshift=-0.1cm,below]{$\mathcal{O}_{k_4}$};
\draw[-] (4,0)++(-30:1)--++(30:1)--+(90:1) node[above]{$\mathcal{O}_{i_7}$};
\draw[-] (4,0)++(-30:1)--++(30:1)--+(-30:1) node[right]{$\mathcal{O}_{i_1}$};
%\draw[dashed] (4,0)--+(-30:2);
\draw[dashed] (4,0)--+(150:2);
\path[->]
(1) edge node[left]{} (2)
(2) edge node[right]{} (1);
\end{scope}
\begin{scope}[xshift=6cm]
\draw[-] (4,0)--+(-150:1) node[pos=0.6,above]{$\mathcal{O}_{k_2}$};
\draw[-] (4,0)++(-150:1)--+(-90:1) node[below] (1) {$\mathcal{O}_{i_3}$};
\draw[-] (4,0)++(-150:1)--+(150:1) node[left] (2) {$\mathcal{O}_{i_4}$};
\draw[-] (4,0)--+(90:1) node[pos=0.5,right]{$\mathcal{O}_{k_3}$};
\draw[-] (4,0)++(90:1)--+(30:1) node[above]{$\mathcal{O}_{i_6}$};
\draw[-] (4,0)++(90:1)--+(150:1) node[above]{$\mathcal{O}_{i_5}$};
\draw[-] (4,0)--+(-30:1) node[pos=0.4,below]{$\mathcal{O}_{k_1}$};
\draw[-] (4,0)++(-30:1)--+(-90:1) node[below]{$\mathcal{O}_{i_2}$};
\draw[-] (4,0)++(-30:1)--+(30:1) node[pos=0.8,yshift=-0.1cm,below]{$\mathcal{O}_{k_4}$};
\draw[-] (4,0)++(-30:1)--++(30:1)--+(90:1) node[above]{$\mathcal{O}_{i_7}$};
\draw[-] (4,0)++(-30:1)--++(30:1)--+(-30:1) node[right]{$\mathcal{O}_{i_1}$};
\path[->]
(1) edge[bend left] node[right]{} (2)
(2) edge[bend right] node[left]{} (1);
\end{scope}
\begin{scope}[xshift=12cm]
\draw[-] (4,0)--+(-150:1) node[pos=0.6,above]{$\mathcal{O}_{k_2}$};
\draw[-] (4,0)++(-150:1)--+(-90:1) node[below]{$\mathcal{O}_{i_3}$};
\draw[-] (4,0)++(-150:1)--+(150:1) node[left]{$\mathcal{O}_{i_4}$};
\draw[-] (4,0)--+(90:1) node[pos=0.5,right]{$\mathcal{O}_{k_3}$};
\draw[-] (4,0)++(90:1)--+(30:1) node[above]{$\mathcal{O}_{i_6}$};
\draw[-] (4,0)++(90:1)--+(150:1) node[above]{$\mathcal{O}_{i_5}$};
\draw[-] (4,0)--+(-30:1) node[pos=0.4,below]{$\mathcal{O}_{k_1}$};
\draw[-] (4,0)++(-30:1)--+(-90:1) node[below]{$\mathcal{O}_{i_2}$};
\draw[-] (4,0)++(-30:1)--+(30:1) node[pos=0.8,yshift=-0.1cm,below]{$\mathcal{O}_{k_4}$};
\draw[-] (4,0)++(-30:1)--++(30:1)--+(90:1) node[above] (1) {$\mathcal{O}_{i_7}$};
\draw[-] (4,0)++(-30:1)--++(30:1)--+(-30:1) node[right] (2) {$\mathcal{O}_{i_1}$};
\path[->]
(1) edge[bend left] node[right]{} (2)
(2) edge[bend right] node[left]{} (1);
\end{scope}
\end{tikzpicture}
}
\caption{Symmetries of the scalar seven-point conformal blocks in the extended snowflake channel.  The figure shows the three generators with reflection (left), dendrite permutation of the first kind (middle), and dendrite permutation of the second kind (right).}
\label{FigSym}
\end{figure}

For seven-point correlation functions in the extended snowflake channel, the symmetry group is $H_{7|\substack{\text{extended}\\\text{snowflake}}}=\mathbb{Z}_2\times\left((\mathbb{Z}_2)^2\rtimes\mathbb{Z}_2\right)$ since the extended snowflake topology is invariant under the three generators shown in Figure \ref{FigSym}.
The first $\mathbb{Z}_2$ subgroup corresponds to dendrite (or OPE) permutations of the second kind, \textit{i.e.} permutations $\mathcal{O}_{i_1}\leftrightarrow\mathcal{O}_{i_7}$.  The second and third cyclic groups, denoted by $(\mathbb{Z}_2)^2$ in the symmetry group $H_{7|\substack{\text{extended}\\\text{snowflake}}}$, correspond to dendrite permutations of the first kind, \textit{i.e.} permutations $\mathcal{O}_{i_3}\leftrightarrow\mathcal{O}_{i_4}$ as well as $\mathcal{O}_{i_5}\leftrightarrow\mathcal{O}_{i_6}$.  Finally, the last $\mathbb{Z}_2$ subgroup corresponds to simultaneous reflections $\mathcal{O}_{i_3}\leftrightarrow\mathcal{O}_{i_5}$ and $\mathcal{O}_{i_4}\leftrightarrow\mathcal{O}_{i_6}$.  The latter can be seen as an internal dendrite permutation, and its effect is to switch the two external dendrites associated to the $(\mathbb{Z}_2)^2$ subgroup.  This observation implies the semi-direct nature of the product $(\mathbb{Z}_2)^2\rtimes\mathbb{Z}_2$ and distinguishes between the two different kinds (first and second) of external dendrite permutations.

Since the order of the symmetry group for the extended snowflake topology is $|H_{7|\substack{\text{extended}\\\text{snowflake}}}|=16$, there are sixteen equivalent ways of writing the contributions to the scalar seven-point correlation functions $I_7$.  Hence, there are fifteen identities that the scalar seven-point conformal blocks in the extended snowflake channel $G_7$ must verify.  We now present the three identities associated to the three generators of Figure \ref{FigSym} from which all remaining identities can be obtained.  The proofs are left to Appendix \ref{SAppSym}.

For the reflection generator, where we choose
\eqn{\mathcal{O}_{i_3}(\eta_3)\leftrightarrow\mathcal{O}_{i_5}(\eta_5),\qquad\mathcal{O}_{i_4}(\eta_4)\leftrightarrow\mathcal{O}_{i_6}(\eta_6),\qquad\Delta_{k_2}\leftrightarrow\Delta_{k_3},}
the legs and conformal cross-ratios \eqref{EqCB7CR} transform as
\eqn{
\begin{gathered}
L_7\prod_{1\leq a\leq4}(u_a^7)^{\frac{\Delta_{k_a}}{2}}\to(v_{11}^7)^{h_3-h_6}(v_{12}^7)^{h_4}(v_{23}^7)^{h_5}(v_{34}^7)^{h_6}L_7\prod_{1\leq a\leq4}(u_a^7)^{\frac{\Delta_{k_a}}{2}},\\
u_1^7\to\frac{u_1^7}{v_{11}^7},\qquad u_2^7\to\frac{u_3^7}{v_{23}^7},\qquad u_3^7\to\frac{u_2^7}{v_{12}^7},\qquad u_4^7\to\frac{u_4^7}{v_{34}^7},\\
v_{11}^7\to\frac{1}{v_{11}^7},\qquad v_{12}^7\to\frac{1}{v_{23}^7},\qquad v_{13}^7\to\frac{v_{22}^7}{v_{11}^7v_{12}^7},\qquad v_{14}^7\to\frac{v_{11}^7v_{14}^7}{v_{34}^7},\\
v_{22}^7\to\frac{v_{13}^7}{v_{11}^7v_{23}^7},\qquad v_{23}^7\to\frac{1}{v_{12}^7},\qquad v_{24}^7\to\frac{v_{44}^7}{v_{23}^7v_{34}^7},\qquad v_{33}^7\to\frac{v_{33}^7}{v_{12}^7v_{23}^7},\\
v_{34}^7\to\frac{1}{v_{34}^7},\qquad v_{44}^7\to\frac{v_{24}^7}{v_{12}^7v_{34}^7},
\end{gathered}
}
which leads to the following identity,
\eqna{
&G_{7|\substack{\text{extended}\\\text{snowflake}}}^{(d,h_2,h_3,h_4,h_5,h_6;p_2,p_3,p_4,p_5,p_6,p_7)}(u_1^7,u_2^7,u_3^7,u_4^7;v_{11}^7,v_{12}^7,v_{13}^7,v_{14}^7,v_{22}^7,v_{23}^7,v_{24}^7,v_{33}^7,v_{34}^7,v_{44}^7)\\
&\qquad=(v_{11}^7)^{h_3-h_6}(v_{12}^7)^{h_4}(v_{23}^7)^{h_5}(v_{34}^7)^{h_6}\\
&\qquad\phantom{=}\qquad\times G_{7|\substack{\text{extended}\\\text{snowflake}}}^{(d,-p_2-h_2,h_3,h_5,h_4,h_6;p_2,p_3,p_4,p_6,p_5,p_7)}\left(\frac{u_1^7}{v_{11}^7},\frac{u_3^7}{v_{23}^7},\frac{u_2^7}{v_{12}^7},\frac{u_4^7}{v_{34}^7};\right.\\
&\qquad\phantom{=}\qquad\phantom{\times}\qquad\left.\frac{1}{v_{11}^7},\frac{1}{v_{23}^7},\frac{v_{22}^7}{v_{11}^7v_{12}^7},\frac{v_{11}^7v_{14}^7}{v_{34}^7},\frac{v_{13}^7}{v_{11}^7v_{23}^7},\frac{1}{v_{12}^7},\frac{v_{44}^7}{v_{23}^7v_{34}^7},\frac{v_{33}^7}{v_{12}^7v_{23}^7},\frac{1}{v_{34}^7},\frac{v_{24}^7}{v_{12}^7v_{34}^7}\right).
}[EqSymRef]
With the help of \eqref{EqG} and using \eqref{EqCB7C} and \eqref{EqCB7F}, \eqref{EqSymRef} implies that $F_7$ does not change under the reflection generator.  As such, there is an associated identity for $C_7$ which we prove in the appendix.

The generator of dendrite permutations of the first kind that we choose acts as $\mathcal{O}_{i_3}(\eta_3)\leftrightarrow\mathcal{O}_{i_4}(\eta_4)$.  Hence, the legs and conformal cross-ratios \eqref{EqCB7CR} change under this generator as
\eqn{
\begin{gathered}
L_7\prod_{1\leq a\leq4}(u_a^7)^{\frac{\Delta_{k_a}}{2}}\to(v_{12}^7)^{-p_3-h_5}(v_{33}^7)^{h_5}L_7\prod_{1\leq a\leq4}(u_a^7)^{\frac{\Delta_{k_a}}{2}},\\
u_1^7\to u_1^7v_{12}^7,\qquad u_2^7\to\frac{u_2^7}{v_{12}^7},\qquad u_3^7\to\frac{u_3^7}{v_{33}^7},\qquad u_4^7\to u_4^7,\\
v_{11}^7\to v_{22}^7,\qquad v_{12}^7\to\frac{1}{v_{12}^7},\qquad v_{13}^7\to\frac{v_{12}^7v_{13}^7}{v_{33}^7},\qquad v_{14}^7\to v_{14}^7,\\
v_{22}^7\to v_{11}^7,\qquad v_{23}^7\to\frac{v_{12}^7v_{23}^7}{v_{33}^7},\qquad v_{24}^7\to v_{34}^7,\qquad v_{33}^7\to\frac{1}{v_{33}^7},\\
v_{34}^7\to v_{24}^7,\qquad v_{44}^7\to\frac{v_{12}^7v_{44}^7}{v_{33}^7},
\end{gathered}
}
and that translates into the identity
\eqna{
&G_{7|\substack{\text{extended}\\\text{snowflake}}}^{(d,h_2,h_3,h_4,h_5,h_6;p_2,p_3,p_4,p_5,p_6,p_7)}(u_1^7,u_2^7,u_3^7,u_4^7;v_{11}^7,v_{12}^7,v_{13}^7,v_{14}^7,v_{22}^7,v_{23}^7,v_{24}^7,v_{33}^7,v_{34}^7,v_{44}^7)\\
&\qquad=(v_{12}^7)^{-p_3-h_5}(v_{33}^7)^{h_5}\\
&\qquad\phantom{=}\qquad\times G_{7|\substack{\text{extended}\\\text{snowflake}}}^{(d,h_2,h_3,-p_3+h_2-h_4,h_5,h_6;p_2,p_3,p_4,p_5,p_6,p_7)}\left(u_1^7v_{12}^7,\frac{u_2^7}{v_{12}^7},\frac{u_3^7}{v_{33}^7},u_4^7;\right.\\
&\qquad\phantom{=}\qquad\phantom{\times}\qquad\left.v_{22}^7,\frac{1}{v_{12}^7},\frac{v_{12}^7v_{13}^7}{v_{33}^7},v_{14}^7,v_{11}^7,\frac{v_{12}^7v_{23}^7}{v_{33}^7},v_{34}^7,\frac{1}{v_{33}^7},v_{24}^7,\frac{v_{12}^7v_{44}^7}{v_{33}^7}\right).
}[EqSymPerm1]
Again, \eqref{EqG} with \eqref{EqCB6C} and \eqref{EqCB6F} imply that $F_7$ is invariant under dendrite permutations of the first kind.  Therefore, there exist a second identity for $C_7$.

The last generator $\mathcal{O}_{i_1}(\eta_1)\leftrightarrow\mathcal{O}_{i_7}(\eta_7)$ is the generator for dendrite permutations of the second kind and it leads to
\eqn{
\begin{gathered}
L_7\prod_{1\leq a\leq4}(u_a^7)^{\frac{\Delta_{k_a}}{2}}\to(v_{14}^7)^{-p_4}L_7\prod_{1\leq a\leq4}(u_a^7)^{\frac{\Delta_{k_a}}{2}},\\
u_1^7\to u_1^7v_{14}^7,\qquad u_2^7\to u_2^7,\qquad u_3^7\to u_3^7,\qquad u_4^7\to\frac{u_4^7}{v_{14}^7},\\
v_{11}^7\to v_{34}^7,\qquad v_{12}^7\to v_{12}^7,\qquad v_{13}^7\to v_{44}^7,\qquad v_{14}^7\to\frac{1}{v_{14}^7},\\
v_{22}^7\to v_{24}^7,\qquad v_{23}^7\to v_{23}^7,\qquad v_{24}^7\to v_{22}^7,\qquad v_{33}^7\to v_{33}^7,\\
v_{34}^7\to v_{11}^7,\qquad v_{44}^7\to v_{13}^7,
\end{gathered}
}
for the legs and conformal cross-ratios \eqref{EqCB7CR}.  The corresponding identity for the conformal blocks is therefore
\eqna{
&G_{7|\substack{\text{extended}\\\text{snowflake}}}^{(d,h_2,h_3,h_4,h_5,h_6;p_2,p_3,p_4,p_5,p_6,p_7)}(u_1^7,u_2^7,u_3^7,u_4^7;v_{11}^7,v_{12}^7,v_{13}^7,v_{14}^7,v_{22}^7,v_{23}^7,v_{24}^7,v_{33}^7,v_{34}^7,v_{44}^7)\\
&\qquad=(v_{14}^7)^{-p_4}\\
&\qquad\phantom{=}\qquad\times G_{7|\substack{\text{extended}\\\text{snowflake}}}^{(d,h_2,h_3,h_4,h_5,-p_4+h_3-h_6;p_2,p_3,p_4,p_5,p_6,p_7)}\left(u_1^7v_{14}^7,u_2^7,u_3^7,\frac{u_4^7}{v_{14}^7};\right.\\
&\qquad\phantom{=}\qquad\phantom{\times}\qquad\left.v_{34}^7,v_{12}^7,v_{44}^7,\frac{1}{v_{14}^7},v_{24}^7,v_{23}^7,v_{22}^7,v_{33}^7,v_{11}^7,v_{13}^7\right).
}[EqSymPerm2]
The decomposition \eqref{EqG} with \eqref{EqCB7C} and \eqref{EqCB7F} shows once again that $F_7$ given by \eqref{EqCB7F} does not transform under this last generator, resulting in a third identity for $C_7$.

To summarize, the symmetry group of the extended snowflake topology is $H_{7|\substack{\text{extended}\\\text{snowflake}}}=\mathbb{Z}_2\times\left((\mathbb{Z}_2)^2\rtimes\mathbb{Z}_2\right)$, its order is sixteen and it is generated by the three elements shown in Figure \ref{FigSym}.  These three generators lead to identities for the scalar seven-point conformal blocks \eqref{EqG} in the extended snowflake channel with \eqref{EqCB7C} and \eqref{EqCB7F}.  The three identities are \eqref{EqSymRef}, \eqref{EqSymPerm1} and \eqref{EqSymPerm2} which represent reflections, dendrite permutations of the first kind, and dendrite permutations of the second kind, respectively.  It is straightforward to produce the remaining twelve equivalent representations for $I_7$ in the extended snowflake channel by composing the three identities above.

In the following, we investigate two limits: the OPE limit and the limit of unit operator.  Due to the symmetry properties of the extended snowflake topology mentioned above, there are only two (three) limits to verify for the OPE (unit operator) case, the remaining limits being related by the action of the symmetry group.

%%%%%%%%%%%%%%%%%%%%%%%%%%%%%%%%%%%%%%%%%%%%%%%%%%

\subsection{OPE Limit}

For two external operators that emerge from the same OPE vertex in a given topology, we can define the OPE limit.  In this limit, the embedding space coordinates of these two operators coincide.  Therefore, the OPE limit reduces the original $M$-point correlation function to the appropriate $(M-1)$-point correlation function, up to a pre-factor originating from the OPE \eqref{EqOPE}.  In the scalar case, the pre-factor is obtained directly from \eqref{EqIfromOPE}.

For scalar seven-point correlation functions in the extended snowflake channel, as depicted in Figure \ref{Fig7pt}, there are only three possible OPE limits.  They are $\eta_3\to\eta_4$, $\eta_5\to\eta_6$, and $\eta_7\to\eta_1$, respectively.  With the invariance of $I_7$ under the symmetry group $H_{7|\substack{\text{extended}\\\text{snowflake}}}=\mathbb{Z}_2\times\left((\mathbb{Z}_2)^2\rtimes\mathbb{Z}_2\right)$, we only need to investigate $I_7$ under two OPE limits.  Here, we choose $\eta_7\to\eta_1$ (which is not related by symmetry to another pair) and $\eta_3\to\eta_4$.

For the limit $\eta_7\to\eta_1$, we have
\eqn{\left.I_{7(\Delta_{k_1},\Delta_{k_2},\Delta_{k_3},\Delta_{k_4})}^{(\Delta_{i_2},\Delta_{i_3},\Delta_{i_4},\Delta_{i_5},\Delta_{i_6},\Delta_{i_7},\Delta_{i_1})}\right|_{\substack{\text{extended}\\\text{snowflake}}}\underset{\eta_7\to\eta_1}{\to}(\eta_{17})^{-p_7}\left.I_{6(\Delta_{k_1},\Delta_{k_2},\Delta_{k_3})}^{(\Delta_{i_2},\Delta_{i_3},\Delta_{i_4},\Delta_{i_5},\Delta_{i_6},\Delta_{i_1})}\right|_{\text{snowflake}},}[EqOPELim71]
as well as
\eqn{
\begin{gathered}
L_7\prod_{1\leq a\leq4}(u_a^7)^{\frac{\Delta_{k_a}}{2}}\to(\eta_{17})^{-p_7}L_6\prod_{1\leq a\leq3}(u_a^6)^{\frac{\Delta_{k_a}}{2}},\\
u_1^7\to u_1^6,\qquad u_2^7\to u_2^6,\qquad u_3^7\to u_3^6,\qquad u_4^7\to0, \\
v_{11}^7\to v_{11}^6,\qquad v_{12}^7\to v_{12}^6,\qquad v_{13}^7\to v_{13}^6,\qquad v_{14}^7\to1,\\
v_{22}^7\to v_{22}^6,\qquad v_{23}^7\to v_{23}^6,\qquad v_{24}^7\to v_{22}^6,\qquad v_{33}^7\to v_{33}^6,\\
v_{34}^7\to v_{11}^6,\qquad v_{44}^7\to v_{13}^6.
\end{gathered}
}
Here, the quantities and the conformal dimensions on the RHS of the limits are the ones relevant for the six-point correlation functions in the snowflake channel, \textit{i.e.} \eqref{EqCB6CR}, \eqref{EqCB6C}, \eqref{EqCB6F} and trivial substitutions for the vectors $\boldsymbol{h}$ and $\boldsymbol{p}$.

As a consequence, the OPE limit \eqref{EqOPELim71} leads to the identity $G_{7|\eta_7\to\eta_1}=G_6$, or more precisely
\eqna{
&G_{7|\substack{\text{extended}\\\text{snowflake}}}^{(d,h_2,h_3,h_4,h_5,h_6;p_2,p_3,p_4,p_5,p_6,p_7)}(u_1^6,u_2^6,u_3^6,0;v_{11}^6,v_{12}^6,v_{13}^6,1,v_{22}^6,v_{23}^6,v_{22}^6,v_{33}^6,v_{11}^6,v_{13}^6)\\
&\qquad=G_{6|\text{snowflake}}^{(d,h_2,h_3,h_4,h_5;p_2,p_3,p_4,p_5,p_6)}(u_1^6,u_2^6,u_3^6;v_{11}^6,v_{12}^6,v_{13}^6,v_{22}^6,v_{23}^6,v_{33}^6).
}
Since $F_7\to F_6$ in that limit, we have
\eqna{
G_{7|\eta_7\to\eta_1}&=\sum_{m_a,m_{ab}\geq0}\frac{(p_2+h_3)_{m_1+m_{23}}(p_3)_{-m_1+m_2+m_3+m_{12}+m_{33}}(-h_3)_{m_1+m_{11}+m_{13}+m_{22}}}{(p_2)_{2m_1+m_{11}+m_{13}+m_{22}+m_{23}}(p_2+1-d/2)_{m_1}}\\
&\qquad\phantom{=}\times\frac{(p_3-h_2+h_4)_{m_{2}+m_{11}}(p_2+h_2)_{m_1-m_2+m_3+m_{13}+m_{23}}(-h_4)_{m_2+m_{12}+m_{22}+m_{33}}}{(p_3-h_2)_{2m_2+m_{11}+m_{12}+m_{22}+m_{33}}(p_3-h_2+1-d/2)_{m_2}}\\
&\qquad\phantom{=}\times\frac{(\bar{p}_3+h_2+h_5)_{m_3+m_{12}}(-h_2)_{m_1+m_2-m_3+m_{11}+m_{22}}(-h_5)_{m_3+m_{13}+m_{23}+m_{33}}}{(\bar{p}_3+h_2)_{2m_3+m_{12}+m_{13}+m_{23}+m_{33}}(\bar{p}_3+h_2+1-d/2)_{m_3}}\\
&\qquad\phantom{=}\times\frac{(-h_6)_{m_{24}+m_{34}+m_{44}}(p_4-h_3+h_6)_{m_{11}+m_{13}+m_{22}-m_{24}-m_{34}-m_{44}}}{(p_4-h_3)_{m_{11}+m_{13}+m_{22}}}\\
&\qquad\phantom{=}\times F_6\frac{m_{11}!m_{13}!m_{22}!}{(m_{11}-m_{34})!(m_{13}-m_{44})!(m_{22}-m_{24})!m_{24}!m_{34}!m_{44}!}\\
&\qquad\phantom{=}\times\prod_{1\leq a\leq3}\frac{(u_a^6)^{m_a}}{m_a!}\prod_{1\leq a\leq b\leq3}\frac{(1-v_{ab}^6)^{m_{ab}}}{m_{ab}!},
}
where we performed the following change of variables,
\eqn{m_{11}\to m_{11}-m_{34},\qquad m_{13}\to m_{13}-m_{44},\qquad m_{22}\to m_{22}-m_{24}.}
Evaluating the sums over $m_{24}$, $m_{34}$, and finally $m_{44}$ using standard hypergeometric re-summation formula straightforwardly leads to $G_{7|\eta_7\to\eta_1}=G_6$, proving \eqref{EqOPELim71}.

Since the proof for the last independent OPE limit $\eta_3\to\eta_4$ is more intricate, it is left for Appendix \ref{SAppLim}.

%%%%%%%%%%%%%%%%%%%%%%%%%%%%%%%%%%%%%%%%%%%%%%%%%%

\subsection{Limit of Unit Operator}

The limit of unit operator consists in setting one external scalar quasi-primary operator to the identity operator.  As a consequence, a given $M$-point correlation function morphs into the appropriate $(M-1)$-point correlation function.

For scalar seven-point correlation functions in the extended snowflake channel, there are only three limits of unit operator to check, thanks to the symmetry group.  They are $\mathcal{O}_{i_2}(\eta_2)\to\1$, $\mathcal{O}_{i_3}(\eta_3)\to\1$ and $\mathcal{O}_{i_7}(\eta_7)\to\1$, respectively.

Focusing first on the limit $\mathcal{O}_{i_7}(\eta_7)\to\1$, we have $\Delta_{i_7}=0$, $\Delta_{k_4}=\Delta_{i_1}$, and
\eqn{\left.I_{7(\Delta_{k_1},\Delta_{k_2},\Delta_{k_3},\Delta_{k_4})}^{(\Delta_{i_2},\Delta_{i_3},\Delta_{i_4},\Delta_{i_5},\Delta_{i_6},\Delta_{i_7},\Delta_{i_1})}\right|_{\substack{\text{extended}\\\text{snowflake}}}\underset{\mathcal{O}_{i_7}(\eta_7)\to\1}{\to}\left.I_{6(\Delta_{k_1},\Delta_{k_2},\Delta_{k_3})}^{(\Delta_{i_1},\Delta_{i_2},\Delta_{i_3},\Delta_{i_4},\Delta_{i_5},\Delta_{i_6})}\right|_{\text{snowflake}}.}[EqLimUnit7Id]
In this limit, it is straightforward to see that $h_6=p_7=0$.  Moreover, the remaining components of the vectors $\boldsymbol{h}$ and $\boldsymbol{p}$ \eqref{EqCB7hp} become the appropriate components of the scalar six-point correlation functions in the snowflake channel \eqref{EqCB6hp}.  Since the conformal cross-ratios transform as
\eqn{u_a^7\to u_a^6\qquad(1\leq a\leq3),\qquad v_{ab}^7\to v_{ab}^6,\qquad(1\leq a\leq b\leq3),}
and the remaining conformal cross-ratios disappear due to the sums over $m_4$, $m_{14}$, $m_{24}$, $m_{34}$ and $m_{44}$ in \eqref{EqCB7C} being trivial [thanks to the Pochhammer symbol $(-h_6)_{m_4+m_{14}+m_{24}+m_{34}+m_{44}}$ forcing $m_4=m_{14}=m_{24}=m_{34}=m_{44}=0$], we obtain
\eqn{L_7\prod_{1\leq a\leq3}(u_a^7)^{\frac{\Delta_{k_a}}{2}}\to L_6\prod_{1\leq a\leq3}(u_a^6)^{\frac{\Delta_{k_a}}{2}},}
in the appropriate channels.

Furthermore, the limit of unit operator \eqref{EqLimUnit7Id} leads to
\eqna{
F_{7|\mathcal{O}_{i_7}\to\1}&=\frac{(-\bar{p}_3-h_2+d/2-m_3)_{m_3}(-p_3+h_2+d/2-m_2)_{m_2}}{(p_3)_{-m_1}}\sum_{r_1,r_2\geq0}\frac{(-m_2)_{t_1}(-m_3)_{t_2}}{t_1!t_2!}\\
&\phantom{=}\qquad\times\frac{(\bar{p}_3-d/2)_{t_1+t_2}}{(\bar{p}_3+h_2+1-d/2)_{t_2}(p_3-m_1)_{t_1+t_2}}\frac{(p_3)_{t_1+t_2}}{(p_3-h_2+1-d/2)_{t_1}},
}
which implies $F_7\to F_6$ with the help of the standard ${}_3F_2$-hypergeometric function identity \eqref{Eq3F2}.  Taking into account that $C_7\to C_6$ trivially, we thus prove that $G_7\to G_6$ in the limit of unit operator \eqref{EqLimUnit7Id}.

The two remaining limits of unit operator being longer, their proofs are left for Appendix \ref{SAppLim}.

%%%%%%%%%%%%%%%%%%%%%%%%%%%%%%%%%%%%%%%%%%%%%%%%%%
%%%%%%%%%%%%%%%%%%%%%%%%%%%%%%%%%%%%%%%%%%%%%%%%%%

\section{Discussion and Conclusion}\label{SecConc}

With the knowledge of the action of the OPE differential operator on a general product of conformal cross-ratios, the embedding space OPE formalism introduced in \cite{Fortin:2019fvx,Fortin:2019dnq} leads to explicit results for any conformal correlation function.  Therefore, it is perfectly suited to investigate $d$-dimensional higher-point correlation functions, a nascent field of research in CFT.  With the tools provided by the embedding space OPE formalism, we already determined scalar higher-point correlation functions in the comb channel in \cite{Fortin:2019zkm} and scalar six-point correlation functions in the snowflake channel in \cite{Fortin:2020yjz}.

In this paper we computed the scalar seven-point conformal blocks in the extended snowflake channel, see \eqref{EqCB7CR}, \eqref{EqCB7C}, \eqref{EqCB7F}, and \eqref{EqCB7hp}.  With that result and the scalar seven-point conformal blocks in the comb channel, the scalar conformal blocks for seven-point correlation functions are known in all topologies.  We stress that all our results are derived from the OPE, as such they are exact and do not need to be verified \textit{e.g.} from the Casimir equations.  Nevertheless, to check the algebra, we did observe that all the symmetries [namely, the $\mathbb{Z}_2\times\left((\mathbb{Z}_2)^2\rtimes\mathbb{Z}_2\right)$ symmetry group of the extended snowflake topology] and appropriate limits (namely, the OPE limit where two embedding space coordinates coincide, and the limit of unit operator where one external quasi-primary operator is set to the identity operator) are satisfied.

As we did in prior work, we defined the scalar conformal blocks in terms of coefficients of the power series in the conformal cross-ratios.  These coefficients are products of two functions, the $C$-function which is a product of terms, and the $F$-function that contains multiple sums, see \eqref{EqG}.

In \cite{Fortin:2019zkm}, we argued that using OPE limits and limits of unit operators repetitively allows one to construct the $C$-function.  It seems this argument can be extended to all channels.  Indeed, by rewriting the vectors $\boldsymbol{h}$ and $\boldsymbol{p}$ explicitly in terms of the conformal dimensions, it is clear that a pattern emerges.  Focusing on the scalar $M$-point $C$-functions in the comb channel \eqref{EqCBCcomb} and the scalar six- and seven-point $C$-functions in the snowflake and extended snowflake channels \eqref{EqCB6C} and \eqref{EqCB7C}, we have
\eqna{
&C_{M|\text{comb}}^{(d,\boldsymbol{h};\boldsymbol{p})}(\boldsymbol{m},\textbf{m})=\frac{\left(\frac{\Delta_{k_1}-\Delta_{i_2}+\Delta_{i_3}}{2}\right)_{m_1+\text{tr}_1\textbf{m}}\left(\frac{\Delta_{k_1}+\Delta_{i_2}-\Delta_{i_3}}{2}\right)_{m_1+\text{tr}_0\textbf{m}}}{(\Delta_{k_1})_{2m_1+\bar{m}_1+\bar{\bar{m}}_1}(\Delta_{k_1}+1-d/2)_{m_1}}\\
&\qquad\times\frac{\left(\frac{\Delta_{k_{M-3}}-\Delta_{i_1}+\Delta_{i_M}}{2}\right)_{m_{M-3}+\bar{m}_{M-4}}\left(\frac{\Delta_{k_{M-3}}+\Delta_{i_1}-\Delta_{i_M}}{2}\right)_{m_{M-3}+\bar{m}_{M-3}+\bar{\bar{m}}_{M-3}}}{(\Delta_{k_{M-3}})_{2m_{M-3}+\bar{m}_{M-4}+\bar{m}_{M-3}+\bar{\bar{m}}_{M-3}}(\Delta_{k_{M-3}}+1-d/2)_{m_{M-3}}}\\
&\qquad\times\frac{\left(\frac{\Delta_{k_2}-\Delta_{k_1}+\Delta_{i_4}}{2}\right)_{m_2-m_1+\text{tr}_2\textbf{m}}\left(\frac{\Delta_{k_2}+\Delta_{k_1}-\Delta_{i_4}}{2}\right)_{m_2+m_1+\bar{m}_1+\bar{\bar{m}}_1}}{(\Delta_{k_2})_{2m_2+\bar{m}_1+\bar{m}_2+\bar{\bar{m}}_2}(\Delta_{k_2}+1-d/2)_{m_2}}\\
&\qquad\qquad\qquad\qquad\times\frac{\left(\frac{\Delta_{i_4}-\Delta_{k_2}+\Delta_{k_1}}{2}\right)_{m_1}\left(\frac{\Delta_{i_4}-\Delta_{k_2}+\Delta_{k_1}}{2}\right)_{m_1-m_2}}{\left(\frac{\Delta_{i_4}+\Delta_{k_2}-\Delta_{k_1}}{2}\right)_{-m_1}\left(\frac{\Delta_{i_4}-\Delta_{k_2}+\Delta_{k_1}}{2}\right)_{m_1-m_2}}\\
&\qquad\times\frac{\left(\frac{\Delta_{k_3}-\Delta_{k_2}+\Delta_{i_5}}{2}\right)_{m_3-m_2+\text{tr}_3\textbf{m}}\left(\frac{\Delta_{k_3}+\Delta_{k_2}-\Delta_{i_5}}{2}\right)_{m_3+m_2+\bar{m}_2+\bar{\bar{m}}_2}}{(\Delta_{k_3})_{2m_3+\bar{m}_2+\bar{m}_3+\bar{\bar{m}}_3}(\Delta_{k_3}+1-d/2)_{m_3}}\\
&\qquad\qquad\qquad\qquad\times\frac{\left(\frac{\Delta_{i_5}-\Delta_{k_3}+\Delta_{k_2}}{2}\right)_{m_2}\left(\frac{\Delta_{i_5}-\Delta_{k_3}+\Delta_{k_2}}{2}\right)_{m_2-m_3+\bar{m}_1}}{\left(\frac{\Delta_{i_5}+\Delta_{k_3}-\Delta_{k_2}}{2}\right)_{-m_2}\left(\frac{\Delta_{i_5}-\Delta_{k_3}+\Delta_{k_2}}{2}\right)_{m_2-m_3}}\\
&\qquad\qquad\qquad\qquad\qquad\qquad\qquad\qquad\qquad\qquad\vdots\\
&\qquad\times\frac{\left(\frac{\Delta_{k_{M-3}}-\Delta_{k_{M-4}}+\Delta_{i_{M-1}}}{2}\right)_{m_{M-3}-m_{M-4}+\text{tr}_{M-3}\textbf{m}}\left(\frac{\Delta_{k_{M-3}}+\Delta_{k_{M-4}}-\Delta_{i_{M-1}}}{2}\right)_{m_{M-3}+m_{M-4}+\bar{m}_{M-4}+\bar{\bar{m}}_{M-4}}}{(\Delta_{k_{M-4}})_{2m_{M-4}+\bar{m}_{M-5}+\bar{m}_{M-4}+\bar{\bar{m}}_{M-4}}(\Delta_{k_{M-4}}+1-d/2)_{m_{M-4}}}\\
&\qquad\qquad\qquad\qquad\times\frac{\left(\frac{\Delta_{i_{M-1}}-\Delta_{k_{M-3}}+\Delta_{k_{M-4}}}{2}\right)_{m_{M-4}}\left(\frac{\Delta_{i_{M-1}}-\Delta_{k_{M-3}}+\Delta_{k_{M-4}}}{2}\right)_{m_{M-4}-m_{M-3}+\bar{m}_{M-5}}}{\left(\frac{\Delta_{i_{M-1}}+\Delta_{k_{M-3}}-\Delta_{k_{M-4}}}{2}\right)_{-m_{M-4}}\left(\frac{\Delta_{i_{M-1}}-\Delta_{k_{M-3}}+\Delta_{k_{M-4}}}{2}\right)_{m_{M-4}-m_{M-3}}},
}
as well as
\eqna{
&C_{6|\text{snowflake}}^{(d,\boldsymbol{h};\boldsymbol{p})}(\boldsymbol{m},\textbf{m})=\frac{\left(\frac{\Delta_{k_1}-\Delta_{i_2}+\Delta_{i_3}}{2}\right)_{m_1+m_{23}}\left(\frac{\Delta_{k_1}+\Delta_{i_2}-\Delta_{i_3}}{2}\right)_{m_1+m_{11}+m_{22}+m_{13}}}{(\Delta_{k_1})_{2m_1+m_{11}+m_{13}+m_{22}+m_{23}}(\Delta_{k_1}+1-d/2)_{m_1}}\\
&\qquad\times\frac{\left(\frac{\Delta_{k_2}-\Delta_{i_4}+\Delta_{i_5}}{2}\right)_{m_2+m_{11}}\left(\frac{\Delta_{k_2}+\Delta_{i_4}-\Delta_{i_5}}{2}\right)_{m_2+m_{12}+m_{22}+m_{33}}}{(\Delta_{k_2})_{2m_2+m_{11}+m_{12}+m_{22}+m_{33}}(\Delta_{k_2}+1-d/2)_{m_2}}\\
&\qquad\times\frac{\left(\frac{\Delta_{k_3}-\Delta_{i_6}+\Delta_{i_1}}{2}\right)_{m_3+m_{12}}\left(\frac{\Delta_{k_3}+\Delta_{i_6}-\Delta_{i_1}}{2}\right)_{m_3+m_{13}+m_{23}+m_{33}}}{(\Delta_{k_3})_{2m_3+m_{12}+m_{13}+m_{23}+m_{33}}(\Delta_{k_3}+1-d/2)_{m_3}}\\
&\qquad\times\left(\frac{-\Delta_{k_1}+\Delta_{k_2}+\Delta_{k_3}}{2}\right)_{-m_1+m_2+m_3+m_{12}+m_{33}}\\
&\qquad\times\left(\frac{-\Delta_{k_2}+\Delta_{k_3}+\Delta_{k_1}}{2}\right)_{-m_2+m_3+m_1+m_{13}+m_{23}}\\
&\qquad\times\left(\frac{-\Delta_{k_3}+\Delta_{k_1}+\Delta_{k_2}}{2}\right)_{-m_3+m_1+m_2+m_{11}+m_{22}},
}
and
\eqna{
&C_{7|\substack{\text{extended}\\\text{snowflake}}}^{(d,\boldsymbol{h};\boldsymbol{p})}(\boldsymbol{m},\textbf{m})=\frac{\left(\frac{\Delta_{k_2}-\Delta_{i_3}+\Delta_{i_4}}{2}\right)_{m_{2}+m_{11}+m_{34}}\left(\frac{\Delta_{k_2}+\Delta_{i_3}-\Delta_{i_4}}{2}\right)_{m_2+m_{12}+m_{22}+m_{24}+m_{33}}}{(\Delta_{k_2})_{2m_2+m_{11}+m_{12}+m_{22}+m_{24}+m_{33}+m_{34}}(\Delta_{k_2}+1-d/2)_{m_2}}\\
&\qquad\times\frac{\left(\frac{\Delta_{k_3}-\Delta_{i_5}+\Delta_{i_6}}{2}\right)_{m_3+m_{12}}\left(\frac{\Delta_{k_3}+\Delta_{i_5}-\Delta_{i_6}}{2}\right)_{m_3+m_{13}+m_{23}+m_{33}+m_{44}}}{(\Delta_{k_3})_{2m_3+m_{12}+m_{13}+m_{23}+m_{33}+m_{44}}(\Delta_{k_3}+1-d/2)_{m_3}}\\
&\qquad\times\frac{\left(\frac{\Delta_{k_4}-\Delta_{i_7}+\Delta_{i_1}}{2}\right)_{m_4+m_{11}+m_{13}+m_{22}}\left(\frac{\Delta_{k_4}+\Delta_{i_7}-\Delta_{i_1}}{2}\right)_{m_4+m_{14}+m_{24}+m_{34}+m_{44}}}{(\Delta_{k_4})_{2m_4+m_{11}+m_{13}+m_{14}+m_{22}+m_{24}+m_{34}+m_{44}}(\Delta_{k_4}+1-d/2)_{m_4}}\\
&\qquad\times\frac{\left(\frac{\Delta_{k_1}-\Delta_{k_4}+\Delta_{i_2}}{2}\right)_{m_1-m_4+m_{23}}\left(\frac{\Delta_{k_1}+\Delta_{k_4}-\Delta_{i_2}}{2}\right)_{m_1+m_4+m_{11}+m_{13}+m_{22}+m_{24}+m_{34}+m_{44}}}{(\Delta_{k_1})_{2m_1+m_{11}+m_{13}+m_{22}+m_{23}+m_{24}+m_{34}+m_{44}}(\Delta_{k_1}+1-d/2)_{m_1}}\\
&\qquad\qquad\qquad\qquad\times\frac{\left(\frac{\Delta_{i_2}+\Delta_{k_1}-\Delta_{k_4}}{2}\right)_{m_1}\left(\frac{\Delta_{i_2}-\Delta_{k_1}+\Delta_{k_4}}{2}\right)_{-m_1+m_4+m_{14}}}{\left(\frac{\Delta_{i_2}-\Delta_{k_1}+\Delta_{k_4}}{2}\right)_{-m_1}\left(\frac{\Delta_{i_2}+\Delta_{k_1}-\Delta_{k_4}}{2}\right)_{m_1-m_4}}\\
&\qquad\times\left(\frac{-\Delta_{k_1}+\Delta_{k_2}+\Delta_{k_3}}{2}\right)_{-m_1+m_2+m_3+m_{12}+m_{33}}\\
&\qquad\times\left(\frac{-\Delta_{k_2}+\Delta_{k_3}+\Delta_{k_1}}{2}\right)_{-m_2+m_3+m_1+m_{13}+m_{23}+m_{44}}\\
&\qquad\times\left(\frac{-\Delta_{k_3}+\Delta_{k_1}+\Delta_{k_2}}{2}\right)_{-m_3+m_1+m_2+m_{11}+m_{22}+m_{24}+m_{34}},
}
respectively.  Hence, for a proper choice of $\boldsymbol{u}$ conformal cross-ratios behaving accordingly under the OPE limit,\footnote{The $u$ conformal cross-ratios must be chosen such that only one of them vanishes for a given OPE limit.} the $C$-function is given by multiplying factors of
\eqn{\frac{\left(\frac{\Delta_{k_a}-\Delta_i+\Delta_j}{2}\right)_{m_a+\ldots}\left(\frac{\Delta_{k_a}+\Delta_i-\Delta_j}{2}\right)_{m_a+\ldots}}{(\Delta_{k_a})_{2m_a+\ldots}(\Delta_{k_a}+1-d/2)_{m_a}},}[EqEEI]
for each pair of external quasi-primary operators $\mathcal{O}_i$ and $\mathcal{O}_j$ connected to an internal quasi-primary operator through the OPE,
\eqn{\frac{\left(\frac{\Delta_{k_b}-\Delta_{k_a}+\Delta_i}{2}\right)_{m_b-m_a+\ldots}\left(\frac{\Delta_{k_b}+\Delta_{k_a}-\Delta_i}{2}\right)_{m_b+m_a+\ldots}}{(\Delta_{k_b})_{2m_b+\ldots}(\Delta_{k_b}+1-d/2)_{m_b}}\frac{\left(\frac{\Delta_i-\Delta_{k_b}+\Delta_{k_a}}{2}\right)_{m_a}\left(\frac{\Delta_i-\Delta_{k_b}+\Delta_{k_a}}{2}\right)_{m_a-m_b+\ldots}}{\left(\frac{\Delta_i+\Delta_{k_b}-\Delta_{k_a}}{2}\right)_{-m_a}\left(\frac{\Delta_i-\Delta_{k_b}+\Delta_{k_a}}{2}\right)_{m_a-m_b}},}[EqEII]
for each external quasi-primary operator $\mathcal{O}_i$ connected to a pair of internal quasi-primary operators $\mathcal{O}_{k_b}$ and $\mathcal{O}_{k_a}$, and
\eqna{
&\left(\frac{-\Delta_{k_a}+\Delta_{k_b}+\Delta_{k_c}}{2}\right)_{-m_a+m_b+m_c+\ldots}\left(\frac{-\Delta_{k_b}+\Delta_{k_c}+\Delta_{k_a}}{2}\right)_{-m_b+m_c+m_a+\ldots}\\
&\qquad\qquad\qquad\times\left(\frac{-\Delta_{k_c}+\Delta_{k_a}+\Delta_{k_b}}{2}\right)_{-m_c+m_a+m_b+\ldots},
}[EqIII]
for each internal OPE with three internal quasi-primary operators $\mathcal{O}_{k_c}$, $\mathcal{O}_{k_b}$, and $\mathcal{O}_{k_a}$.  Here, internal quasi-primary operators appear only once and the second factor in the rule \eqref{EqEII} is not completely fixed yet (as can be seen by comparing the comb to the extended snowflake).  Moreover, the dependence on the matrix of indices $\textbf{m}$ (denoted by ellipses in the rules above) is not determined.  For a proper choice of $\textbf{v}$ conformal cross-ratios, we conjecture that the remaining ambiguities can be fixed by the OPE limit and the limit of unit operator, although it is probable that a recurring process must be used [$(M-1)$-point correlation functions must be known to verify these limits for $M$-point correlation functions].

Moreover, we observed that the $F$-functions \eqref{EqCBFcomb}, \eqref{EqCB6F}, and \eqref{EqCB7F}, when expressed in terms of the conformal dimensions
\eqna{
F_{M|\text{comb}}^{(d,\boldsymbol{h};\boldsymbol{p})}(\boldsymbol{m})&={}_3F_2\left[\begin{array}{c}-m_1,-m_2,-\Delta_{k_1}+d/2-m_1\\\frac{\Delta_{k_2}-\Delta_{k_1}+\Delta_{i_4}}{2}-m_1,\frac{\Delta_{k_2}-\Delta_{k_1}-\Delta_{i_4}}{2}+1-m_1\end{array};1\right]\\
&\phantom{=}\qquad\times{}_3F_2\left[\begin{array}{c}-m_2,-m_3,-\Delta_{k_2}+d/2-m_2\\\frac{\Delta_{k_3}-\Delta_{k_2}+\Delta_{i_5}}{2}-m_2,\frac{\Delta_{k_3}-\Delta_{k_2}-\Delta_{i_5}}{2}+1-m_2\end{array};1\right]\\
&\phantom{=}\qquad\qquad\qquad\qquad\qquad\qquad\qquad\vdots\\
&\phantom{=}\qquad\times{}_3F_2\left[\begin{array}{c}-m_{M-4},-m_{M-3},-\Delta_{k_{M-4}}+d/2-m_{M-4}\\\frac{\Delta_{k_{M-3}}-\Delta_{k_{M-4}}+\Delta_{i_{M-1}}}{2}-m_{M-4},\frac{\Delta_{k_{M-3}}-\Delta_{k_{M-4}}-\Delta_{i_{M-1}}}{2}+1-m_{M-4}\end{array};1\right]
}
as well as
\eqna{
F_{6|\text{snowflake}}^{(d,\boldsymbol{h};\boldsymbol{p})}(\boldsymbol{m})&=\frac{1}{\left(\frac{-\Delta_{k_1}+\Delta_{k_2}+\Delta_{k_3}}{2}\right)_{-m_1}(-\Delta_{k_2}+d/2)_{-m_2}(-\Delta_{k_3}+d/2)_{-m_3}}\\
&\phantom{=}\qquad\times F_{1,1,1}^{2,1,1}\left[\left.\begin{array}{c}\frac{\Delta_{k_1}+\Delta_{k_2}+\Delta_{k_3}-d}{2},\frac{-\Delta_{k_1}+\Delta_{k_2}+\Delta_{k_3}}{2};-m_2;-m_3\\\frac{-\Delta_{k_1}+\Delta_{k_2}+\Delta_{k_3}}{2}-m_1;\Delta_{k_2}+1-d/2;\Delta_{k_3}+1-d/2\end{array}\right|1,1\right],
}
and finally
\eqna{
F_{7|\substack{\text{extended}\\\text{snowflake}}}^{(d,\boldsymbol{h};\boldsymbol{p})}(\boldsymbol{m})&=\frac{1}{\left(\frac{-\Delta_{k_1}+\Delta_{k_2}+\Delta_{k_3}}{2}\right)_{-m_1}(-\Delta_{k_2}+d/2)_{-m_2}(-\Delta_{k_3}+d/2)_{-m_3}}\\
&\phantom{=}\qquad\times F_{1,1,1}^{2,1,1}\left[\left.\begin{array}{c}\frac{\Delta_{k_1}+\Delta_{k_2}+\Delta_{k_3}-d}{2},\frac{-\Delta_{k_1}+\Delta_{k_2}+\Delta_{k_3}}{2};-m_2;-m_3\\\frac{-\Delta_{k_1}+\Delta_{k_2}+\Delta_{k_3}}{2}-m_1;\Delta_{k_2}+1-d/2;\Delta_{k_3}+1-d/2\end{array}\right|1,1\right]\\
&\phantom{=}\qquad\times{}_3F_2\left[\begin{array}{c}-m_1,-m_4,-\Delta_{k_1}+d/2-m_1\\\frac{\Delta_{k_4}-\Delta_{k_1}+\Delta_{i_2}}{2}-m_1,\frac{\Delta_{k_4}-\Delta_{k_1}-\Delta_{i_2}}{2}+1-m_1\end{array};1\right],
}
satisfy interesting factorization properties.  Indeed, in the comb channel it was already pointed out in \cite{Fortin:2019zkm} that the $F$-function is a product of hypergeometric functions.  This can be stated schematically as $F_{M|\text{comb}}=(F_{5|\text{comb}})^{M-4}$ with $F_{5|\text{comb}}$ a building block for the $F$-function.  For the snowflake channel, no such factorization occurs, implying that $F_{6|\text{snowflake}}$ is another building block for the $F$-function.  We now argue that $F_{5|\text{comb}}$ and $F_{6|\text{snowflake}}$, together with $F_{4|\text{comb}}=1$, are the only building blocks necessary to construct the $F$-function in any topology, with the rules that the $F$-function is obtained by multiplication of factors of
\eqn{{}_3F_2\left[\begin{array}{c}-m_a,-m_b,-\Delta_{k_a}+d/2-m_a\\\frac{\Delta_{k_b}-\Delta_{k_a}+\Delta_i}{2}-m_a,\frac{\Delta_{k_b}-\Delta_{k_a}-\Delta_i}{2}+1-m_a\end{array};1\right],}[EqFcomb]
for each external quasi-primary operator $\mathcal{O}_i$ connected to two internal quasi-primary operators $\mathcal{O}_{k_b}$ and $\mathcal{O}_{k_a}$, and factors of
\eqna{
&\frac{1}{\left(\frac{-\Delta_{k_a}+\Delta_{k_b}+\Delta_{k_c}}{2}\right)_{-m_a}(-\Delta_{k_b}+d/2)_{-m_b}(-\Delta_{k_c}+d/2)_{-m_c}}\\
&\qquad\qquad\qquad\times F_{1,1,1}^{2,1,1}\left[\left.\begin{array}{c}\frac{\Delta_{k_a}+\Delta_{k_b}+\Delta_{k_c}-d}{2},\frac{-\Delta_{k_a}+\Delta_{k_b}+\Delta_{k_c}}{2};-m_b;-m_c\\\frac{-\Delta_{k_a}+\Delta_{k_b}+\Delta_{k_c}}{2}-m_1;\Delta_{k_b}+1-d/2;\Delta_{k_c}+1-d/2\end{array}\right|1,1\right],
}[EqFsnowflake]
for each internal OPE connecting three internal quasi-primary operators $\mathcal{O}_{k_c}$, $\mathcal{O}_{k_b}$ and $\mathcal{O}_{k_a}$.

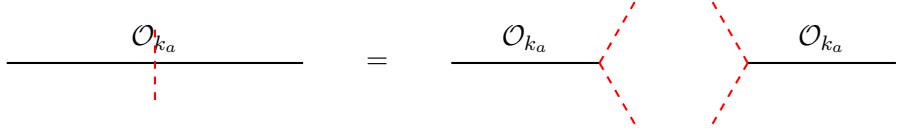
\begin{figure}[t]
\centering
\resizebox{12cm}{!}{%
\begin{tikzpicture}[thick]
\begin{scope}
\draw[-] (0,0)--+(0:4) node[pos=0.5,above]{$\mathcal{O}_{k_a}$};
\draw[dashed,red] (2,-0.5)--+(90:1);
\node at (5,0) {$=$};
\draw[-] (6,0)--+(0:2) node[pos=0.5,above]{$\mathcal{O}_{k_a}$};
\draw[dashed,red] (8,0)--+(60:1);
\draw[dashed,red] (8,0)--+(-60:1);
\draw[-] (12,0)--+(180:2) node[pos=0.5,above]{$\mathcal{O}_{k_a}$};
\draw[dashed,red] (10,0)--+(120:1);
\draw[dashed,red] (10,0)--+(-120:1);
\end{scope}
\end{tikzpicture}
}
\caption{Cutting procedure for the computation of the $F$-function.  For a given topology, an internal line is cut and the remaining segments are both dressed as one OPE.}
\label{FigCutting}
\end{figure}

First, we observe that this rule is respected when we consider the extended snowflake for which $F_{7|\substack{\text{extended}\\\text{snowflake}}}=F_{6|\text{snowflake}}F_{5|\text{comb}}$.  The main reason why this factorization should occur for the extended snowflake again originates in the OPE limit and the limit of unit operator.  It can also be understood diagrammatically directly from the topology by cutting an internal line and replacing it by a double OPE as in Figure \ref{FigCutting}.

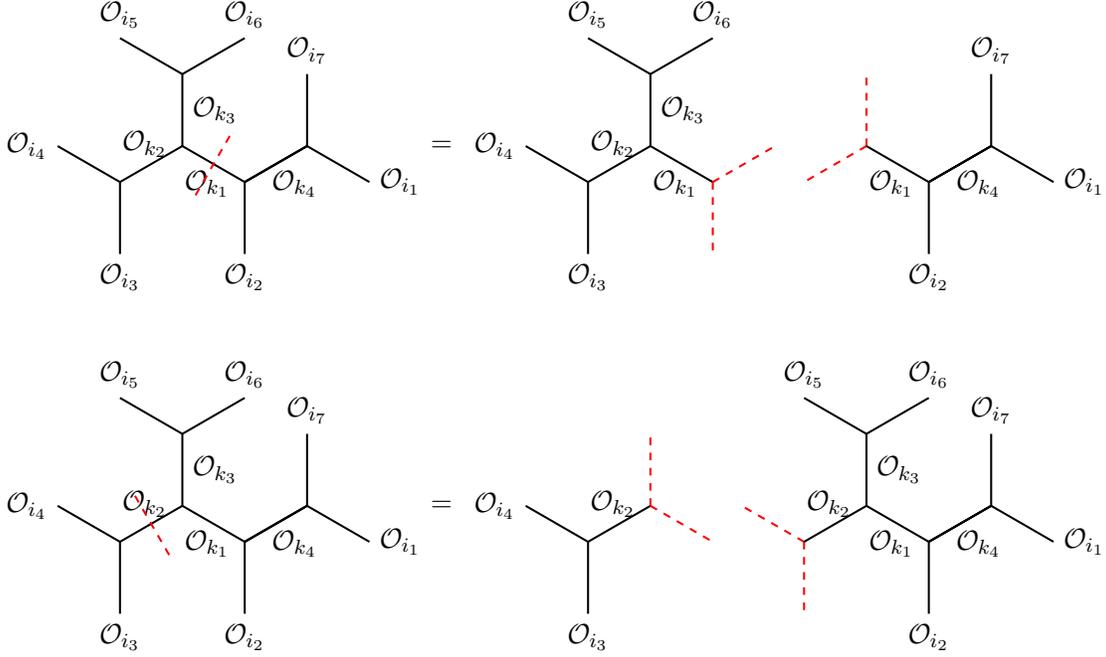
\begin{figure}[t]
\centering
\resizebox{15cm}{!}{%
\begin{tikzpicture}[thick]
\begin{scope}
\draw[-] (4,0)--+(-150:1) node[pos=0.6,above]{$\mathcal{O}_{k_2}$};
\draw[-] (4,0)++(-150:1)--+(-90:1) node[below]{$\mathcal{O}_{i_3}$};
\draw[-] (4,0)++(-150:1)--+(150:1) node[left]{$\mathcal{O}_{i_4}$};
\draw[-] (4,0)--+(90:1) node[pos=0.5,right]{$\mathcal{O}_{k_3}$};
\draw[-] (4,0)++(90:1)--+(30:1) node[above]{$\mathcal{O}_{i_6}$};
\draw[-] (4,0)++(90:1)--+(150:1) node[above]{$\mathcal{O}_{i_5}$};
\draw[-] (4,0)--+(-30:1) node[pos=0.4,below]{$\mathcal{O}_{k_1}$};
\draw[-] (4,0)++(-30:1)--+(-90:1) node[below]{$\mathcal{O}_{i_2}$};
\draw[-] (4,0)++(-30:1)--+(30:1) node[pos=0.8,yshift=-0.1cm,below]{$\mathcal{O}_{k_4}$};
\draw[-] (4,0)++(-30:1)--++(30:1)--+(90:1) node[above]{$\mathcal{O}_{i_7}$};
\draw[-] (4,0)++(-30:1)--++(30:1)--+(-30:1) node[right]{$\mathcal{O}_{i_1}$};
\draw[dashed,red] (4,0)++(-30:0.5)++(-120:0.5)--+(60:1);
\node at (7.6,0) {$=$};
\end{scope}
\begin{scope}[xshift=6.5cm]
\draw[-] (4,0)--+(-150:1) node[pos=0.6,above]{$\mathcal{O}_{k_2}$};
\draw[-] (4,0)++(-150:1)--+(-90:1) node[below]{$\mathcal{O}_{i_3}$};
\draw[-] (4,0)++(-150:1)--+(150:1) node[left]{$\mathcal{O}_{i_4}$};
\draw[-] (4,0)--+(90:1) node[pos=0.5,right]{$\mathcal{O}_{k_3}$};
\draw[-] (4,0)++(90:1)--+(30:1) node[above]{$\mathcal{O}_{i_6}$};
\draw[-] (4,0)++(90:1)--+(150:1) node[above]{$\mathcal{O}_{i_5}$};
\draw[-] (4,0)--+(-30:1) node[pos=0.4,below]{$\mathcal{O}_{k_1}$};
\draw[dashed,red] (4,0)++(-30:1)--+(-90:1);
\draw[dashed,red] (4,0)++(-30:1)--+(30:1);
\draw[dashed,red] (7,0)--+(90:1);
\draw[dashed,red] (7,0)--+(-150:1);
\draw[-] (7,0)--+(-30:1) node[pos=0.4,below]{$\mathcal{O}_{k_1}$};
\draw[-] (7,0)++(-30:1)--+(-90:1) node[below]{$\mathcal{O}_{i_2}$};
\draw[-] (7,0)++(-30:1)--+(30:1) node[pos=0.8,yshift=-0.1cm,below]{$\mathcal{O}_{k_4}$};
\draw[-] (7,0)++(-30:1)--++(30:1)--+(90:1) node[above]{$\mathcal{O}_{i_7}$};
\draw[-] (7,0)++(-30:1)--++(30:1)--+(-30:1) node[right]{$\mathcal{O}_{i_1}$};
\end{scope}
\begin{scope}[yshift=-5cm]
\draw[-] (4,0)--+(-150:1) node[pos=0.6,above]{$\mathcal{O}_{k_2}$};
\draw[-] (4,0)++(-150:1)--+(-90:1) node[below]{$\mathcal{O}_{i_3}$};
\draw[-] (4,0)++(-150:1)--+(150:1) node[left]{$\mathcal{O}_{i_4}$};
\draw[-] (4,0)--+(90:1) node[pos=0.5,right]{$\mathcal{O}_{k_3}$};
\draw[-] (4,0)++(90:1)--+(30:1) node[above]{$\mathcal{O}_{i_6}$};
\draw[-] (4,0)++(90:1)--+(150:1) node[above]{$\mathcal{O}_{i_5}$};
\draw[-] (4,0)--+(-30:1) node[pos=0.4,below]{$\mathcal{O}_{k_1}$};
\draw[-] (4,0)++(-30:1)--+(-90:1) node[below]{$\mathcal{O}_{i_2}$};
\draw[-] (4,0)++(-30:1)--+(30:1) node[pos=0.8,yshift=-0.1cm,below]{$\mathcal{O}_{k_4}$};
\draw[-] (4,0)++(-30:1)--++(30:1)--+(90:1) node[above]{$\mathcal{O}_{i_7}$};
\draw[-] (4,0)++(-30:1)--++(30:1)--+(-30:1) node[right]{$\mathcal{O}_{i_1}$};
\draw[dashed,red] (4,0)++(-150:0.5)++(-60:0.5)--+(120:1);
\node at (7.6,0) {$=$};
\end{scope}
\begin{scope}[xshift=6.5cm,yshift=-5cm]
\draw[-] (4,0)--+(-150:1) node[pos=0.6,above]{$\mathcal{O}_{k_2}$};
\draw[-] (4,0)++(-150:1)--+(-90:1) node[below]{$\mathcal{O}_{i_3}$};
\draw[-] (4,0)++(-150:1)--+(150:1) node[left]{$\mathcal{O}_{i_4}$};
\draw[dashed,red] (4,0)--+(-30:1);
\draw[dashed,red] (4,0)--+(90:1);
\draw[dashed,red] (7,0)++(-150:1)--+(-90:1);
\draw[dashed,red] (7,0)++(-150:1)--+(150:1);
\draw[-] (7,0)--+(-150:1) node[pos=0.6,above]{$\mathcal{O}_{k_2}$};
\draw[-] (7,0)--+(90:1) node[pos=0.5,right]{$\mathcal{O}_{k_3}$};
\draw[-] (7,0)++(90:1)--+(30:1) node[above]{$\mathcal{O}_{i_6}$};
\draw[-] (7,0)++(90:1)--+(150:1) node[above]{$\mathcal{O}_{i_5}$};
\draw[-] (7,0)--+(-30:1) node[pos=0.4,below]{$\mathcal{O}_{k_1}$};
\draw[-] (7,0)++(-30:1)--+(-90:1) node[below]{$\mathcal{O}_{i_2}$};
\draw[-] (7,0)++(-30:1)--+(30:1) node[pos=0.8,yshift=-0.1cm,below]{$\mathcal{O}_{k_4}$};
\draw[-] (7,0)++(-30:1)--++(30:1)--+(90:1) node[above]{$\mathcal{O}_{i_7}$};
\draw[-] (7,0)++(-30:1)--++(30:1)--+(-30:1) node[right]{$\mathcal{O}_{i_1}$};
\end{scope}
\end{tikzpicture}
}
\caption{Cutting procedure implemented on two different internal lines of the extended snowflake topology, leading to two factorization properties of its $F$-function.}
\label{Fig7=65}
\end{figure}

Indeed, concentrating on the extended snowflake, Figure \ref{Fig7=65} leads to several identities for its $F$-function.
The two identities depicted in Figure \ref{Fig7=65} are
\eqn{F_{7|\substack{\text{extended}\\\text{snowflake}}}=F_{6|\text{snowflake}}F_{5|\text{comb}}\qquad\text{and}\qquad F_{7|\substack{\text{extended}\\\text{snowflake}}}=F_{4|\text{comb}}F_{7|\substack{\text{extended}\\\text{snowflake}}},}
where the second identity is trivial due to $F_{4|\text{comb}}=1$.  Obviously, there are two extra trivial identities when the cutting procedure is performed on the internal lines with $\mathcal{O}_{k_3}$ and $\mathcal{O}_{k_4}$, respectively.

To verify that the cutting procedure of Figure \ref{FigCutting} makes sense, we investigate its implications for the other topologies, starting from the building blocks $F_{4|\text{comb}}=1$, $F_{5|\text{comb}}$ and $F_{6|\text{snowflake}}$.
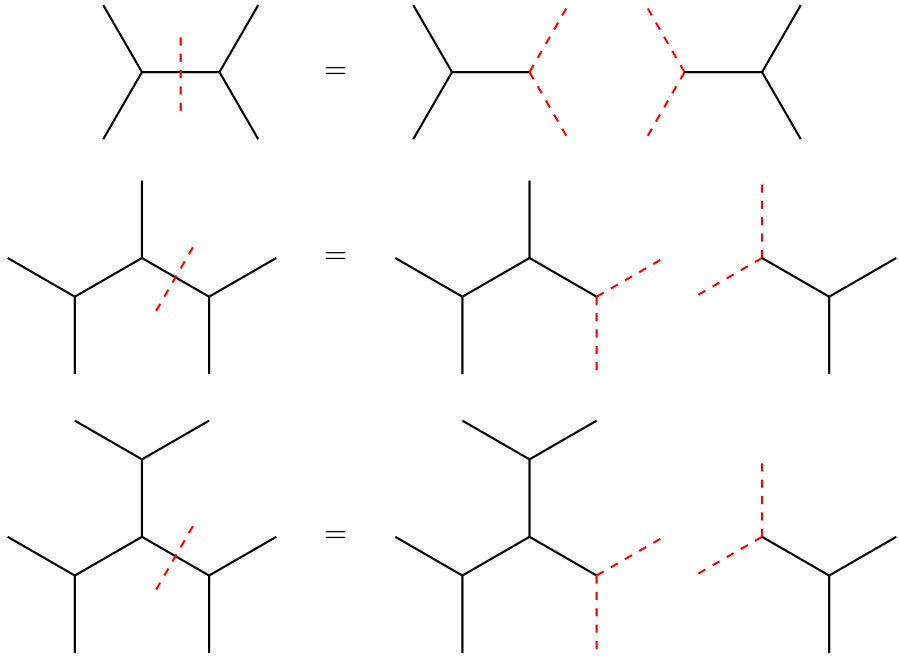
\begin{figure}[t]
\centering
\resizebox{12cm}{!}{%
\begin{tikzpicture}[thick]
\begin{scope}
\draw[-] (0,0)--+(120:1);
\draw[-] (0,0)--+(-120:1);
\draw[-] (0,0)--+(0:1);
\draw[-] (1,0)--+(60:1);
\draw[-] (1,0)--+(-60:1);
\draw[dashed,red] (0.5,-0.5)--+(90:1);
\node at (2.5,0) {$=$};
\draw[-] (4,0)--+(120:1);
\draw[-] (4,0)--+(-120:1);
\draw[-] (4,0)--+(0:1);
\draw[dashed,red] (5,0)--+(60:1);
\draw[dashed,red] (5,0)--+(-60:1);
\draw[dashed,red] (7,0)--+(120:1);
\draw[dashed,red] (7,0)--+(-120:1);
\draw[-] (7,0)--+(0:1);
\draw[-] (8,0)--+(60:1);
\draw[-] (8,0)--+(-60:1);
\end{scope}
\begin{scope}[yshift=-2.4cm]
\draw[-] (0,0)--+(-150:1);
\draw[-] (0,0)++(-150:1)--+(-90:1);
\draw[-] (0,0)++(-150:1)--+(150:1);
\draw[-] (0,0)--+(90:1);
\draw[-] (0,0)--+(-30:1);
\draw[-] (0,0)++(-30:1)--+(-90:1);
\draw[-] (0,0)++(-30:1)--+(30:1);
\draw[dashed,red] (0,0)++(-30:0.5)++(-120:0.5)--+(60:1);
\node at (2.5,0) {$=$};
\draw[-] (5,0)--+(-150:1);
\draw[-] (5,0)++(-150:1)--+(-90:1);
\draw[-] (5,0)++(-150:1)--+(150:1);
\draw[-] (5,0)--+(90:1);
\draw[-] (5,0)--+(-30:1);
\draw[dashed,red] (5,0)++(-30:1)--+(-90:1);
\draw[dashed,red] (5,0)++(-30:1)--+(30:1);
\draw[dashed,red] (8,0)--+(90:1);
\draw[dashed,red] (8,0)--+(210:1);
\draw[-] (8,0)--+(-30:1);
\draw[-] (8,0)++(-30:1)--+(-90:1);
\draw[-] (8,0)++(-30:1)--+(30:1);
\end{scope}
\begin{scope}[yshift=-6cm]
\draw[-] (0,0)--+(-150:1);
\draw[-] (0,0)++(-150:1)--+(-90:1);
\draw[-] (0,0)++(-150:1)--+(150:1);
\draw[-] (0,0)--+(90:1);
\draw[-] (0,0)++(90:1)--+(30:1);
\draw[-] (0,0)++(90:1)--+(150:1);
\draw[-] (0,0)--+(-30:1);
\draw[-] (0,0)++(-30:1)--+(-90:1);
\draw[-] (0,0)++(-30:1)--+(30:1);
\draw[dashed,red] (0,0)++(-30:0.5)++(-120:0.5)--+(60:1);
\node at (2.5,0) {$=$};
\draw[-] (5,0)--+(-150:1);
\draw[-] (5,0)++(-150:1)--+(-90:1);
\draw[-] (5,0)++(-150:1)--+(150:1);
\draw[-] (5,0)--+(90:1);
\draw[-] (5,0)++(90:1)--+(30:1);
\draw[-] (5,0)++(90:1)--+(150:1);
\draw[-] (5,0)--+(-30:1);
\draw[dashed,red] (5,0)++(-30:1)--+(-90:1);
\draw[dashed,red] (5,0)++(-30:1)--+(30:1);
\draw[dashed,red] (8,0)--+(90:1);
\draw[dashed,red] (8,0)--+(210:1);
\draw[-] (8,0)--+(-30:1);
\draw[-] (8,0)++(-30:1)--+(-90:1);
\draw[-] (8,0)++(-30:1)--+(30:1);
\end{scope}
\end{tikzpicture}
}
\caption{Cutting procedure for the $F$-function building blocks.}
\label{Fig456}
\end{figure}
In these cases, we get the results shown in Figure \ref{Fig456} which imply
\eqn{F_{4|\text{comb}}=F_{4|\text{comb}}F_{4|\text{comb}},\qquad F_{5|\text{comb}}=F_{5|\text{comb}}F_{4|\text{comb}},\qquad F_{6|\text{snowflake}}=F_{6|\text{snowflake}}F_{4|\text{comb}}.}
These identities lead to $F_{4|\text{comb}}=1$ as expected, and suggest that $F_{5|\text{comb}}$ and $F_{6|\text{snowflake}}$ cannot be determined by the cutting procedure---they have to be computed independently, for example from the embedding space OPE formalism \cite{Fortin:2019zkm,Fortin:2020yjz}.  More complicated topologies always generate non-trivial factorization properties leading to the full determination of their $F$-functions from the building blocks (with the total number of extra sums appearing in $F_M$ always fixed to $M-4$).

\begin{figure}[t]
\centering
\resizebox{6cm}{!}{%
\begin{tikzpicture}[thick]
\begin{scope}
\draw[-] (0,0)--+(-150:1);
\draw[-] (0,0)++(-150:1)--+(-90:1);
\draw[-] (0,0)++(-150:1)--+(150:1);
\draw[-] (0,0)++(-150:1)++(150:1)--+(90:1);
\draw[-] (0,0)++(-150:1)++(150:1)--+(-150:1);
\draw[-] (0,0)--+(90:1);
\draw[-] (0,0)++(90:1)--+(30:1);
\draw[-] (0,0)++(90:1)--+(150:1);
\draw[-] (0,0)--+(-30:1);
\draw[-] (0,0)++(-30:1)--+(-90:1);
\draw[-] (0,0)++(-30:1)++(-90:1)--+(-30:1);
\draw[-] (0,0)++(-30:1)++(-90:1)--+(-150:1);
\draw[-] (0,0)++(-30:1)--+(30:1);
\draw[-] (0,0)++(-30:1)--++(30:1)--+(90:1);
\draw[-] (0,0)++(-30:1)--++(30:1)--+(-30:1);
\draw[-] (0,0)++(-30:1)--++(30:1)++(-30:1)--+(30:1);
\draw[-] (0,0)++(-30:1)--++(30:1)++(-30:1)--+(-90:1);
\end{scope}
\end{tikzpicture}
}
\caption{Cutting procedure for one specific topology appearing in ten-point correlation functions.}
\label{Fig10pt}
\end{figure}
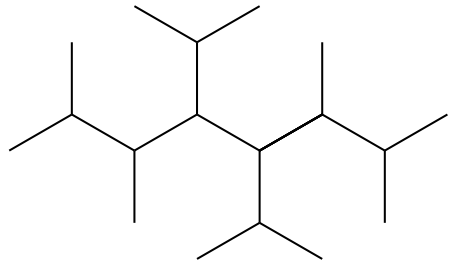

For example, for $F_{M|\text{comb}}$, cutting the internal line with quasi-primary operator $\mathcal{O}_{k_a}$ in the topology depicted in Figure \ref{FigComb} leads to $F_{M|\text{comb}}=F_{3+a|\text{comb}}F_{M+1-a|\text{comb}}$ for any $1\leq a\leq M-3$.  From its definition \eqref{EqCBFcomb}, this identity is verified for any $a$.  Moreover, repetitively cutting the comb topology to extract factors of $F_{5|\text{comb}}$ leads to the identity $F_{M|\text{comb}}=(F_{5|\text{comb}})^{M-4}$ mentioned above.  Considering the ten-point correlation function in the topology of Figure \ref{Fig10pt} as another example,
we expect schematically that
\eqn{F_{10|\text{topology of Figure \ref{Fig10pt}}}=(F_{5|\text{comb}}F_{6|\text{snowflake}})^2,}
where the proper parameters are determined by the quasi-primary operators as in \eqref{EqFcomb} and \eqref{EqFsnowflake}.

In general, from the counting of the number of extra sums, the factorization property should imply that
\eqn{F_{M|\text{any topology}}=(F_{5|\text{comb}})^{M-4-2n}(F_{6|\text{snowflake}})^n,}[EqFactor]
for $0\leq n\leq\lfloor\frac{M-4}{2}\rfloor$.  From Figure \ref{Fig456}, the factorization \eqref{EqFactor} should also be unique, meaning a given topology should have only one value of $n$ in the factorization that depends on the topology.  Going the opposite way, one should be able to reconstruct the possible topologies by gluing the $F$-function building blocks $F_{5|\text{comb}}$ and $F_{6|\text{snowflake}}$ ($F_{4|\text{comb}}$ being trivial, it can be discarded) following a gluing procedure analog to the opposite of the cutting procedure shown in Figure \ref{FigCutting}.  Indeed, for a fixed $M$-point correlation function and $n$-factorization \eqref{EqFactor}, the gluing procedure should lead to all inequivalent topologies for fixed $M$ and $n$.  Denoting the number of inequivalent $M$-point topologies with $n$-factorization as $T_0(M;n)$ and taking into account the uniqueness of the factorization, this observation implies that the number of unrooted binary trees with $M$ unlabeled leaves should be expressible as
\eqn{T_0(M)=\sum_{n\geq0}^{\lfloor\frac{M-4}{2}\rfloor}T_0(M;n).}
Hence, if $T_0(M;n)$ can be obtained by the gluing procedure, it should lead to an expression for $T_0(M)$.

Due to their symmetries, the building blocks $F_{5|\text{comb}}$ and $F_{6|\text{snowflake}}$ can be glued in only one independent way, however they can be glued at several different locations on an existing topology.  For example, for $n=0$, the factorization \eqref{EqFactor} leads to $(F_{5|\text{comb}})^{M-4}$ and since the $F_{5|\text{comb}}$ building blocks can only be glued together in one inequivalent topology associated to the comb, we find that $T_0(M;0)=1$ corresponding to the comb channel.  This is not the case when $n>0$ since the $F_{6|\text{snowflake}}$ building blocks can be glued together in different topologies.  Indeed, in the case of $M=8$ we have $T_0(8;1)=2$ since the two $F_{5|\text{comb}}$ building blocks can be glued in two inequivalent ways with $F_{6|\text{snowflake}}$, while for $M=12$ we have $T_0(12;4)=2$ because the four $F_{6|\text{snowflake}}$ building blocks can be glued in two inequivalent topologies.  For $n=1$, we find that $T_0(M;1)$ is given by the number of partitions of $M-6$ with at most three parts, which is given by $\left[\frac{(M-3)^2}{12}\right]$ where $[\,]$ means rounding to the nearest integer.  It would be interesting to study higher $n$-factorizations in general.

To summarize, we conjecture that there exist rules as in \eqref{EqEEI}, \eqref{EqEII} and \eqref{EqIII} to construct the $C$-function as well as \eqref{EqFcomb} and \eqref{EqFsnowflake} to construct the $F$-function, for any topology.  These rules rely on the OPE limit and the limit of unit operator for consistency, and they necessitate a proper set of conformal cross-ratios which is not necessarily simple to find.  We hope to verify and eventually prove such rules in future work.

Finally, other research avenues worth pursuing include the conformal bootstrap from higher-point correlation functions (where the knowledge of the correlation functions with external quasi-primary operators in scalar representations only can replace the usual conformal bootstrap of four-point correlation functions when fermions are discarded), higher-point correlation functions of quasi-primary operators in arbitrary irreducible representations, and their use in the AdS/CFT correspondence (for example with respect to geodesic Witten diagrams, see for instance \cite{Jepsen:2019svc}).

%%%%%%%%%%%%%%%%%%%%%%%%%%%%%%%%%%%%%%%%%%%%%%%%%%
%%%%%%%%%%%%%%%%%%%%%%%%%%%%%%%%%%%%%%%%%%%%%%%%%%

\ack{
The authors would like to thank Valentina Prilepina for useful discussions.  The work of JFF is supported by NSERC.  WJM is supported by the China Scholarship Council and in part by NSERC.  The work of WS is supported in part by DOE HEP grant DE-SC00-17660.
}

%%%%%%%%%%%%%%%%%%%%%%%%%%%%%%%%%%%%%%%%%%%%%%%%%%
%%%%%%%%%%%%%%%%%%%%%%%%%%%%%%%%%%%%%%%%%%%%%%%%%%

\setcounter{section}{0}
\renewcommand{\thesection}{\Alph{section}}
	
\section{Extended Snowflake and the OPE}\label{SAppCB7}

This appendix presents the proof leading to the scalar seven-point conformal blocks in the extended snowflake channel, starting from the OPE acting on the scalar six-point conformal blocks in the snowflake channel found in \cite{Fortin:2020yjz}.

The proof consists in several re-summations of the hypergeometric type.  For example, we use the binomial identity
\eqn{(1-v)^{a+b}=\sum_{i\geq0}(-1)^i\binom{a+b}{i}v^i=\sum_{i,j\geq0}(-1)^{i+j}\binom{a}{i}\binom{b}{j}v^{i+j},}[EqBinom]
as well as
\eqn{
\begin{gathered}
{}_2F_1(a,b;c;z)=(1-z)^{c-a-b}{}_2F_1(c-a,c-b;c;z),\\
{}_2F_1\left[\begin{array}{c}-n,b\\c\end{array};1\right]=\frac{(c-b)_n}{(c)_n},
\end{gathered}
}[Eq2F1]
and
\eqn{
\begin{gathered}
{}_3F_2\left[\begin{array}{c}-n,b,c\\d,e\end{array};1\right]=\frac{(d-b)_n}{(d)_n}{}_3F_2\left[\begin{array}{c}-n,b,e-c\\b-d-n+1,e\end{array};1\right],\\
{}_3F_2\left[\begin{array}{c}-n,b,c\\d,1+b+c-d-n\end{array};1\right]=\frac{(d-b)_n(d-c)_n}{(d)_n(d-b-c)_n},
\end{gathered}
}[Eq3F2]
for $n$ a non-negative integer.

%%%%%%%%%%%%%%%%%%%%%%%%%%%%%%%%%%%%%%%%%%%%%%%%%%
	
\subsection{Proof of the Extended Snowflake}

In order to use \eqref{EqIfromOPE}, it is necessary to first shift the quasi-primary operators in the scalar six-point correlation functions \eqref{EqCB6CR}, \eqref{EqCB6C}, and \eqref{EqCB6F}, such that $\mathcal{O}_{i_a}(\eta_a)\to\mathcal{O}_{i_{a-1}}(\eta_{a-1})$ with $\mathcal{O}_{i_0}(\eta_0)\equiv\mathcal{O}_{i_6}(\eta_6)$.  Doing so, the legs and the conformal cross-ratios transform as
\eqn{
\begin{gathered}
L_{6|\text{snowflake}}^{(\Delta_{i_2},\ldots,\Delta_{i_6},\Delta_{k_4})}=\left(\frac{\eta_{16}}{\eta_{12}\eta_{26}}\right)^{\frac{\Delta_{i_2}}{2}}\left(\frac{\eta_{24}}{\eta_{23}\eta_{34}}\right)^{\frac{\Delta_{i_3}}{2}}\left(\frac{\eta_{23}}{\eta_{24}\eta_{34}}\right)^{\frac{\Delta_{i_4}}{2}}\\
\qquad\qquad\qquad\qquad\qquad\qquad\times\left(\frac{\eta_{46}}{\eta_{45}\eta_{56}}\right)^{\frac{\Delta_{i_5}}{2}}\left(\frac{\eta_{45}}{\eta_{46}\eta_{56}}\right)^{\frac{\Delta_{i_6}}{2}}\left(\frac{\eta_{26}}{\eta_{12}\eta_{16}}\right)^{\frac{\Delta_{k_4}}{2}},\\
u_1^6=\frac{\eta_{12}\eta_{46}}{\eta_{16}\eta_{24}},\qquad u_2^6=\frac{\eta_{26}\eta_{34}}{\eta_{23}\eta_{46}},\qquad u_3^6=\frac{\eta_{24}\eta_{56}}{\eta_{26}\eta_{45}},\\
v_{11}^6=\frac{\eta_{14}\eta_{26}}{\eta_{16}\eta_{24}},\qquad v_{12}^6=\frac{\eta_{24}\eta_{36}}{\eta_{23}\eta_{46}},\qquad v_{13}^6=\frac{\eta_{15}\eta_{46}}{\eta_{16}\eta_{45}},\\
v_{22}^6=\frac{\eta_{13}\eta_{26}}{\eta_{16}\eta_{23}},\qquad v_{23}^6=\frac{\eta_{25}\eta_{46}}{\eta_{26}\eta_{45}},\qquad v_{33}^6=\frac{\eta_{24}\eta_{35}}{\eta_{23}\eta_{45}},
\end{gathered}
}[EqLCR]
where we also changed $\mathcal{O}_{i_1}(\eta_1)\to\mathcal{O}_{k_4}(\eta_1)$ to implement the recurrence relation \eqref{EqIfromOPE}.  The shift also implies that the vectors $\boldsymbol{h}$ and $\boldsymbol{p}$ become the vectors \eqref{EqCB7hp} when the new components $h_6$ and $p_7$ appearing in \eqref{EqIfromOPE} are included.  In addition, we note that $C_6$ \eqref{EqCB6C} and $F_6$ \eqref{EqCB6F} are the same functions but of the new vectors $\boldsymbol{h}$ and $\boldsymbol{p}$ \eqref{EqCB7hp}, thus $G_6$ stays the same function but of the new vectors and conformal cross-ratios \eqref{EqLCR}.

We can now use the recurrence relation \eqref{EqIfromOPE} (choosing $k=5$, $l=6$, and $m=6$) to reach the scalar seven-point conformal blocks in the extended snowflake channel.  To this end, we apply the OPE differential operator following \eqref{EqD} on the conformal cross-ratios \eqref{EqCROPE}, \textit{i.e.}
\eqn{
\begin{gathered}
x_6^7=\frac{\eta_{17}\eta_{56}}{\eta_{57}\eta_{16}},\qquad y_2^7=1-\frac{\eta_{12}\eta_{67}}{\eta_{16}\eta_{27}},\qquad y_3^7=1-\frac{\eta_{13}\eta_{67}}{\eta_{16}\eta_{37}},\qquad y_4^7=1-\frac{\eta_{14}\eta_{67}}{\eta_{16}\eta_{47}},\qquad y_5^7=1-\frac{\eta_{15}\eta_{67}}{\eta_{16}\eta_{57}},\\
z_{23}^7=\frac{\eta_{57}\eta_{67}\eta_{23}}{\eta_{27}\eta_{37}\eta_{56}},\qquad z_{24}^7=\frac{\eta_{24}\eta_{57}\eta_{67}}{\eta_{27}\eta_{47}\eta_{56}},\qquad z_{25}^7=\frac{\eta_{25}\eta_{67}}{\eta_{27}\eta_{56}},\qquad z_{26}^7=\frac{\eta_{26}\eta_{57}}{\eta_{27}\eta_{56}},\\
z_{34}^7=\frac{\eta_{34}\eta_{57}\eta_{67}}{\eta_{37}\eta_{47}\eta_{56}},\qquad z_{35}^7=\frac{\eta_{35}\eta_{67}}{\eta_{37}\eta_{56}},\qquad z_{36}^7=\frac{\eta_{36}\eta_{57}}{\eta_{37}\eta_{56}},\qquad z_{45}^7=\frac{\eta_{45}\eta_{67}}{\eta_{47}\eta_{56}},\qquad z_{46}^7=\frac{\eta_{46}\eta_{57}}{\eta_{47}\eta_{56}}.
\end{gathered}
}[Eqxyz]
Therefore, we must first re-express the conformal cross-ratios \eqref{EqLCR} in terms of the conformal cross-ratios \eqref{Eqxyz}, which leads to
\eqn{u_1^6=\frac{1-y_2^7}{1-y_6^7}\frac{z_{46}^7}{z_{24}^7},\qquad v_{11}^6=\frac{1-y_4^7}{1-y_6^7}\frac{z_{26}^7}{z_{24}^7},\qquad v_{13}^6=\frac{1-y_5^7}{1-y_6^7}\frac{z_{46}^7}{z_{45}^7},\qquad v_{22}^6=\frac{1-y_3^7}{1-y_6^7}\frac{z_{26}^7}{z_{23}^7}.}
After acting with the OPE differential operator, we then substitute the conformal cross-ratios \eqref{Eqxyz} with the conformal cross-ratios \eqref{EqCB7CR} using
\eqn{
\begin{gathered}
x_6^7=\frac{u_1^7u_3^7u_4^7}{v_{44}^7},\qquad y_2^7=1-\frac{1}{v_{14}^7},\qquad y_3^7=1-\frac{v_{22}^7}{v_{24}^7},\qquad y_4^7=1-\frac{v_{11}^7}{v_{34}^7},\qquad y_5^7=1-\frac{v_{13}^7}{v_{44}^7},\\
z_{23}^7=\frac{v_{44}^7}{v_{14}^7v_{24}^7u_1^7u_3^7},\qquad z_{24}^7=\frac{v_{44}^7}{v_{14}^7v_{34}^7u_1^7u_3^7},\qquad z_{25}^7=\frac{v_{23}^7}{v_{14}^7u_1^7u_3^7},\qquad z_{26}^7=\frac{v_{44}^7}{v_{14}^7u_1^7u_3^7},\\
z_{34}^7=\frac{v_{44}^7u_2^7}{v_{24}^7v_{34}^7u_3^7},\qquad z_{35}^7=\frac{v_{33}^7}{v_{24}^7u_3^7},\qquad z_{36}^7=\frac{v_{12}^7v_{44}^7}{v_{24}^7u_3^7},\qquad z_{45}^7=\frac{1}{v_{34}^7u_3^7},\qquad z_{46}^7=\frac{v_{44}^7}{v_{34}^7u_3^7},
\end{gathered}
}
which results in
\eqna{
G_7&=\sum(-1)^{l_2+l_3+l_4+l_5+m_{11}+m_{13}+m_{14}+m_{22}+m_{24}+m_{34}+m_{44}+k_{12}+k_{23}+k_{33}+s_{11}+s_{22}+s_{13}}\\
&\phantom{=}\qquad\times\frac{(-h_6)_{m_4}(-h_3+n_1+s_{11}+s_{22}+s_{13})_{\bar{r}_6}(p_4-h_3+h_6)_{m_4+\sum_{a=2}^5\sigma_a}}{(p_4-h_3)_{2m_4+\sum_{a=2}^5\sigma_a}(p_4-h_3+1-d/2)_{m_4}}\\
&\phantom{=}\qquad\times\frac{(p_4-n_1)_{\sigma_2+r_{26}+\bar{r}_2}(-s_{22})_{\sigma_3+r_{36}+\bar{r}_3}}{\sigma_2!\sigma_3!}\frac{(-s_{11})_{\sigma_4+r_{46}+\bar{r}_4}(-s_{13})_{\sigma_5+r_{56}+\bar{r}_5}}{\sigma_4!\sigma_5!}\binom{n_{11}}{s_{11}}\binom{n_{22}}{s_{22}}\binom{n_{13}}{s_{13}}\\
&\phantom{=}\qquad\times\binom{\sigma_2}{l_2}\binom{\sigma_3}{l_3}\binom{\sigma_4}{l_4}\binom{\sigma_5}{l_5}\binom{l_5}{m_{13}}\binom{-p_4+n_1-r_{26}-\bar{r}_2-l_2}{m_{14}}\binom{l_4}{m_{11}}\binom{r_{36}}{k_{12}}\\
&\phantom{=}\qquad\times\binom{l_3}{m_{22}}\binom{r_{25}}{k_{23}}\binom{r_{35}}{k_{33}}\binom{s_{22}-r_{36}-\bar{r}_3-l_3}{m_{24}}\binom{s_{11}-r_{46}-\bar{r}_4-l_4}{m_{34}}\binom{s_{13}-r_{56}-\bar{r}_5-l_5}{m_{44}}\\
&\phantom{=}\qquad\times\frac{ (u_1^7)^{n_1+\sum_{a,b\neq 2}r_{ab}}(u_2^7)^{n_2+r_{34}}(u_3^7)^{n_3+r_{56}}(u_4^7)^{\bar{r}_{6}+\bar{\bar{r}}}\prod_{1\leq a\leq b\leq4}(1-v_{ab}^7)^{m_{ab}}}{r_{23}!r_{24}!r_{25}!r_{26}!r_{34}!r_{35}!r_{36}!r_{45}!r_{46}!r_{56}!n_1!n_2!n_3!n_{11}!m_{12}!n_{13}!n_{22}!m_{23}!n_{33}!}C_6F_6,
}[EqG7extrasums]
with the appropriate legs \eqref{EqCB7CR} and also $n_{12}=m_{12}$, $n_{23}=m_{23}$, $m_{12}=n_{12}+k_{12}$, $m_{23}=n_{23}+k_{23}$, and $n_{33}+k_{33}=m_{33}$ after some simple manipulations (here the new indices of summation originating from the OPE are denoted by $r$).  In \eqref{EqG7extrasums} and most of the appendixes, we omit the indices under the summation sign to avoid cluttering the equations.  The scalar seven-point conformal blocks in the extended snowflake channel are thus given explicitly by \eqref{EqG7extrasums}, and the rest of this appendix is dedicated to re-summing the extra sums [using \eqref{EqBinom}, \eqref{Eq2F1} and \eqref{Eq3F2} repetitively] to reach the results stated in \eqref{EqCB7C} and \eqref{EqCB7F}.

We first evaluate the sums over $l_a$ and $\sigma_a$ by changing variables to
\eqn{
\begin{gathered}
l_3\to l_3+m_{22},\qquad  l_4\to l_4+m_{11},\qquad l_5\to l_5+m_{13},\\
\sigma_3\to\sigma_3+m_{22},\qquad\sigma_4\to\sigma_4+m_{11},\qquad\sigma_5\to\sigma_5+m_{13},
\end{gathered}
}
which results in
\eqna{
&G_7=\sum(-1)^{k_{12}+k_{23}+k_{33}+s_{11}+s_{22}+s_{13}}\binom{n_{11}}{s_{11}}\binom{n_{22}}{s_{22}}\binom{n_{13}}{s_{13}}\binom{r_{36}}{k_{12}}\binom{r_{25}}{k_{23}}\binom{r_{35}}{k_{33}}\\
&\qquad\times\frac{(p_4-n_1)_{m_{14}+r_{26}+\bar{r}_2}(-s_{22})_{m_{22}+m_{24}+r_{36}+\bar{r}_3}}{r_{23}!r_{24}!r_{25}!r_{26}!r_{34}!r_{35}!r_{36}!r_{45}!r_{46}!r_{56}!}\frac{m_{33}!(-s_{11})_{m_{11}+m_{34}+r_{46}+\bar{r}_4}(-s_{13})_{m_{13}+m_{44}+r_{56}+\bar{r}_5}}{n_{11}!n_{13}!n_{22}!n_{33}!}\\
&\qquad\times\frac{(-h_6)_{m_4+m_{14}+m_{24}+m_{34}+m_{44}}(-h_3+n_1+s_{11}+s_{22}+s_{13})_{\bar{r}_6}(p_4-h_3+h_6)_{m_4+m_{11}+m_{13}+m_{22}}}{(p_4-h_3)_{2m_4+m_{11}+m_{13}+m_{14}+m_{22}+m_{24}+m_{34}+m_{44}}(p_4-h_3+1-d/2)_{m_4}}\\
&\qquad\times\frac{(u_1^7)^{n_1+\sum_{a,b\neq 2}r_{ab}}(u_2^7)^{n_2+r_{34}}(u_3^7)^{n_3+r_{56}}(u_4^7)^{\bar{r}_6+\bar{\bar{r}}}}{n_1!n_2!n_3!}\prod_{1\leq a\leq b\leq4}\frac{(1-v_{ab}^7)^{m_{ab}}}{m_{ab}!}C_6F_6.
}
We then change variables as
\eqn{
\begin{gathered}
s_{11}\to s_{11}+m_{11}+m_{34}+r_{46}+\bar{r}_4,\\
s_{22}\to s_{22}+m_{22}+m_{24}+r_{36}+\bar{r}_3,\\
s_{13}\to s_{13}+m_{13}+m_{44}+r_{56}+\bar{r}_5,
\end{gathered}
}
and define a new set of summation indices by
\eqn{
\begin{gathered}
r_{23}=r_{23},\qquad r_{34}=r_{34},\qquad r_{46}=r_{46},\qquad r_{56}=r_{56},\\
r_1^*=r_{23}+r_{24},\qquad r_2^*=r_1^*+r_{25},\qquad r_1=r_{34}+r_{35},\qquad r_2=r_1+r_{45},\\
r_{26}=m_4-r-r_2^*,r_{36}=r-r_{46}-r_{56}-r_2,
\end{gathered}
}
such that
\eqna{
&G_7=\sum\frac{(-h_6)_{m_4+m_{14}+m_{24}+m_{34}+m_{44}}(p_4-h_3+h_6)_{m_4+m_{11}+m_{13}+m_{22}}}{(p_4-h_3)_{2m_4+m_{11}+m_{13}+m_{14}+m_{22}+m_{24}+m_{34}+m_{44}}(p_4-h_3+1-d/2)_{m_4}}\\
&\qquad\times\frac{(p_4-n_1)_{m_4+m_{14}-r}}{r_{23}!(r_1^*-r_{23})!(r_2^*-r_1^*)!(m_4-r-r_2^*)!r_{34}!(r_1-r_{34})!(r-r_{46}-r_{56}-r_2)!(r_2-r_1)!r_{46}!r_{56}!}\\
&\qquad\times\frac{m_{33}!(-1)^{k_{12}+k_{23}+k_{33}+s_{11}+s_{22}+s_{13}}}{(n_{13}-s_{13}-m_{13}-m_{44}-r_{56}-r_2-r_2^*+r_1^*+r_{34})!n_{33}!}\binom{r-r_{46}-r_{56}-r_2}{k_{12}}\binom{r_2^*-r_1^*}{k_{23}}\binom{r_1-r_{34}}{k_{33}}\\
&\qquad\times\frac{(-h_3+n_1+m_{11}+m_{13}+m_{22}+m_{24}+m_{34}+m_{44}+s_{11}+s_{22}+s_{13}+r+r_2+r_2^*)_{m_4-r_2-r_2^*}}{(n_{22}-s_{22}-m_{22}-m_{24}-r-r_1-r_{23}+r_2+r_{46}+r_{56})!}\\
&\qquad\times\frac{1}{(n_{11}-s_{11}-m_{11}-m_{34}-r_{46}-r_1^*-r_2-r_{34}+r_1+r_{23})!s_{11}!s_{13}!s_{22}!}\\
&\qquad\times\frac{(u_1^7)^{n_1+r}(u_2^7)^{n_2+r_{34}}(u_3^7)^{n_3+r_{56}}(u_4^7)^{m_4}}{n_1!n_2!n_3!}\prod_{1\leq a\leq b\leq4}\frac{(1-v_{ab}^7)^{m_{ab}}}{m_{ab}!}C_6F_6.
}

We can now perform the summations over $r_{23}$, $r_1^*$, $r_2^*$, $r_{46}$, $r_1$, $r_2$, and finally $s_{13}$,\footnote{To evaluate the sums over $r_2^*$, $r_1$ and $r_2$, we first change variables by
\eqn{r_2^*\to r_2^*+k_{23},\qquad r_1\to r_1+r_{34}+k_{33},\qquad r_2\to r_2+r_{34}+k_{33}.}
}
 leading to
\eqna{
&G_7=\sum\frac{(-h_6)_{m_4+m_{14}+m_{24}+m_{34}+m_{44}}(p_4-h_3+h_6)_{m_4+m_{11}+m_{13}+m_{22}}}{(p_4-h_3)_{2m_4+m_{11}+m_{13}+m_{14}+m_{22}+m_{24}+m_{34}+m_{44}}(p_4-h_3+1-d/2)_{m_4}}\\
&\qquad\times\frac{(-h_3+n_1+n_{11}+n_{13}+n_{22})_{m_4-r-k_{23}}(p_4-n_1)_{m_4+m_{14}-r}}{(m_4-r-k_{23})!r_{34}!(r-r_{56}-r_{34}-k_{12}-k_{33})!r_{56}!k_{12}!k_{23}!k_{33}!}\\
&\qquad\times\frac{m_{33}!(-1)^{r-r_{34}-r_{56}+k_{23}+s_{11}+s_{22}}(-r_{56}-k_{12})_{n_{13}-m_{13}-m_{44}-r_{56}-k_{23}-k_{33}}}{(n_{13}-m_{13}-m_{44}-r_{56}-k_{23}-k_{33})!n_{33}!}\\
&\qquad\times\frac{(-h_3+n_1+n_{13}+m_4+m_{11}+m_{22}+m_{24}+m_{34}+s_{11}+s_{22}+r_{34}-r_{56}-k_{23})_{m_{13}+m_{44}-n_{13}+r+r_{56}+k_{23}-r_{34}}}{(n_{22}-s_{22}-m_{22}-m_{24}-r_{34}-k_{12}-k_{33})!}\\
&\qquad\times\frac{(-n_{11}-n_{22}+s_{11}+s_{22}+m_{11}+m_{22}+m_{24}+m_{34}+2r_{34}+k_{12}+k_{33})_{r-r_{34}-r_{56}-k_{12}-k_{33}}}{(n_{11}-s_{11}-m_{11}-m_{34}-r_{34})!s_{11}!s_{22}!}\\
&\qquad\times\frac{(u_1^7)^{n_1+r}(u_2^7)^{n_2+r_{34}}(u_3^7)^{n_3+r_{56}}(u_4^7)^{m_4}}{n_1!n_2!n_3!}\prod_{1\leq a\leq b\leq4}\frac{(1-v_{ab}^7)^{m_{ab}}}{m_{ab}!}C_6F_6.
}

After defining $s_{22}=s-s_{11}$, we evaluate the sum over $s_{11}$ and $s$, giving
\eqna{
&G_7=\sum\frac{(-h_6)_{m_4+m_{14}+m_{24}+m_{34}+m_{44}}(p_4-h_3+h_6)_{m_4+m_{11}+m_{13}+m_{22}}}{(p_4-h_3)_{2m_4+m_{11}+m_{13}+m_{14}+m_{22}+m_{24}+m_{34}+m_{44}}(p_4-h_3+1-d/2)_{m_4}}\\
&\qquad\times\frac{(-m_{13}-m_{44}+n_{13}+r_{34}-r-r_{56}-k_{23})_{n_{11}+n_{22}-m_{11}-m_{22}-m_{24}-m_{34}-r-r_{34}+r_{56}}(p_4-n_1)_{m_4+m_{14}-r}}{(m_4-r-k_{23})!r_{34}!(r-r_{56}-r_{34}-k_{12}-k_{33})!r_{56}!k_{12}!k_{23}!k_{33}!}\\
&\qquad\times\frac{m_{33}!(-1)^{r-r_{34}-r_{56}+k_{23}}(-r_{56}-k_{12})_{n_{13}-m_{13}-m_{44}-r_{56}-k_{23}-k_{33}}}{(n_{13}-m_{13}-m_{44}-r_{56}-k_{23}-k_{33})!(n_{22}-m_{22}-m_{24}-r_{34}-k_{12}-k_{33})!n_{33}!}\\
&\qquad\times\frac{(-n_{11}-n_{22}+m_{11}+m_{22}+m_{24}+m_{34}+2r_{34}+k_{12}+k_{33})_{r-r_{34}-r_{56}-k_{12}-k_{33}}}{(n_{11}-m_{11}-m_{34}-r_{34})!}\\
&\qquad\times\frac{(p_2+h_3)_{n_1+n_{23}}(p_3-h_2+h_4)_{n_2+n_{11}}(\bar{p}_3+h_2+h_5)_{n_3+n_{12}}}{(\bar{p}_3+h_2)_{2n_3+n_{12}+n_{13}+n_{23}+n_{33}}(\bar{p}_3+h_2+1-d/2)_{n_3}}\\
&\qquad\times\frac{(p_3)_{n_2+n_3-n_1+n_{12}+n_{33}}(p_2+h_2)_{n_1-n_2+n_3+n_{13}+n_{23}}(-h_2)_{n_1+n_2-n_3+n_{11}+n_{22}}}{(p_2)_{2n_1+n_{11}+n_{22}+n_{13}+n_{23}}(p_2+1-d/2)_{n_1}}\\
&\qquad\times\frac{(-h_3)_{m_1+m_4+m_{11}+m_{13}+m_{22}+m_{24}+m_{34}+m_{44}}(-h_4)_{n_2+n_{12}+n_{22}+n_{33}}(-h_5)_{n_3+n_{13}+n_{23}+n_{33}}}{(p_3-h_2)_{2n_2+n_{11}+n_{12}+n_{22}+n_{33}}(p_3-h_2+1-d/2)_{n_2}}\\
&\qquad\times\frac{(u_1^7)^{n_1+r}(u_2^7)^{n_2+r_{34}}(u_3^7)^{n_3+r_{56}}(u_4^7)^{m_4}}{n_1!n_2!n_3!}\prod_{1\leq a\leq b\leq4}\frac{(1-v_{ab}^7)^{m_{ab}}}{m_{ab}!}F_6,
}
where we explicitly expanded $C_6$ \eqref{EqCB6C}.

We then define $n_{22}=n-n_{11}$ and change variables by $n_{11}\to n_{11}+m_{11}+m_{34}+r_{34}$.  The sum over $n_{11}$ thus leads to
\eqna{
&G_7=\sum\frac{(-h_6)_{m_4+m_{14}+m_{24}+m_{34}+m_{44}}(p_4-h_3+h_6)_{m_4+m_{11}+m_{13}+m_{22}}}{(p_4-h_3)_{2m_4+m_{11}+m_{13}+m_{14}+m_{22}+m_{24}+m_{34}+m_{44}}(p_4-h_3+1-d/2)_{m_4}}\\
&\qquad\times\frac{(-m_{13}-m_{44}+n_{13}+r_{34}-r-r_{56}-k_{23})_{n-m_{11}-m_{22}-m_{24}-m_{34}-r-r_{34}+r_{56}}(p_4-n_1)_{m_4+m_{14}-r}}{(m_4-r-k_{23})!r_{34}!(r-r_{56}-r_{34}-k_{12}-k_{33})!r_{56}!k_{12}!k_{23}!k_{33}!}\\
&\qquad\times\frac{m_{33}!(-1)^{r-r_{34}-r_{56}+k_{23}}(-r_{56}-k_{12})_{n_{13}-m_{13}-m_{44}-r_{56}-k_{23}-k_{33}}}{(n_{13}-m_{13}-m_{44}-r_{56}-k_{23}-k_{33})!n_{33}!}\\
&\qquad\times\frac{(-n+m_{11}+m_{22}+m_{24}+m_{34}+2r_{34}+k_{12}+k_{33})_{r-r_{34}-r_{56}-k_{12}-k_{33}}}{(n-m_{11}-m_{34}-m_{22}-m_{24}-2r_{34}-k_{12}-k_{33})!}\\
&\qquad\times\frac{(p_2+h_3)_{n_1+n_{23}}(p_3-h_2+h_4)_{m_2+m_{11}+m_{34}}(\bar{p}_3+h_2+h_{5})_{n_3+n_{12}}}{(\bar{p}_3+h_2)_{2n_3+n_{12}+n_{13}+n_{23}+n_{33}}(\bar{p}_3+h_2+1-d/2)_{n_3}}\\
&\qquad\times\frac{(p_3)_{n_2+n_3-n_1+n_{12}+n_{33}}(p_2+h_2)_{n_1-n_2+n_3+n_{13}+n_{23}}(-h_2)_{n_1+n_2-n_3+n}}{(p_2)_{2n_1+n+n_{13}+n_{23}}(p_2+1-d/2)_{n_1}}\\
&\qquad\times\frac{(-h_3)_{m_1+m_4+m_{11}+m_{13}+m_{22}+m_{24}+m_{34}+m_{44}}(-h_4)_{m_2+m_{12}+m_{22}+m_{24}+m_{33}}(-h_5)_{n_3+n_{13}+n_{23}+n_{33}}}{(p_3-h_2)_{2m_2+m_{11}+m_{12}+m_{22}+m_{24}+m_{33}+m_{34}}(p_3-h_2+1-d/2)_{n_2}}\\
&\qquad\times\frac{(u_1^7)^{n_1+r}(u_2^7)^{n_2+r_{34}}(u_3^7)^{n_3+r_{56}}(u_4^7)^{m_4}}{n_1!n_2!n_3!}\prod_{1\leq a\leq b\leq4}\frac{(1-v_{ab}^7)^{m_{ab}}}{m_{ab}!}F_6.
}

After shifting $n$ such that $n\to n+m_{11}+m_{22}+m_{24}+m_{34}+r+r_{34}-r_{56}$, we can evaluate the sum over $n$, giving
\eqna{
&G_7=\sum\frac{(-h_6)_{m_4+m_{14}+m_{24}+m_{34}+m_{44}}(p_4-h_3+h_6)_{m_4+m_{11}+m_{13}+m_{22}}}{(p_4-h_3)_{2m_4+m_{11}+m_{13}+m_{14}+m_{22}+m_{24}+m_{34}+m_{44}}(p_4-h_3+1-d/2)_{m_4}}\\
&\qquad\times\frac{(p_4-n_1)_{m_4+m_{14}-r}}{(m_4-r-k_{23})!r_{34}!(r-r_{56}-r_{34}-k_{12}-k_{33})!r_{56}!k_{12}!k_{23}!k_{33}!}\\
&\qquad\times\frac{m_{33}!(-1)^{k_{12}+k_{23}+k_{33}}(-r_{56}-k_{12})_{n_{13}-m_{13}-m_{44}-r_{56}-k_{23}-k_{33}}}{(n_{13}-m_{13}-m_{44}-r_{56}-k_{23}-k_{33})!n_{33}!}\\
&\qquad\times\frac{(p_2+h_3)_{n_1+n_{23}}(p_3-h_2+h_4)_{m_2+m_{11}+m_{34}}(\bar{p}_3+h_2+h_{5})_{n_3+n_{12}}}{(\bar{p}_3+h_2)_{2n_3+n_{12}+n_{13}+n_{23}+n_{33}}(\bar{p}_3+h_2+1-d/2)_{n_3}}\\
&\qquad\times\frac{(p_3)_{n_{2}+n_3-n_1+n_{12}+n_{33}}(p_2+h_2)_{m_1-m_2+m_3+m_{13}+m_{23}+m_{44}}(-h_2)_{m_1+m_2-m_3+m_{11}+m_{22}+m_{24}+m_{34}}}{(p_2)_{2m_1+m_{11}+m_{13}+m_{22}+m_{23}+m_{24}+m_{34}+m_{44}}(p_2+1-d/2)_{n_1}}\\
&\qquad\times\frac{(-h_3)_{m_1+m_4+m_{11}+m_{13}+m_{22}+m_{24}+m_{34}+m_{44}}(-h_4)_{m_2+m_{12}+m_{22}+m_{24}+m_{33}}(-h_5)_{n_3+n_{13}+n_{23}+n_{33}}}{(p_3-h_2)_{2m_2+m_{11}+m_{12}+m_{22}+m_{24}+m_{33}+m_{34}}(p_3-h_2+1-d/2)_{n_2}}\\
&\qquad\times\frac{(u_1^7)^{n_1+r}(u_2^7)^{n_2+r_{34}}(u_3^7)^{n_3+r_{56}}(u_4^7)^{m_4}}{n_1!n_2!n_3!}\prod_{1\leq a\leq b\leq4}\frac{(1-v_{ab}^7)^{m_{ab}}}{m_{ab}!}F_6.
}
We then change variables again, this time $n_{13}\to n_{13}+m_{13}+m_{44}+r_{56}+k_{23}+k_{33}$, and sum over $n_{13}$, which leads to
\eqna{
&G_7=\sum\frac{(-h_6)_{m_4+m_{14}+m_{24}+m_{34}+m_{44}}(p_4-h_3+h_6)_{m_4+m_{11}+m_{13}+m_{22}}}{(p_4-h_3)_{2m_4+m_{11}+m_{13}+m_{14}+m_{22}+m_{24}+m_{34}+m_{44}}(p_4-h_3+1-d/2)_{m_4}}\\
&\qquad\times\frac{(p_4-n_1)_{m_4+m_{14}-r}(-h_5)_{m_3+m_{13}+m_{23}+m_{33}+m_{44}}}{(m_4-r-k_{23})!r_{34}!(r-r_{56}-r_{34}-k_{12}-k_{33})!r_{56}!k_{12}!k_{23}!k_{33}!}\\
&\qquad\times\frac{(p_2+h_3)_{n_1+n_{23}}(p_3-h_2+h_4)_{m_2+m_{11}+m_{34}}(\bar{p}_3+h_2+h_5)_{m_3+m_{12}}}{(\bar{p}_3+h_2)_{2m_3+m_{12}+m_{13}+m_{23}+m_{33}+m_{44}}(\bar{p}_3+h_2+1-d/2)_{n_3}}\\
&\qquad\times\frac{(p_3)_{n_2+n_3-n_1+n_{12}+n_{33}}(p_2+h_2)_{m_1-m_2+m_3+m_{13}+m_{23}+m_{44}}(-h_2)_{m_1+m_2-m_3+m_{11}+m_{22}+m_{24}+m_{34}}}{(p_2)_{2m_1+m_{11}+m_{13}+m_{22}+m_{23}+m_{24}+m_{34}+m_{44}}(p_2+1-d/2)_{n_1}}\\
&\qquad\times\frac{(-h_3)_{m_1+m_4+m_{11}+m_{13}+m_{22}+m_{24}+m_{34}+m_{44}}(-h_4)_{m_2+m_{12}+m_{22}+m_{24}+m_{33}}}{(p_3-h_2)_{2m_2+m_{11}+m_{12}+m_{22}+m_{24}+m_{33}+m_{34}}(p_3-h_2+1-d/2)_{n_2}}\\
&\qquad\times\frac{m_{12}!m_{23}!m_{33}!(-1)^{k_{12}+k_{23}+k_{33}}}{n_{12}!n_{23}!n_{33}!}\frac{(u_1^7)^{n_1+r}(u_2^7)^{n_2+r_{34}}(u_3^7)^{n_3+r_{56}}(u_4^7)^{m_4}}{n_1!n_2!n_3!}\prod_{1\leq a\leq b\leq 4}\frac{(1-v_{ab}^7)^{m_{ab}}}{m_{ab}!}F_6.
}

Using the fact that $m_{12}=n_{12}+k_{12}$, $m_{23}=n_{13}+k_{23}$, and $m_{33}=k_{33}+n_{33}$, we evaluate the sums over $k_{12}$, $k_{23}$, and $k_{33}$, to obtain
\eqna{
&G_7=\sum\frac{(p_4-m_1)_{m_4+m_{14}}(-h_6)_{m_4+m_{14}+m_{24}+m_{34}+m_{44}}(p_4-h_3+h_6)_{m_4+m_{11}+m_{13}+m_{22}}}{(p_2+h_3+m_1)_{-m_4}(p_4-h_3)_{2m_4+m_{11}+m_{13}+m_{14}+m_{22}+m_{24}+m_{34}+m_{44}}(p_4-h_3+1-d/2)_{m_4}}\\
&\qquad\times\frac{(p_2+h_3)_{m_1-m_4+m_{23}}(p_3)_{-m_1+m_2+m_3+m_{12}+m_{33}}(p_3-h_2+h_4)_{m_2+m_{11}+m_{34}}(\bar{p}_3+h_2+h_5)_{m_3+m_{12}}}{(\bar{p}_3+h_2)_{2m_3+m_{12}+m_{13}+m_{23}+m_{33}+m_{44}}(\bar{p}_3+h_2+1-d/2)_{m_3}}\\
&\qquad\times\frac{(p_2+h_2)_{m_1-m_2+m_3+m_{13}+m_{23}+m_{44}}(-h_2)_{m_1+m_2-m_3+m_{11}+m_{22}+m_{24}+m_{34}}}{(p_2)_{2m_1+m_{11}+m_{13}+m_{22}+m_{23}+m_{24}+m_{34}+m_{44}}(p_2+1-d/2)_{m_1}}\\
&\qquad\times\frac{(-h_3)_{m_1+m_4+m_{11}+m_{13}+m_{22}+m_{24}+m_{34}+m_{44}}(-h_4)_{m_2+m_{12}+m_{22}+m_{24}+m_{33}}(-h_5)_{m_3+m_{13}+m_{23}+m_{33}+m_{44}}}{(p_3-h_2)_{2m_2+m_{11}+m_{12}+m_{22}+m_{24}+m_{33}+m_{34}}(p_3-h_2+1-d/2)_{m_2}}\\
&\qquad\times\frac{(-m_1)_r(-m_4)_r(-p_2+d/2-m_1)_r(p_3-m_1+m_2+m_3)_{r-r_{34}-r_{56}}}{(1-p_2-h_3-m_1)_r(p_4-m_1)_r(r-r_{56}-r_{34})!}\\
&\qquad\times\frac{(-m_2)_{r_{34}}(-m_3)_{r_{56}}(-p_3+h_2+d/2-m_2)_{r_{34}}(-\bar{p}_3-h_2+d/2-m_3)_{r_{56}}}{r_{34}!r_{56}!}\\
&\qquad\times F_6\prod_{1\leq a\leq4}\frac{(u_a^7)^{m_a}}{m_a!}\prod_{1\leq a\leq b\leq4}\frac{(1-v_{ab}^7)^{m_{ab}}}{m_{ab}!}.
}

Expanding $F_6$ with the help of the first result in \eqref{EqCB6F} leads to
\eqna{
&G_7=\sum\frac{(p_4-m_1)_{m_4+m_{14}}(-h_6)_{m_4+m_{14}+m_{24}+m_{34}+m_{44}}(p_4-h_3+h_6)_{m_4+m_{11}+m_{13}+m_{22}}}{(p_2+h_3+m_1)_{-m_4}(p_4-h_3)_{2m_4+m_{11}+m_{13}+m_{14}+m_{22}+m_{24}+m_{34}+m_{44}}(p_4-h_3+1-d/2)_{m_4}}\\
&\qquad\times\frac{(p_2+h_3)_{m_1-m_4+m_{23}}(p_3)_{-m_1+m_2+m_3+m_{12}+m_{33}}(p_3-h_2+h_4)_{m_2+m_{11}+m_{34}}(\bar{p}_3+h_2+h_5)_{m_3+m_{12}}}{(\bar{p}_3+h_2)_{2m_3+m_{12}+m_{13}+m_{23}+m_{33}+m_{44}}(\bar{p}_3+h_2+1-d/2)_{m_3}}\\
&\qquad\times\frac{(p_2+h_2)_{m_1-m_2+m_3+m_{13}+m_{23}+m_{44}}(-h_2)_{m_1+m_2-m_3+m_{11}+m_{22}+m_{24}+m_{34}}}{(p_2)_{2m_1+m_{11}+m_{13}+m_{22}+m_{23}+m_{24}+m_{34}+m_{44}}(p_2+1-d/2)_{m_1}}\\
&\qquad\times\frac{(-h_3)_{m_1+m_4+m_{11}+m_{13}+m_{22}+m_{24}+m_{34}+m_{44}}(-h_4)_{m_2+m_{12}+m_{22}+m_{24}+m_{33}}(-h_5)_{m_3+m_{13}+m_{23}+m_{33}+m_{44}}}{(p_3-h_2)_{2m_2+m_{11}+m_{12}+m_{22}+m_{24}+m_{33}+m_{34}}(p_3-h_2+1-d/2)_{m_2}}\\
&\qquad\times F_7\prod_{1\leq a\leq4}\frac{(u_a^7)^{m_a}}{m_a!}\prod_{1\leq a\leq b\leq4}\frac{(1-v_{ab}^7)^{m_{ab}}}{m_{ab}!},
}
which results in the appropriate $C_7$ presented in \eqref{EqCB7C} and implies that $F_7$ is given by
\eqna{
F_7&=\frac{(-p_3+h_2+d/2-m_2)_{m_2}}{(p_3)_{-m_1}(-h_2)_{-m_3}}\sum\frac{(-m_2)_{r_{34}+t_1}(-m_3)_{r_{56}+t_2}(-\bar{p}_3-h_2+d/2-m_3)_{r_{56}}}{(r-r_{56}-r_{34})!r_{34}!r_{56}!}\\
&\qquad\times\frac{(-m_1)_{r+t_2}(-m_4)_r(-p_2+d/2-m_1)_r(p_3-m_1+m_2+m_3)_{r-r_{56}-r_{34}}}{(1-p_2-h_3-m_1)_r(p_4-m_1)_r}\\
&\qquad\times\frac{(\bar{p}_3-d/2)_{t_1+t_2}}{(-h_2-m_3)_{r_{56}+t_1+t_2}(p_3-m_1)_{r+t_1+t_2}}\frac{(-h_2)_{t_1}(p_3)_{t_1}}{(p_3-h_2+1-d/2)_{t_1}t_1!t_2!}.
}
To reach the solution introduced in \eqref{EqCB7F}, we first evaluate the sum over $r_{34}$, leading to
\eqna{
F_7&=\frac{(-p_3+h_2+d/2-m_2)_{m_2}}{(p_3)_{-m_1}(-h_2)_{-m_3}}\sum\frac{(-m_2)_{t_1}(-m_3)_{r_{56}+t_2}(-\bar{p}_3-h_2+d/2-m_3)_{r_{56}}}{(r-r_{56})!r_{56}!}\\
&\qquad\times\frac{(-m_1)_{r+t_2}(-m_4)_r(-p_2+d/2-m_1)_r(p_3-m_1+m_3+t_1)_{r-r_{56}}}{(1-p_2-h_3-m_1)_r(p_4-m_1)_r}\\
&\qquad\times\frac{(\bar{p}_3-d/2)_{t_1+t_2}}{(-h_2-m_3)_{r_{56}+t_1+t_2}(p_3-m_1)_{r+t_1+t_2}}\frac{(-h_2)_{t_1}(p_3)_{t_1}}{(p_3-h_2+1-d/2)_{t_1}t_1!t_2!}.
}
We then use the first identity in \eqref{Eq3F2} and get
\eqna{
F_7&=\frac{(-p_3+h_2+d/2-m_2)_{m_2}}{(p_3)_{-m_1}(-h_2)_{-m_3}}\sum\frac{(-m_2)_{t_1}(-m_3)_{r_{56}+t_2}(-\bar{p}_3-h_2+d/2-m_3)_{m_3}}{(r-r_{56})!r_{56}!}\\
&\qquad\times\frac{(-m_1)_{r}(-m_4)_r(-p_2+d/2-m_1)_r(p_3-m_1+m_3+t_1)_{r-r_{56}}}{(1-p_2-h_3-m_1)_r(p_4-m_1)_r(\bar{p}_3+h_2+1-d/2)_{t_2}}\\
&\qquad\times\frac{(\bar{p}_3-d/2)_{t_1+t_2}}{(-h_2-m_3)_{m_3+t_1}(p_3-m_1)_{r+t_1+t_2}}\frac{(-h_2)_{t_1}(p_3)_{t_1+t_2}}{(p_3-h_2+1-d/2)_{t_1}t_1!t_2!}.
}
We finally sum over $r_{56}$, leading to $\eqref{EqCB7F}$ with the second equality for $F_6$ in \eqref{EqCB6F}, which concludes our proof of the scalar seven-point conformal blocks in the extended snowflake channel.

%%%%%%%%%%%%%%%%%%%%%%%%%%%%%%%%%%%%%%%%%%%%%%%%%%
%%%%%%%%%%%%%%%%%%%%%%%%%%%%%%%%%%%%%%%%%%%%%%%%%%

\section{Symmetry Properties}\label{SAppSym}

This appendix is dedicated to the proofs of the symmetry properties of the scalar seven-point conformal blocks in the extended snowflake channel.  The three generators to investigate are the reflections as well as the dendrite permutations of the first and second kinds introduced in \eqref{EqSymRef}, \eqref{EqSymPerm1}, and \eqref{EqSymPerm2}, respectively.

%%%%%%%%%%%%%%%%%%%%%%%%%%%%%%%%%%%%%%%%%%%%%%%%%%

\subsection{Reflections}

We aim to prove \eqref{EqSymRef}, which corresponds to the invariance of the scalar seven-point conformal blocks under reflections.  For simplicity, we denote \eqref{EqSymRef} by $G_7=G_{7R}$.  We first rewrite the conformal cross-ratios in \eqref{EqSymRef} in terms of the original conformal cross-ratios and then re-sum the extra sums to obtain $G_7$ back, taking into account that $F_7$ is invariant under reflections as dictated by \eqref{EqCB7F}.

We therefore start with
\eqna{
&G_{7R}=\sum\frac{(p_4-m^{\prime}_{1})_{m^{\prime}_4+m_{14}}(-h_{6})_{m^{\prime}_4+m_{14}+m_{24}+m_{34}+m_{44}}(p_{4}-h_3+h_{6})_{m^{\prime}_4+m_{11}+m_{13}+m_{22}}}{(p_2+h_3+m^{\prime}_1)_{-m^{\prime}_4}(p_{4}-h_3)_{2m^{\prime}_4+m_{11}+m_{13}+m_{14}+m_{22}+m_{24}+m_{34}+m_{44}}(p_4-h_3+1-d/2)_{m^{\prime}_4}}\\
&\qquad\times\frac{(p_2+h_3)_{m^{\prime}_1-m^{\prime}_4+m_{23}}(p_3)_{-m^{\prime}_1+m^{\prime}_2+m^{\prime}_3+m_{12}+m_{33}}(\bar{p}_3+h_2+h_5)_{m^{\prime}_{3}+m_{11}+m_{34}}(p_{3}-h_2+h_{4})_{m^{\prime}_2+m_{12}}}{(p_{3}-h_2)_{2m^{\prime}_2+m_{12}+m_{13}+m_{23}+m_{33}+m_{44}}(p_3-h_2+1-d/2)_{m^{\prime}_2}}\\
&\qquad\times\frac{(-h_2)_{m^{\prime}_1+m^{\prime}_2-m^{\prime}_3+m_{13}+m_{23}+m_{44}}(p_2+h_2)_{m^{\prime}_1-m^{\prime}_2+m^{\prime}_3+m_{11}+m_{22}+m_{24}+m_{34}}}{(p_2)_{2m^{\prime}_{1}+m_{11}+m_{13}+m_{22}+m_{23}+m_{24}+m_{34}+m_{44}}(p_2+1-d/2)_{m^{\prime}_1}}\\
&\qquad\times\frac{(-h_3)_{m^{\prime}_1+m^{\prime}_4+m_{11}+m_{13}+m_{22}+m_{24}+m_{34}+m_{44}}(-h_{5})_{m^{\prime}_{3}+m_{12}+m_{22}+m_{24}+m_{33}}(-h_{4})_{m^{\prime}_2+m_{13}+m_{23}+m_{33}+m_{44}}}{(\bar{p}_{3}+h_{2})_{2m^{\prime}_{3}+m_{11}+m_{12}+m_{22}+m_{24}+m_{33}+m_{34}}(\bar{p}_3+h_2+1-d/2)_{m^{\prime}_3}}\\
&\qquad\times\prod_{1\leq a\leq b\leq 4}\binom{m_{ab}}{k_{ab}}\binom{h_3-h_6-m^{\prime}_1-k_{11}-k_{13}+k_{14}-k_{22}}{m^{\prime}_{11}}\binom{h_4-m^{\prime}_2-k_{13}-k_{23}-k_{33}-k_{44}}{m^{\prime}_{12}}\\
&\qquad\times\binom{k_{22}}{m^{\prime}_{13}}\binom{k_{14}}{m^{\prime}_{14}}\binom{k_{13}}{m^{\prime}_{22}}\binom{h_5-m^{\prime}_3-k_{12}-k_{22}-k_{24}-k_{33}}{m^{\prime}_{23}}\binom{k_{44}}{m^{\prime}_{24}}\binom{k_{33}}{m^{\prime}_{33}}\\
&\qquad\times\binom{h_6-m^{\prime}_4-k_{14}-k_{24}-k_{34}-k_{44}}{m^{\prime}_{34}}\binom{k_{24}}{m^{\prime}_{44}}(-1)^{\sum_{1\leq a\leq b\leq 4}(k_{ab}+m^{\prime}_{ab})}\\
&\qquad\times F_7\prod_{1\leq a\leq4}\frac{(u_a^7)^{m^{\prime}_a}}{m^{\prime}_a!}\prod_{1\leq a\leq b\leq4}\frac{(1-v_{ab}^7)^{m^{\prime}_{ab}}}{m_{ab}!},
}
and evaluate the sums over $k_{11}$, $k_{12}$, $k_{23}$ and $k_{34}$ to get
\eqna{
&G_{7R}=\sum\frac{(p_4-m^{\prime}_{1})_{m^{\prime}_4+m_{14}}(-h_{6})_{m^{\prime}_4+m_{14}+m_{24}+m_{34}+m_{44}}(p_{4}-h_3+h_{6})_{m^{\prime}_4+m_{11}+m_{13}+m_{22}}}{(p_2+h_3+m^{\prime}_1)_{-m^{\prime}_4}(p_{4}-h_3)_{2m^{\prime}_4+m_{11}+m_{13}+m_{14}+m_{22}+m_{24}+m_{34}+m_{44}}(p_4-h_3+1-d/2)_{m^{\prime}_4}}\\
&\qquad\times\frac{(p_2+h_3)_{m^{\prime}_1-m^{\prime}_4+m_{23}}(p_3)_{-m^{\prime}_1+m^{\prime}_2+m^{\prime}_3+m_{12}+m_{33}}(\bar{p}_3+h_2+h_5)_{m^{\prime}_{3}+m_{11}+m_{34}}(p_{3}-h_2+h_{4})_{m^{\prime}_2+m_{12}}}{(p_{3}-h_2)_{2m^{\prime}_2+m_{12}+m_{13}+m_{23}+m_{33}+m_{44}}(p_3-h_2+1-d/2)_{m^{\prime}_2}}\\
&\qquad\times\frac{(-h_2)_{m^{\prime}_1+m^{\prime}_2-m^{\prime}_3+m_{13}+m_{23}+m_{44}}(p_2+h_2)_{m^{\prime}_1-m^{\prime}_2+m^{\prime}_3+m_{11}+m_{22}+m_{24}+m_{34}}}{(p_2)_{2m^{\prime}_{1}+m_{11}+m_{13}+m_{22}+m_{23}+m_{24}+m_{34}+m_{44}}(p_2+1-d/2)_{m^{\prime}_1}}\\
&\qquad\times\frac{(-h_3)_{m^{\prime}_1+m^{\prime}_4+m_{11}+m_{13}+m_{22}+m_{24}+m_{34}+m_{44}}(-h_{5})_{m^{\prime}_{3}+m_{12}+m_{22}+m_{24}+m_{33}}(-h_{4})_{m^{\prime}_2+m_{13}+m_{23}+m_{33}+m_{44}}}{(\bar{p}_{3}+h_{2})_{2m^{\prime}_{3}+m_{11}+m_{12}+m_{22}+m_{24}+m_{33}+m_{34}}(\bar{p}_3+h_2+1-d/2)_{m^{\prime}_3}}\\
&\qquad\times\binom{m_{13}}{k_{13}}\binom{m_{14}}{k_{14}}\binom{m_{22}}{k_{22}}\binom{m_{24}}{k_{24}}\binom{m_{33}}{k_{33}}\binom{m_{44}}{k_{44}}\binom{k_{22}}{m^{\prime}_{13}}\binom{k_{14}}{m^{\prime}_{14}}\binom{k_{13}}{m^{\prime}_{22}}\binom{k_{44}}{m^{\prime}_{24}}\binom{k_{33}}{m^{\prime}_{33}}\binom{k_{24}}{m^{\prime}_{44}}\\
&\qquad\times(-1)^{k_{13}+k_{14}+k_{22}+k_{24}+k_{33}+k_{44}+m^{\prime}_{13}+m^{\prime}_{14}+m^{\prime}_{22}+m^{\prime}_{24}+m^{\prime}_{33}+m^{\prime}_{44}}\\
&\qquad\times\frac{(-m_{11}^{\prime})_{m_{11}}(-h_3+h_6+m^{\prime}_1+m_{11}+k_{13}-k_{14}+k_{22})_{m^{\prime}_{11}-m_{11}}}{m^{\prime}_{11}!}\\
&\qquad\times\frac{(-m_{23}^{\prime})_{m_{12}}(-h_5+m^{\prime}_3+m_{12}+k_{22}+k_{24}+k_{33})_{m^{\prime}_{23}-m_{12}}}{m^{\prime}_{23}!}\\
&\qquad\times\frac{(-m_{12}^{\prime})_{m_{23}}(-h_4+m^{\prime}_2+m_{23}+k_{13}+k_{33}+k_{44})_{m^{\prime}_{12}-m_{23}}}{m^{\prime}_{12}!}\\
&\qquad\times\frac{(-m_{34}^{\prime})_{m_{34}}(-h_6+m^{\prime}_4+m_{34}+k_{14}+k_{24}+k_{44})_{m^{\prime}_{34}-m_{34}}}{m^{\prime}_{34}!}\\
&\qquad\times F_7\prod_{1\leq a\leq4}\frac{(u_a^7)^{m^{\prime}_a}}{m^{\prime}_a!}\prod_{1\leq a\leq b\leq4}\frac{(1-v_{ab}^7)^{m^{\prime}_{ab}}}{m_{ab}!}.
}

We then change variables by
\eqn{
\begin{gathered}
k_{13}\to k_{13}+m^{\prime}_{22},\qquad k_{14}\to k_{14}+m^{\prime}_{14},\qquad k_{22}\to k_{22}+m^{\prime}_{13},\\
k_{24}\to k_{24}+m^{\prime}_{44},\qquad k_{33}\to k_{33}+m^{\prime}_{33},\qquad k_{44}\to k_{44}+m^{\prime}_{24},
\end{gathered}
}
and use the following identity
\eqna{
&(-h_3+h_6+m^{\prime}_1+m^{\prime}_{13}-m^{\prime}_{14}+m^{\prime}_{22}+m_{11}+k_{13}-k_{14}+k_{22})_{m^{\prime}_{11}-m_{11}}\\
&\qquad=\sum_{j_1\geq 0}\binom{m_{11}^{\prime}-m_{11}}{j_1}(-m_{13}+m^{\prime}_{22}+k_{13})_{j_1}\\
&\qquad\phantom{=}\times(-h_3+h_6+m^{\prime}_1+m^{\prime}_{13}-m^{\prime}_{14}+m_{11}+m_{13}-k_{14}+k_{22})_{m^{\prime}_{11}-m_{11}-j_1},
}
to compute the sum over $k_{13}$.  With the help of a similar trick, we evaluate the sums over all other $k_{ab}$'s except $k_{14}$, to get
\eqna{
&G_{7R}=\sum\frac{(p_4-m^{\prime}_{1})_{m^{\prime}_4+m_{14}}(-h_{6})_{m^{\prime}_4+m_{14}+m_{24}+m_{34}+m_{44}}(p_{4}-h_3+h_{6})_{m^{\prime}_4+m_{11}+m_{13}+m_{22}}}{(p_2+h_3+m^{\prime}_1)_{-m^{\prime}_4}(p_{4}-h_3)_{2m^{\prime}_4+m_{11}+m_{13}+m_{14}+m_{22}+m_{24}+m_{34}+m_{44}}(p_4-h_3+1-d/2)_{m^{\prime}_4}}\\
&\qquad\times\frac{(p_2+h_3)_{m^{\prime}_1-m^{\prime}_4+m_{23}}(p_3)_{-m^{\prime}_1+m^{\prime}_2+m^{\prime}_3+m_{12}+m_{33}}(\bar{p}_3+h_2+h_5)_{m^{\prime}_{3}+m_{11}+m_{34}}(p_{3}-h_2+h_{4})_{m^{\prime}_2+m_{12}}}{(p_{3}-h_2)_{2m^{\prime}_2+m_{12}+m_{13}+m_{23}+m_{33}+m_{44}}(p_3-h_2+1-d/2)_{m^{\prime}_2}}\\
&\qquad\times\frac{(-h_2)_{m^{\prime}_1+m^{\prime}_2-m^{\prime}_3+m_{13}+m_{23}+m_{44}}(p_2+h_2)_{m^{\prime}_1-m^{\prime}_2+m^{\prime}_3+m_{11}+m_{22}+m_{24}+m_{34}}}{(p_2)_{2m^{\prime}_{1}+m_{11}+m_{13}+m_{22}+m_{23}+m_{24}+m_{34}+m_{44}}(p_2+1-d/2)_{m^{\prime}_1}}\\
&\qquad\times\frac{(-h_3)_{m^{\prime}_1+m^{\prime}_4+m_{11}+m_{13}+m_{22}+m_{24}+m_{34}+m_{44}}(-h_{5})_{m^{\prime}_{3}+m^{\prime}_{13}+m^{\prime}_{23}+m^{\prime}_{33}+m^{\prime}_{44}}(-h_{4})_{m^{\prime}_2+m^{\prime}_{12}+m^{\prime}_{22}+m^{\prime}_{24}+m_{33}-j_4}}{(\bar{p}_{3}+h_{2})_{2m^{\prime}_{3}+m_{11}+m_{12}+m_{22}+m_{24}+m_{33}+m_{34}}(\bar{p}_3+h_2+1-d/2)_{m^{\prime}_3}}\\
&\qquad\times\frac{(-m_{11}^{\prime})_{m_{11}+j_1+j_2}(-h_3+h_6+m^{\prime}_1-m^{\prime}_{14}+m_{11}+m_{13}+m_{22}-k_{14})_{m^{\prime}_{11}-m_{11}-j_1-j_2}}{(m_{22}-m^{\prime}_{13}-j_2)!(m_{13}-m^{\prime}_{22}-j_1)!}\\
&\qquad\times\frac{(-m_{23}^{\prime})_{m_{12}+m_{22}+m_{24}+m_{33}-m^{\prime}_{13}-m^{\prime}_{33}-m^{\prime}_{44}-j_2-j_3-j_4}}{(-h_5+m^{\prime}_3+m_{12}+m_{22}+m_{24}+m_{33})_{-j_2-j_3-j_4}(m_{24}-m^{\prime}_{44}-j_3)!(m_{44}-m^{\prime}_{24}-j_5)!}\\
&\qquad\times\frac{(-m_{34}^{\prime})_{m_{34}+j_3+j_5}(-m_{12}^{\prime})_{m_{13}+m_{23}+m_{44}-m^{\prime}_{22}-m^{\prime}_{24}-j_1+j_4-j_5}}{(-h_4+m^{\prime}_2+m_{13}+m_{23}+m_{33}+m_{44})_{-j_1-j_5}j_1!j_2!j_3!j_4!j_5!(m_{33}-m^{\prime}_{33}-j_4)!}\\
&\qquad\times(-1)^{k_{14}}\frac{(-h_6+m^{\prime}_4+m^{\prime}_{14}+m_{24}+m_{34}+m_{44}+k_{14})_{m^{\prime}_{34}-m_{34}-j_3-j_5}}{(m_{14}-m^{\prime}_{14}-k_{14})!k_{14}!m_{11}!m_{12}!m_{23}!m_{34}!}\\
&\qquad\times F_7\prod_{1\leq a\leq4}\frac{(u_a^7)^{m^{\prime}_a}}{m^{\prime}_a!}\prod_{1\leq a\leq b\leq4}\frac{(1-v_{ab}^7)^{m^{\prime}_{ab}}}{m^{\prime}_{ab}!}.
}

We now use the identity
\eqna{
&(-h_3+h_6+m^{\prime}_1-m^{\prime}_{14}+m_{11}+m_{13}+m_{22}-k_{14})_{m^{\prime}_{11}-m_{11}-j_1-j_2}\\
&\qquad=\sum_{j_6\geq 0}\binom{m^{\prime}_{11}-m_{11}-j_1-j_2}{j_6}(-k_{14})_{j_6}\\
&\qquad\phantom{=}\times(-h_3+h_6+m^{\prime}_1-m^{\prime}_{14}+m_{11}+m_{13}+m_{22})_{m^{\prime}_{11}-m_{11}-j_1-j_2-j_6},
}
and then change $k_{14}$ by $k_{14}\to k_{14}+j_6$ to re-sum over $k_{14}$.  This procedure leads to
\eqna{
&G_{7R}=\sum\frac{(p_4-m^{\prime}_{1})_{m^{\prime}_4+m_{14}}(-h_{6})_{m^{\prime}_4+m^{\prime}_{14}+m_{24}+m^{\prime}_{34}+m_{44}-j_3-j_5+j_6}(p_{4}-h_3+h_{6})_{m^{\prime}_4+m_{11}+m_{13}+m_{22}}}{(p_2+h_3+m^{\prime}_1)_{-m^{\prime}_4}(p_{4}-h_3)_{2m^{\prime}_4+m_{11}+m_{13}+m_{14}+m_{22}+m_{24}+m_{34}+m_{44}}(p_4-h_3+1-d/2)_{m^{\prime}_4}}\\
&\qquad\times\frac{(p_2+h_3)_{m^{\prime}_1-m^{\prime}_4+m_{23}}(p_3)_{-m^{\prime}_1+m^{\prime}_2+m^{\prime}_3+m_{12}+m_{33}}(\bar{p}_3+h_2+h_5)_{m^{\prime}_{3}+m_{11}+m_{34}}(p_{3}-h_2+h_{4})_{m^{\prime}_2+m_{12}}}{(p_{3}-h_2)_{2m^{\prime}_2+m_{12}+m_{13}+m_{23}+m_{33}+m_{44}}(p_3-h_2+1-d/2)_{m^{\prime}_2}}\\
&\qquad\times\frac{(-h_2)_{m^{\prime}_1+m^{\prime}_2-m^{\prime}_3+m_{13}+m_{23}+m_{44}}(p_2+h_2)_{m^{\prime}_1-m^{\prime}_2+m^{\prime}_3+m_{11}+m_{22}+m_{24}+m_{34}}}{(p_2)_{2m^{\prime}_{1}+m_{11}+m_{13}+m_{22}+m_{23}+m_{24}+m_{34}+m_{44}}(p_2+1-d/2)_{m^{\prime}_1}}\\
&\qquad\times\frac{(-h_3)_{m^{\prime}_1+m^{\prime}_4+m_{11}+m_{13}+m_{22}+m_{24}+m_{34}+m_{44}}(-h_{5})_{m^{\prime}_{3}+m^{\prime}_{13}+m^{\prime}_{23}+m^{\prime}_{33}+m^{\prime}_{44}}(-h_{4})_{m^{\prime}_2+m^{\prime}_{12}+m^{\prime}_{22}+m^{\prime}_{24}+m_{33}-j_4}}{(\bar{p}_{3}+h_{2})_{2m^{\prime}_{3}+m_{11}+m_{12}+m_{22}+m_{24}+m_{33}+m_{34}}(\bar{p}_3+h_2+1-d/2)_{m^{\prime}_3}}\\
&\qquad\times\frac{(-m_{11}^{\prime})_{m_{11}+j_1+j_2+j_6}(-h_3+h_6+m^{\prime}_1-m^{\prime}_{14}+m_{11}+m_{13}+m_{22})_{m^{\prime}_{11}-m_{11}-j_1-j_2-j_6}}{(m_{22}-m^{\prime}_{13}-j_2)!(m_{13}-m^{\prime}_{22}-j_1)!}\\
&\qquad\times\frac{(-m_{23}^{\prime})_{m_{12}+m_{22}+m_{24}+m_{33}-m^{\prime}_{13}-m^{\prime}_{33}-m^{\prime}_{44}-j_2-j_3-j_4}}{(-h_5+m^{\prime}_3+m_{12}+m_{22}+m_{24}+m_{33})_{-j_2-j_3-j_4}(m_{24}-m^{\prime}_{44}-j_3)!(m_{44}-m^{\prime}_{24}-j_5)!}\\
&\qquad\times\frac{(-m_{34}^{\prime})_{m_{14}+m_{34}-m^{\prime}_{14}+j_3+j_5-j_6}(-m_{12}^{\prime})_{m_{13}+m_{23}+m_{44}-m^{\prime}_{22}-m^{\prime}_{24}-j_1+j_4-j_5}}{(-h_4+m^{\prime}_2+m_{13}+m_{23}+m_{33}+m_{44})_{-j_1-j_5}j_1!j_2!j_3!j_4!j_5!j_6!(m_{33}-m^{\prime}_{33}-j_4)!}\\
&\qquad\times\frac{(-1)^{j_6}}{(m_{14}-m^{\prime}_{14}-j_6)!m_{11}!m_{12}!m_{23}!m_{34}!}F_7\prod_{1\leq a\leq4}\frac{(u_a^7)^{m^{\prime}_a}}{m^{\prime}_a!}\prod_{1\leq a\leq b\leq4}\frac{(1-v_{ab}^7)^{m^{\prime}_{ab}}}{m^{\prime}_{ab}!}.
}

It is thus possible to evaluate the sum over $m_{14}$ (after shifting $m_{14}$ by $m_{14}+m^{\prime}_{14}+j_6$) such that
\eqna{
&G_{7R}=\sum\frac{(p_4-m^{\prime}_{1})_{m^{\prime}_4+m^{\prime}_{14}+j_6}(-h_{6})_{m^{\prime}_4+m^{\prime}_{14}+m_{24}+m^{\prime}_{34}+m_{44}-j_3-j_5+j_6}(p_{4}-h_3+h_{6})_{m^{\prime}_4+m_{11}+m_{13}+m_{22}}}{(p_2+h_3+m^{\prime}_1)_{-m^{\prime}_4}(p_{4}-h_3)_{2m^{\prime}_4+m_{11}+m_{13}+m^{\prime}_{14}+m_{22}+m_{24}+m^{\prime}_{34}+m_{44}-j_3-j_5+j_6}(p_4-h_3+1-d/2)_{m^{\prime}_4}}\\
&\qquad\times\frac{(p_2+h_3)_{m^{\prime}_1-m^{\prime}_4+m_{23}}(p_3)_{-m^{\prime}_1+m^{\prime}_2+m^{\prime}_3+m_{12}+m_{33}}(\bar{p}_3+h_2+h_5)_{m^{\prime}_{3}+m_{11}+m_{34}}(p_{3}-h_2+h_{4})_{m^{\prime}_2+m_{12}}}{(p_{3}-h_2)_{2m^{\prime}_2+m_{12}+m_{13}+m_{23}+m_{33}+m_{44}}(p_3-h_2+1-d/2)_{m^{\prime}_2}}\\
&\qquad\times\frac{(-h_2)_{m^{\prime}_1+m^{\prime}_2-m^{\prime}_3+m_{13}+m_{23}+m_{44}}(p_2+h_2)_{m^{\prime}_1-m^{\prime}_2+m^{\prime}_3+m_{11}+m_{22}+m_{24}+m_{34}}}{(p_2)_{2m^{\prime}_{1}+m_{11}+m_{13}+m_{22}+m_{23}+m_{24}+m_{34}+m_{44}}(p_2+1-d/2)_{m^{\prime}_1}}\\
&\qquad\times\frac{(-h_3)_{m^{\prime}_1+m^{\prime}_4+m_{11}+m_{13}+m_{22}+m_{24}+m^{\prime}_{34}+m_{44}-j_3-j_5}(-h_{5})_{m^{\prime}_{3}+m^{\prime}_{13}+m^{\prime}_{23}+m^{\prime}_{33}+m^{\prime}_{44}}(-h_{4})_{m^{\prime}_2+m^{\prime}_{12}+m^{\prime}_{22}+m^{\prime}_{24}+m_{33}-j_4}}{(\bar{p}_{3}+h_{2})_{2m^{\prime}_{3}+m_{11}+m_{12}+m_{22}+m_{24}+m_{33}+m_{34}}(\bar{p}_3+h_2+1-d/2)_{m^{\prime}_3}}\\
&\qquad\times\frac{(-m_{11}^{\prime})_{m_{11}+j_1+j_2+j_6}(-h_3+h_6+m^{\prime}_1-m^{\prime}_{14}+m_{11}+m_{13}+m_{22})_{m^{\prime}_{11}-m_{11}-j_1-j_2-j_6}}{(m_{22}-m^{\prime}_{13}-j_2)!(m_{13}-m^{\prime}_{22}-j_1)!}\\
&\qquad\times\frac{(-m_{23}^{\prime})_{m_{12}+m_{22}+m_{24}+m_{33}-m^{\prime}_{13}-m^{\prime}_{33}-m^{\prime}_{44}-j_2-j_3-j_4}}{(-h_5+m^{\prime}_3+m_{12}+m_{22}+m_{24}+m_{33})_{-j_2-j_3-j_4}(m_{24}-m^{\prime}_{44}-j_3)!(m_{44}-m^{\prime}_{24}-j_5)!}\\
&\qquad\times\frac{(-m_{34}^{\prime})_{m_{34}+j_3+j_5}(-m_{12}^{\prime})_{m_{13}+m_{23}+m_{44}-m^{\prime}_{22}-m^{\prime}_{24}-j_1+j_4-j_5}}{(-h_4+m^{\prime}_2+m_{13}+m_{23}+m_{33}+m_{44})_{-j_1-j_5}j_1!j_2!j_3!j_4!j_5!j_6!(m_{33}-m^{\prime}_{33}-j_4)!}\\
&\qquad\times\frac{(-1)^{j_6}}{m_{11}!m_{12}!m_{23}!m_{34}!}F_7\prod_{1\leq a\leq4}\frac{(u_a^7)^{m^{\prime}_a}}{m^{\prime}_a!}\prod_{1\leq a\leq b\leq4}\frac{(1-v_{ab}^7)^{m^{\prime}_{ab}}}{m^{\prime}_{ab}!}.
}

With the help of the second identity in \eqref{Eq3F2}, we now eliminate the sum over $j_6$ and get
\eqna{
&G_{7R}=\sum\frac{(p_4-m^{\prime}_{1})_{m^{\prime}_4+m^{\prime}_{14}}(-h_{6})_{m^{\prime}_4+m^{\prime}_{14}+m_{24}+m^{\prime}_{34}+m_{44}-j_3-j_5}(p_{4}-h_3+h_{6})_{m^{\prime}_4+m^{\prime}_{11}+m_{13}+m_{22}-j_1-j_2}}{(p_2+h_3+m^{\prime}_1)_{-m^{\prime}_4}(p_{4}-h_3)_{2m^{\prime}_4+m^{\prime}_{11}+m_{13}+m^{\prime}_{14}+m_{22}+m_{24}+m^{\prime}_{34}+m_{44}-j_1-j_2-j_3-j_5}(p_4-h_3+1-d/2)_{m^{\prime}_4}}\\
&\qquad\times\frac{(p_2+h_3)_{m^{\prime}_1-m^{\prime}_4+m_{23}}(p_3)_{-m^{\prime}_1+m^{\prime}_2+m^{\prime}_3+m_{12}+m_{33}}(\bar{p}_3+h_2+h_5)_{m^{\prime}_{3}+m_{11}+m_{34}}(p_{3}-h_2+h_{4})_{m^{\prime}_2+m_{12}}}{(p_{3}-h_2)_{2m^{\prime}_2+m_{12}+m_{13}+m_{23}+m_{33}+m_{44}}(p_3-h_2+1-d/2)_{m^{\prime}_2}}\\
&\qquad\times\frac{(-h_2)_{m^{\prime}_1+m^{\prime}_2-m^{\prime}_3+m_{13}+m_{23}+m_{44}}(p_2+h_2)_{m^{\prime}_1-m^{\prime}_2+m^{\prime}_3+m_{11}+m_{22}+m_{24}+m_{34}}}{(p_2)_{2m^{\prime}_{1}+m_{11}+m_{13}+m_{22}+m_{23}+m_{24}+m_{34}+m_{44}}(p_2+1-d/2)_{m^{\prime}_1}}\\
&\qquad\times\frac{(-h_3)_{m^{\prime}_1+m^{\prime}_4+m^{\prime}_{11}+m_{13}+m_{22}+m_{24}+m^{\prime}_{34}+m_{44}-j_1-j_2-j_3-j_5}(-h_{5})_{m^{\prime}_{3}+m^{\prime}_{13}+m^{\prime}_{23}+m^{\prime}_{33}+m^{\prime}_{44}}}{(\bar{p}_{3}+h_{2})_{2m^{\prime}_{3}+m_{11}+m_{12}+m_{22}+m_{24}+m_{33}+m_{34}}(\bar{p}_3+h_2+1-d/2)_{m^{\prime}_3}}\\
&\qquad\times\frac{(-h_{4})_{m^{\prime}_2+m^{\prime}_{12}+m^{\prime}_{22}+m^{\prime}_{24}+m_{33}-j_4}(-m_{23}^{\prime})_{m_{12}+m_{22}+m_{24}+m_{33}-m^{\prime}_{13}-m^{\prime}_{33}-m^{\prime}_{44}-j_2-j_3-j_4}}{(-h_5+m^{\prime}_3+m_{12}+m_{22}+m_{24}+m_{33})_{-j_2-j_3-j_4}(m_{24}-m^{\prime}_{44}-j_3)!(m_{44}-m^{\prime}_{24}-j_5)!}\\
&\qquad\times\frac{(-m_{34}^{\prime})_{m_{34}+j_3+j_5}(-m_{12}^{\prime})_{m_{13}+m_{23}+m_{44}-m^{\prime}_{22}-m^{\prime}_{24}-j_1+j_4-j_5}}{(-h_4+m^{\prime}_2+m_{13}+m_{23}+m_{33}+m_{44})_{-j_1-j_5}j_1!j_2!j_3!j_4!j_5!(m_{22}-m^{\prime}_{13}-j_2)!(m_{33}-m^{\prime}_{33}-j_4)!}\\
&\qquad\times\frac{(-m_{11}^{\prime})_{m_{11}+j_1+j_2}}{(m_{13}-m^{\prime}_{22}-j_1)!m_{11}!m_{12}!m_{23}!m_{34}!}F_7\prod_{1\leq a\leq4}\frac{(u_a^7)^{m^{\prime}_a}}{m^{\prime}_a!}\prod_{1\leq a\leq b\leq4}\frac{(1-v_{ab}^7)^{m^{\prime}_{ab}}}{m^{\prime}_{ab}!}.
}

At this point, we introduce $m_{44}=m-m_{13}$ and then change $m_{13}$ by $m_{13}\to m_{13}+m^{\prime}_{22}+j_1$.  This allows us to sum over $m_{13}$,
\eqna{
&G_{7R}=\sum\frac{(p_4-m^{\prime}_{1})_{m^{\prime}_4+m^{\prime}_{14}}(-h_{6})_{m^{\prime}_4+m^{\prime}_{14}+m^{\prime}_{24}+m_{24}+m^{\prime}_{34}-j_3}(p_{4}-h_3+h_{6})_{m^{\prime}_4+m^{\prime}_{11}+m^{\prime}_{22}+m_{22}-j_2}}{(p_2+h_3+m^{\prime}_1)_{-m^{\prime}_4}(p_{4}-h_3)_{2m^{\prime}_4+m^{\prime}_{11}+m^{\prime}_{14}+m^{\prime}_{22}+m_{22}+m^{\prime}_{24}+m_{24}+m^{\prime}_{34}-j_2-j_3}(p_4-h_3+1-d/2)_{m^{\prime}_4}}\\
&\qquad\times\frac{(p_2+h_3)_{m^{\prime}_1-m^{\prime}_4+m_{23}}(p_3)_{-m^{\prime}_1+m^{\prime}_2+m^{\prime}_3+m_{12}+m_{33}}(\bar{p}_3+h_2+h_5)_{m^{\prime}_{3}+m_{11}+m_{34}}(p_{3}-h_2+h_{4})_{m^{\prime}_2+m_{12}}}{(p_{3}-h_2)_{2m^{\prime}_2+m_{12}+m+m_{23}+m_{33}}(p_3-h_2+1-d/2)_{m^{\prime}_2}}\\
&\qquad\times\frac{(-h_2)_{m^{\prime}_1+m^{\prime}_2-m^{\prime}_3+m+m_{23}}(p_2+h_2)_{m^{\prime}_1-m^{\prime}_2+m^{\prime}_3+m_{11}+m_{22}+m_{24}+m_{34}}}{(p_2)_{2m^{\prime}_{1}+m_{11}+m+m_{22}+m_{23}+m_{24}+m_{34}}(p_2+1-d/2)_{m^{\prime}_1}}\\
&\qquad\times\frac{(-h_3)_{m^{\prime}_1+m^{\prime}_4+m^{\prime}_{11}+m+m_{22}+m_{24}+m^{\prime}_{34}-j_1-j_2-j_3-j_5}(-h_{5})_{m^{\prime}_{3}+m^{\prime}_{13}+m^{\prime}_{23}+m^{\prime}_{33}+m^{\prime}_{44}}}{(\bar{p}_{3}+h_{2})_{2m^{\prime}_{3}+m_{11}+m_{12}+m_{22}+m_{24}+m_{33}+m_{34}}(\bar{p}_3+h_2+1-d/2)_{m^{\prime}_3}}\\
&\qquad\times\frac{(-h_{4})_{m^{\prime}_2+m^{\prime}_{12}+m^{\prime}_{22}+m^{\prime}_{24}+m_{33}-j_4}(-m_{23}^{\prime})_{m_{12}+m_{22}+m_{24}+m_{33}-m^{\prime}_{13}-m^{\prime}_{33}-m^{\prime}_{44}-j_2-j_3-j_4}}{(-h_5+m^{\prime}_3+m_{12}+m_{22}+m_{24}+m_{33})_{-j_2-j_3-j_4}(m_{24}-m^{\prime}_{44}-j_3)!(m-m^{\prime}_{22}-m^{\prime}_{24}-j_1-j_5)!}\\
&\qquad\times\frac{(-m_{34}^{\prime})_{m_{34}+j_3+j_5}(-m_{12}^{\prime})_{m+m_{23}-m^{\prime}_{22}-m^{\prime}_{24}-j_1+j_4-j_5}}{(-h_4+m^{\prime}_2+m+m_{23}+m_{33})_{-j_1-j_5}j_1!j_2!j_3!j_4!j_5!(m_{22}-m^{\prime}_{13}-j_2)!(m_{33}-m^{\prime}_{33}-j_4)!}\\
&\qquad\times\frac{(-m_{11}^{\prime})_{m_{11}+j_1+j_2}}{m_{11}!m_{12}!m_{23}!m_{34}!}F_7\prod_{1\leq a\leq4}\frac{(u_a^7)^{m^{\prime}_a}}{m^{\prime}_a!}\prod_{1\leq a\leq b\leq4}\frac{(1-v_{ab}^7)^{m^{\prime}_{ab}}}{m^{\prime}_{ab}!}.
}

In the same spirit, we define $m_{24}=n-m_{22}$ and change $m_{22}$ by $m_{22}\to m_{22}+m^{\prime}_{13}+j_2$ to evaluate the sum over $m_{22}$, which leads to
\eqna{
&G_{7R}=\sum\frac{(p_4-m^{\prime}_{1})_{m^{\prime}_4+m^{\prime}_{14}}(-h_{6})_{m^{\prime}_4+m^{\prime}_{14}+m^{\prime}_{24}+m^{\prime}_{34}+m^{\prime}_{44}}(p_{4}-h_3+h_{6})_{m^{\prime}_4+m^{\prime}_{11}+m^{\prime}_{13}+m^{\prime}_{22}}}{(p_2+h_3+m^{\prime}_1)_{-m^{\prime}_4}(p_{4}-h_3)_{2m^{\prime}_4+m^{\prime}_{11}+m^{\prime}_{13}+m^{\prime}_{14}+m^{\prime}_{22}+m^{\prime}_{24}+m^{\prime}_{34}+m^{\prime}_{44}}(p_4-h_3+1-d/2)_{m^{\prime}_4}}\\
&\qquad\times\frac{(p_2+h_3)_{m^{\prime}_1-m^{\prime}_4+m_{23}}(p_3)_{-m^{\prime}_1+m^{\prime}_2+m^{\prime}_3+m_{12}+m_{33}}(\bar{p}_3+h_2+h_5)_{m^{\prime}_{3}+m_{11}+m_{34}}(p_{3}-h_2+h_{4})_{m^{\prime}_2+m_{12}}}{(p_{3}-h_2)_{2m^{\prime}_2+m_{12}+m+m_{23}+m_{33}}(p_3-h_2+1-d/2)_{m^{\prime}_2}}\\
&\qquad\times\frac{(-h_2)_{m^{\prime}_1+m^{\prime}_2-m^{\prime}_3+m+m_{23}}(p_2+h_2)_{m^{\prime}_1-m^{\prime}_2+m^{\prime}_3+m_{11}+n+m_{34}}}{(p_2)_{2m^{\prime}_{1}+m_{11}+m+n+m_{23}+m_{34}}(p_2+1-d/2)_{m^{\prime}_1}}\\
&\qquad\times\frac{(-h_3)_{m^{\prime}_1+m^{\prime}_4+m^{\prime}_{11}+m+n+m^{\prime}_{34}-j_1-j_2-j_3-j_5}(-h_{5})_{m^{\prime}_{3}+m^{\prime}_{13}+m^{\prime}_{23}+m^{\prime}_{33}+m^{\prime}_{44}}(-h_{4})_{m^{\prime}_2+m^{\prime}_{12}+m^{\prime}_{22}+m^{\prime}_{24}+m_{33}-j_4}}{(\bar{p}_{3}+h_{2})_{2m^{\prime}_{3}+m_{11}+m_{12}+n+m_{33}+m_{34}}(\bar{p}_3+h_2+1-d/2)_{m^{\prime}_3}}\\
&\qquad\times\frac{(-m_{23}^{\prime})_{m_{12}+n+m_{33}-m^{\prime}_{13}-m^{\prime}_{33}-m^{\prime}_{44}-j_2-j_3-j_4}}{(-h_5+m^{\prime}_3+m_{12}+n+m_{33})_{-j_2-j_3-j_4}(n-m^{\prime}_{13}-m^{\prime}_{44}-j_2-j_3)!(m-m^{\prime}_{22}-m^{\prime}_{24}-j_1-j_5)!}\\
&\qquad\times\frac{(-m_{34}^{\prime})_{m_{34}+j_3+j_5}(-m_{12}^{\prime})_{m+m_{23}-m^{\prime}_{22}-m^{\prime}_{24}-j_1+j_4-j_5}}{(-h_4+m^{\prime}_2+m+m_{23}+m_{33})_{-j_1-j_5}j_1!j_2!j_3!j_4!j_5!(m_{33}-m^{\prime}_{33}-j_4)!}\\
&\qquad\times\frac{(-m_{11}^{\prime})_{m_{11}+j_1+j_2}}{m_{11}!m_{12}!m_{23}!m_{34}!}F_7\prod_{1\leq a\leq4}\frac{(u_a^7)^{m^{\prime}_a}}{m^{\prime}_a!}\prod_{1\leq a\leq b\leq4}\frac{(1-v_{ab}^7)^{m^{\prime}_{ab}}}{m^{\prime}_{ab}!}.
}

We now redefine $m_{34}=k_1-m_{11}$, $j_5=k_2-j_1$, $j_3=k_3-j_2$, and compute the sums over $j_1$, $m_{11}$, and $j_2$, to obtain
\eqna{
&G_{7R}=\sum\frac{(p_4-m^{\prime}_{1})_{m^{\prime}_4+m^{\prime}_{14}}(-h_{6})_{m^{\prime}_4+m^{\prime}_{14}+m^{\prime}_{24}+m^{\prime}_{34}+m^{\prime}_{44}}(p_{4}-h_3+h_{6})_{m^{\prime}_4+m^{\prime}_{11}+m^{\prime}_{13}+m^{\prime}_{22}}}{(p_2+h_3+m^{\prime}_1)_{-m^{\prime}_4}(p_{4}-h_3)_{2m^{\prime}_4+m^{\prime}_{11}+m^{\prime}_{13}+m^{\prime}_{14}+m^{\prime}_{22}+m^{\prime}_{24}+m^{\prime}_{34}+m^{\prime}_{44}}(p_4-h_3+1-d/2)_{m^{\prime}_4}}\\
&\qquad\times\frac{(p_2+h_3)_{m^{\prime}_1-m^{\prime}_4+m_{23}}(p_3)_{-m^{\prime}_1+m^{\prime}_2+m^{\prime}_3+m_{12}+m_{33}}(\bar{p}_3+h_2+h_5)_{m^{\prime}_{3}+k_{1}}(p_{3}-h_2+h_{4})_{m^{\prime}_2+m_{12}}}{(p_{3}-h_2)_{2m^{\prime}_2+m_{12}+m+m_{23}+m_{33}}(p_3-h_2+1-d/2)_{m^{\prime}_2}}\\
&\qquad\times\frac{(-h_2)_{m^{\prime}_1+m^{\prime}_2-m^{\prime}_3+m+m_{23}}(p_2+h_2)_{m^{\prime}_1-m^{\prime}_2+m^{\prime}_3+k_{1}+n}}{(p_2)_{2m^{\prime}_{1}+k_{1}+m+n+m_{23}}(p_2+1-d/2)_{m^{\prime}_1}}\\
&\qquad\times\frac{(-h_3)_{m^{\prime}_1+m^{\prime}_4+m^{\prime}_{11}+m+n+m^{\prime}_{34}-k_2-k_3}(-h_{5})_{m^{\prime}_{3}+m^{\prime}_{13}+m^{\prime}_{23}+m^{\prime}_{33}+m^{\prime}_{44}}(-h_{4})_{m^{\prime}_2+m^{\prime}_{12}+m^{\prime}_{22}+m^{\prime}_{24}+m_{33}-j_4}}{(\bar{p}_{3}+h_{2})_{2m^{\prime}_{3}+k_{1}+m_{12}+n+m_{33}}(\bar{p}_3+h_2+1-d/2)_{m^{\prime}_3}}\\
&\qquad\times\frac{(-m_{23}^{\prime})_{m_{12}+n+m_{33}-m^{\prime}_{13}-m^{\prime}_{33}-m^{\prime}_{44}-k_3-j_4}}{(-h_5+m^{\prime}_3+m_{12}+n+m_{33})_{-k_3-j_4}(n-m^{\prime}_{13}-m^{\prime}_{44}-k_3)!(m-m^{\prime}_{22}-m^{\prime}_{24}-k_2)!}\\
&\qquad\times\frac{(-m_{12}^{\prime})_{m+m_{23}-m^{\prime}_{22}-m^{\prime}_{24}-k_2+j_4}}{(-h_4+m^{\prime}_2+m+m_{23}+m_{33})_{-k_2}k_1!k_2!k_3!j_4!(m_{33}-m^{\prime}_{33}-j_4)!}\\
&\qquad\times\frac{(-m_{11}^{\prime}-m^{\prime}_{34})_{k_1+k_2+k_3}}{m_{12}!m_{23}!}F_7\prod_{1\leq a\leq4}\frac{(u_a^7)^{m^{\prime}_a}}{m^{\prime}_a!}\prod_{1\leq a\leq b\leq4}\frac{(1-v_{ab}^7)^{m^{\prime}_{ab}}}{m^{\prime}_{ab}!}.
}

We then shift $m$ by $m\to m+m^{\prime}_{22}+m^{\prime}_{24}+k_{2}$ and use the first identity in \eqref{Eq3F2} to rewrite the sum over $k_2$, leading to
\eqna{
&G_{7R}=\sum\frac{(p_4-m^{\prime}_{1})_{m^{\prime}_4+m^{\prime}_{14}}(-h_{6})_{m^{\prime}_4+m^{\prime}_{14}+m^{\prime}_{24}+m^{\prime}_{34}+m^{\prime}_{44}}(p_{4}-h_3+h_{6})_{m^{\prime}_4+m^{\prime}_{11}+m^{\prime}_{13}+m^{\prime}_{22}}}{(p_2+h_3+m^{\prime}_1)_{-m^{\prime}_4}(p_{4}-h_3)_{2m^{\prime}_4+m^{\prime}_{11}+m^{\prime}_{13}+m^{\prime}_{14}+m^{\prime}_{22}+m^{\prime}_{24}+m^{\prime}_{34}+m^{\prime}_{44}}(p_4-h_3+1-d/2)_{m^{\prime}_4}}\\
&\qquad\times\frac{(p_2+h_3)_{m^{\prime}_1-m^{\prime}_4+m_{23}}(p_3)_{-m^{\prime}_1+m^{\prime}_2+m^{\prime}_3+m_{12}+m_{33}}(\bar{p}_3+h_2+h_5)_{m^{\prime}_{3}+k_{1}}(p_{3}-h_2+h_{4})_{m^{\prime}_2+m_{12}+k_2}}{(p_{3}-h_2)_{2m^{\prime}_2+m_{12}+m^{\prime}_{22}+m^{\prime}_{24}+m+m_{23}+m_{33}+k_2}(p_3-h_2+1-d/2)_{m^{\prime}_2}}\\
&\qquad\times\frac{(-h_2)_{m^{\prime}_1+m^{\prime}_2-m^{\prime}_3+m^{\prime}_{22}+m^{\prime}_{24}+m+m_{23}+k_2}(p_2+h_2)_{m^{\prime}_1-m^{\prime}_2+m^{\prime}_3+m^{\prime}_{11}+m^{\prime}_{34}-k_{2}-k_3+n}}{(p_2)_{2m^{\prime}_{1}+m^{\prime}_{11}+m^{\prime}_{22}+m^{\prime}_{24}+m^{\prime}_{34}+m+n+m_{23}-k_3}(p_2+1-d/2)_{m^{\prime}_1}}\\
&\qquad\times\frac{(-h_3)_{m^{\prime}_1+m^{\prime}_4+m^{\prime}_{11}+m^{\prime}_{22}+m^{\prime}_{24}+m+n+m^{\prime}_{34}-k_3}(-h_{5})_{m^{\prime}_{3}+m^{\prime}_{13}+m^{\prime}_{23}+m^{\prime}_{33}+m^{\prime}_{44}}}{(\bar{p}_{3}+h_{2})_{2m^{\prime}_{3}+k_{1}+m_{12}+n+m_{33}}(\bar{p}_3+h_2+1-d/2)_{m^{\prime}_3}}\\
&\qquad\times\frac{(-h_{4})_{m^{\prime}_2+m^{\prime}_{12}+m^{\prime}_{22}+m^{\prime}_{24}+m_{33}-j_4}(-m_{23}^{\prime})_{m_{12}+n+m_{33}-m^{\prime}_{13}-m^{\prime}_{33}-m^{\prime}_{44}-k_3-j_4}}{(-h_5+m^{\prime}_3+m_{12}+n+m_{33})_{-k_3-j_4}(n-m^{\prime}_{13}-m^{\prime}_{44}-k_3)!m!}\\
&\qquad\times\frac{(-1)^{k_2}(-m_{12}^{\prime})_{m+m_{23}+j_4}}{k_1!k_2!k_3!j_4!(m_{33}-m^{\prime}_{33}-j_4)!}\frac{(-m_{11}^{\prime}-m^{\prime}_{34})_{k_1+k_2+k_3}}{m_{12}!m_{23}!}F_7\prod_{1\leq a\leq4}\frac{(u_a^7)^{m^{\prime}_a}}{m^{\prime}_a!}\prod_{1\leq a\leq b\leq4}\frac{(1-v_{ab}^7)^{m^{\prime}_{ab}}}{m^{\prime}_{ab}!}.
}

With these modifications, we can complete the sum over $k_1$ and obtain
\eqna{
&G_{7R}=\sum\frac{(p_4-m^{\prime}_{1})_{m^{\prime}_4+m^{\prime}_{14}}(-h_{6})_{m^{\prime}_4+m^{\prime}_{14}+m^{\prime}_{24}+m^{\prime}_{34}+m^{\prime}_{44}}(p_{4}-h_3+h_{6})_{m^{\prime}_4+m^{\prime}_{11}+m^{\prime}_{13}+m^{\prime}_{22}}}{(p_2+h_3+m^{\prime}_1)_{-m^{\prime}_4}(p_{4}-h_3)_{2m^{\prime}_4+m^{\prime}_{11}+m^{\prime}_{13}+m^{\prime}_{14}+m^{\prime}_{22}+m^{\prime}_{24}+m^{\prime}_{34}+m^{\prime}_{44}}(p_4-h_3+1-d/2)_{m^{\prime}_4}}\\
&\qquad\times\frac{(p_2+h_3)_{m^{\prime}_1-m^{\prime}_4+m_{23}}(p_3)_{-m^{\prime}_1+m^{\prime}_2+m^{\prime}_3+m_{12}+m_{33}}(\bar{p}_3+h_2+h_5)_{m^{\prime}_{3}}(p_{3}-h_2+h_{4})_{m^{\prime}_2+m_{12}+k_2}}{(p_{3}-h_2)_{2m^{\prime}_2+m_{12}+m^{\prime}_{22}+m^{\prime}_{24}+m+m_{23}+m_{33}+k_2}(p_3-h_2+1-d/2)_{m^{\prime}_2}}\\
&\qquad\times\frac{(-h_2)_{m^{\prime}_1+m^{\prime}_2-m^{\prime}_3+m^{\prime}_{22}+m^{\prime}_{24}+m+m_{23}+k_2}(p_2+h_2)_{m^{\prime}_1-m^{\prime}_2+m^{\prime}_3+m^{\prime}_{11}+m^{\prime}_{34}-k_{2}-k_3+n}}{(p_2)_{2m^{\prime}_{1}+m^{\prime}_{11}+m^{\prime}_{22}+m^{\prime}_{24}+m^{\prime}_{34}+m+n+m_{23}-k_3}(p_2+1-d/2)_{m^{\prime}_1}}\\
&\qquad\times\frac{(-h_3)_{m^{\prime}_1+m^{\prime}_4+m^{\prime}_{11}+m^{\prime}_{22}+m^{\prime}_{24}+m+n+m^{\prime}_{34}-k_3}(-h_{5})_{m^{\prime}_{3}+m^{\prime}_{13}+m^{\prime}_{23}+m^{\prime}_{33}+m^{\prime}_{44}}(-h_{4})_{m^{\prime}_2+m^{\prime}_{12}+m^{\prime}_{22}+m^{\prime}_{24}+m_{33}-j_4}}{(\bar{p}_{3}+h_{2})_{2m^{\prime}_{3}+m^{\prime}_{11}+m^{\prime}_{34}+m_{12}+n+m_{33}-k_2-k_3}(\bar{p}_3+h_2+1-d/2)_{m^{\prime}_3}}\\
&\qquad\times\frac{(-h_5+m^{\prime}_3+m_{12}+n+m_{33})_{m^{\prime}_{11}+m^{\prime}_{34}-k_2-k_3}(-m_{23}^{\prime})_{m_{12}+n+m_{33}-m^{\prime}_{13}-m^{\prime}_{33}-m^{\prime}_{44}-k_3-j_4}}{(-h_5+m^{\prime}_3+m_{12}+n+m_{33})_{-k_3-j_4}(n-m^{\prime}_{13}-m^{\prime}_{44}-k_3)!m!}\\
&\qquad\times\frac{(-1)^{k_2}(-m_{12}^{\prime})_{m+m_{23}+j_4}}{k_2!k_3!j_4!(m_{33}-m^{\prime}_{33}-j_4)!}\frac{(-m_{11}^{\prime}-m^{\prime}_{34})_{k_2+k_3}}{m_{12}!m_{23}!}F_7\prod_{1\leq a\leq4}\frac{(u_a^7)^{m^{\prime}_a}}{m^{\prime}_a!}\prod_{1\leq a\leq b\leq4}\frac{(1-v_{ab}^7)^{m^{\prime}_{ab}}}{m^{\prime}_{ab}!}.
}

It is then possible to shift $n$ by $n\to n+m^{\prime}_{13}+m^{\prime}_{44}+k_{3}$ and evaluate the sum over $k_3$,\footnote{Using the binomial identity as in \eqref{EqBinom} with $v=1$, the sum over $k_3$ forces $k_2=m_{11}^{\prime}+m^{\prime}_{34}$.} leading to
\eqna{
&G_{7R}=\sum\frac{(p_4-m^{\prime}_{1})_{m^{\prime}_4+m^{\prime}_{14}}(-h_{6})_{m^{\prime}_4+m^{\prime}_{14}+m^{\prime}_{24}+m^{\prime}_{34}+m^{\prime}_{44}}(p_{4}-h_3+h_{6})_{m^{\prime}_4+m^{\prime}_{11}+m^{\prime}_{13}+m^{\prime}_{22}}}{(p_2+h_3+m^{\prime}_1)_{-m^{\prime}_4}(p_{4}-h_3)_{2m^{\prime}_4+m^{\prime}_{11}+m^{\prime}_{13}+m^{\prime}_{14}+m^{\prime}_{22}+m^{\prime}_{24}+m^{\prime}_{34}+m^{\prime}_{44}}(p_4-h_3+1-d/2)_{m^{\prime}_4}}\\
&\qquad\times\frac{(p_2+h_3)_{m^{\prime}_1-m^{\prime}_4+m_{23}}(p_3)_{-m^{\prime}_1+m^{\prime}_2+m^{\prime}_3+m_{12}+m_{33}}(\bar{p}_3+h_2+h_5)_{m^{\prime}_{3}}(p_{3}-h_2+h_{4})_{m^{\prime}_2+m^{\prime}_{11}+m^{\prime}_{34}+m_{12}}}{(p_{3}-h_2)_{2m^{\prime}_2+m_{12}+m^{\prime}_{11}+m^{\prime}_{22}+m^{\prime}_{24}+m^{\prime}_{34}+m+m_{23}+m_{33}}(p_3-h_2+1-d/2)_{m^{\prime}_2}}\\
&\qquad\times\frac{(-h_2)_{m^{\prime}_1+m^{\prime}_2-m^{\prime}_3+m^{\prime}_{11}+m^{\prime}_{22}+m^{\prime}_{24}+m^{\prime}_{34}+m+m_{23}}(p_2+h_2)_{m^{\prime}_1-m^{\prime}_2+m^{\prime}_3+m^{\prime}_{13}+m^{\prime}_{44}+n}}{(p_2)_{2m^{\prime}_{1}+m^{\prime}_{11}+m^{\prime}_{13}+m^{\prime}_{22}+m^{\prime}_{24}+m^{\prime}_{34}+m^{\prime}_{44}+m+n+m_{23}}(p_2+1-d/2)_{m^{\prime}_1}}\\
&\qquad\times\frac{(-h_3)_{m^{\prime}_1+m^{\prime}_4+m^{\prime}_{11}+m^{\prime}_{13}+m^{\prime}_{22}+m^{\prime}_{24}+m^{\prime}_{44}+m+n+m^{\prime}_{34}}(-h_{5})_{m^{\prime}_{3}+m^{\prime}_{13}+m^{\prime}_{23}+m^{\prime}_{33}+m^{\prime}_{44}}}{(\bar{p}_{3}+h_{2})_{2m^{\prime}_{3}+m^{\prime}_{13}+m^{\prime}_{44}+m_{12}+n+m_{33}}(\bar{p}_3+h_2+1-d/2)_{m^{\prime}_3}}\\
&\qquad\times\frac{(-h_{4})_{m^{\prime}_2+m^{\prime}_{12}+m^{\prime}_{22}+m^{\prime}_{24}+m_{33}-j_4}(-m_{23}^{\prime})_{m_{12}+n+m_{33}-m^{\prime}_{33}-j_4}}{(-h_5+m^{\prime}_3+m^{\prime}_{13}+m^{\prime}_{44}+m_{12}+n+m_{33})_{-j_4}n!m!}\\
&\qquad\times\frac{(-m_{12}^{\prime})_{m+m_{23}+j_4}}{j_4!(m_{33}-m^{\prime}_{33}-j_4)!m_{12}!m_{23}!}F_7\prod_{1\leq a\leq4}\frac{(u_a^7)^{m^{\prime}_a}}{m^{\prime}_a!}\prod_{1\leq a\leq b\leq4}\frac{(1-v_{ab}^7)^{m^{\prime}_{ab}}}{m^{\prime}_{ab}!}.
}

As before, we shift $m_{33}$ by $m_{33}\to m_{33}+m^{\prime}_{33}+j_{4}$ and rewrite the sum over $m_{23}$ with the help of the first identity in \eqref{Eq3F2}.  This allows us to express $G_{7R}$ as
\eqna{
&G_{7R}=\sum\frac{(p_4-m^{\prime}_{1})_{m^{\prime}_4+m^{\prime}_{14}}(-h_{6})_{m^{\prime}_4+m^{\prime}_{14}+m^{\prime}_{24}+m^{\prime}_{34}+m^{\prime}_{44}}(p_{4}-h_3+h_{6})_{m^{\prime}_4+m^{\prime}_{11}+m^{\prime}_{13}+m^{\prime}_{22}}}{(p_2+h_3+m^{\prime}_1)_{-m^{\prime}_4}(p_{4}-h_3)_{2m^{\prime}_4+m^{\prime}_{11}+m^{\prime}_{13}+m^{\prime}_{14}+m^{\prime}_{22}+m^{\prime}_{24}+m^{\prime}_{34}+m^{\prime}_{44}}(p_4-h_3+1-d/2)_{m^{\prime}_4}}\\
&\qquad\times\frac{(p_2+h_3)_{m^{\prime}_1-m^{\prime}_4}(p_3)_{-m^{\prime}_1+m^{\prime}_2+m^{\prime}_3+m^{\prime}_{12}+m_{12}+m^{\prime}_{33}+m_{33}-m_{23}-m}(\bar{p}_3+h_2+h_5)_{m^{\prime}_{3}}}{(p_{3}-h_2)_{2m^{\prime}_2+m_{12}+m^{\prime}_{11}+m^{\prime}_{12}+m^{\prime}_{22}+m^{\prime}_{24}+m^{\prime}_{33}+m^{\prime}_{34}+m_{33}}(p_3-h_2+1-d/2)_{m^{\prime}_2}}\\
&\qquad\times\frac{(-h_2)_{m^{\prime}_1+m^{\prime}_2-m^{\prime}_3+m^{\prime}_{11}+m^{\prime}_{22}+m^{\prime}_{24}+m^{\prime}_{34}+m+m_{23}}(p_2+h_2)_{m^{\prime}_1-m^{\prime}_2+m^{\prime}_3+m^{\prime}_{13}+m^{\prime}_{44}+n}}{(p_2)_{2m^{\prime}_{1}+m^{\prime}_{11}+m^{\prime}_{13}+m^{\prime}_{22}+m^{\prime}_{24}+m^{\prime}_{34}+m^{\prime}_{44}+m+n+m_{23}}(p_2+1-d/2)_{m^{\prime}_1}}\\
&\qquad\times\frac{(-h_3)_{m^{\prime}_1+m^{\prime}_4+m^{\prime}_{11}+m^{\prime}_{13}+m^{\prime}_{22}+m^{\prime}_{24}+m^{\prime}_{44}+m^{\prime}_{34}+m_{23}+m+n}(-h_{5})_{m^{\prime}_{3}+m^{\prime}_{13}+m^{\prime}_{23}+m^{\prime}_{33}+m^{\prime}_{44}}}{(\bar{p}_{3}+h_{2})_{2m^{\prime}_{3}+m^{\prime}_{13}+m^{\prime}_{33}+m^{\prime}_{44}+m_{12}+n+m_{33}+j_4}(\bar{p}_3+h_2+1-d/2)_{m^{\prime}_3}}\\
&\qquad\times\frac{(p_{3}-h_2+h_{4})_{m^{\prime}_2+m^{\prime}_{11}+m^{\prime}_{34}+m_{12}}(-h_{4})_{m^{\prime}_2+m^{\prime}_{12}+m^{\prime}_{22}+m^{\prime}_{24}+m^{\prime}_{33}+m_{33}}(-m_{23}^{\prime})_{m_{12}+n+m_{33}}}{n!m!}\\
&\qquad\times\frac{(-1)^{m_{23}}(-h_5+m^{\prime}_3+m^{\prime}_{13}+m^{\prime}_{33}+m^{\prime}_{44}+m_{12}+n+m_{33})_{j_4}(-m_{12}^{\prime})_{m+m_{23}+j_4}}{j_4!m_{33}!m_{12}!m_{23}!}\\
&\qquad\times F_7\prod_{1\leq a\leq4}\frac{(u_a^7)^{m^{\prime}_a}}{m^{\prime}_a!}\prod_{1\leq a\leq b\leq4}\frac{(1-v_{ab}^7)^{m^{\prime}_{ab}}}{m^{\prime}_{ab}!}.
}

At this point, it is trivial to see that the sum over $j_4$ gives
\eqna{
&G_{7R}=\sum\frac{(p_4-m^{\prime}_{1})_{m^{\prime}_4+m^{\prime}_{14}}(-h_{6})_{m^{\prime}_4+m^{\prime}_{14}+m^{\prime}_{24}+m^{\prime}_{34}+m^{\prime}_{44}}(p_{4}-h_3+h_{6})_{m^{\prime}_4+m^{\prime}_{11}+m^{\prime}_{13}+m^{\prime}_{22}}}{(p_2+h_3+m^{\prime}_1)_{-m^{\prime}_4}(p_{4}-h_3)_{2m^{\prime}_4+m^{\prime}_{11}+m^{\prime}_{13}+m^{\prime}_{14}+m^{\prime}_{22}+m^{\prime}_{24}+m^{\prime}_{34}+m^{\prime}_{44}}(p_4-h_3+1-d/2)_{m^{\prime}_4}}\\
&\qquad\times\frac{(p_2+h_3)_{m^{\prime}_1-m^{\prime}_4}(p_3)_{-m^{\prime}_1+m^{\prime}_2+m^{\prime}_3+m^{\prime}_{12}+m_{12}+m^{\prime}_{33}+m_{33}-m_{23}-m}(\bar{p}_3+h_2+h_5)_{m^{\prime}_{3}+m_{12}^{\prime}-m-m_{23}}}{(p_{3}-h_2)_{2m^{\prime}_2+m_{12}+m^{\prime}_{11}+m^{\prime}_{12}+m^{\prime}_{22}+m^{\prime}_{24}+m^{\prime}_{33}+m^{\prime}_{34}+m_{33}}(p_3-h_2+1-d/2)_{m^{\prime}_2}}\\
&\qquad\times\frac{(-h_2)_{m^{\prime}_1+m^{\prime}_2-m^{\prime}_3+m^{\prime}_{11}+m^{\prime}_{22}+m^{\prime}_{24}+m^{\prime}_{34}+m+m_{23}}(p_2+h_2)_{m^{\prime}_1-m^{\prime}_2+m^{\prime}_3+m^{\prime}_{13}+m^{\prime}_{44}+n}}{(p_2)_{2m^{\prime}_{1}+m^{\prime}_{11}+m^{\prime}_{13}+m^{\prime}_{22}+m^{\prime}_{24}+m^{\prime}_{34}+m^{\prime}_{44}+m+n+m_{23}}(p_2+1-d/2)_{m^{\prime}_1}}\\
&\qquad\times\frac{(-h_3)_{m^{\prime}_1+m^{\prime}_4+m^{\prime}_{11}+m^{\prime}_{13}+m^{\prime}_{22}+m^{\prime}_{24}+m^{\prime}_{44}+m^{\prime}_{34}+m_{23}+m+n}(-h_{5})_{m^{\prime}_{3}+m^{\prime}_{13}+m^{\prime}_{23}+m^{\prime}_{33}+m^{\prime}_{44}}}{(\bar{p}_{3}+h_{2})_{2m^{\prime}_{3}+m^{\prime}_{12}+m^{\prime}_{13}+m^{\prime}_{33}+m^{\prime}_{44}+m_{12}+n+m_{33}-m_{23}-m}(\bar{p}_3+h_2+1-d/2)_{m^{\prime}_3}}\\
&\qquad\times\frac{(p_{3}-h_2+h_{4})_{m^{\prime}_2+m^{\prime}_{11}+m^{\prime}_{34}+m_{12}}(-h_{4})_{m^{\prime}_2+m^{\prime}_{12}+m^{\prime}_{22}+m^{\prime}_{24}+m^{\prime}_{33}+m_{33}}(-m_{23}^{\prime})_{m_{12}+n+m_{33}}}{n!m!}\\
&\qquad\times\frac{(-1)^{m_{23}}(-m_{12}^{\prime})_{m+m_{23}}}{m_{33}!m_{12}!m_{23}!}F_7\prod_{1\leq a\leq4}\frac{(u_a^7)^{m^{\prime}_a}}{m^{\prime}_a!}\prod_{1\leq a\leq b\leq4}\frac{(1-v_{ab}^7)^{m^{\prime}_{ab}}}{m^{\prime}_{ab}!}.
}

After redefining $m_{33}=k-m_{12}$, we can compute the sums over $m_{12}$, $k$, and $n$, leading to
\eqna{
&G_{7R}=\sum\frac{(p_4-m^{\prime}_{1})_{m^{\prime}_4+m^{\prime}_{14}}(-h_{6})_{m^{\prime}_4+m^{\prime}_{14}+m^{\prime}_{24}+m^{\prime}_{34}+m^{\prime}_{44}}(p_{4}-h_3+h_{6})_{m^{\prime}_4+m^{\prime}_{11}+m^{\prime}_{13}+m^{\prime}_{22}}}{(p_2+h_3+m^{\prime}_1)_{-m^{\prime}_4}(p_{4}-h_3)_{2m^{\prime}_4+m^{\prime}_{11}+m^{\prime}_{13}+m^{\prime}_{14}+m^{\prime}_{22}+m^{\prime}_{24}+m^{\prime}_{34}+m^{\prime}_{44}}(p_4-h_3+1-d/2)_{m^{\prime}_4}}\\
&\qquad\times\frac{(p_2+h_3)_{m^{\prime}_1-m^{\prime}_4+m^{\prime}_{23}}(p_3)_{-m^{\prime}_1+m^{\prime}_2+m^{\prime}_3+m^{\prime}_{12}+m^{\prime}_{33}-m_{23}-m}(\bar{p}_3+h_2+h_5)_{m^{\prime}_{3}+m_{12}^{\prime}-m-m_{23}}}{(p_{3}-h_2)_{2m^{\prime}_2+m^{\prime}_{11}+m^{\prime}_{12}+m^{\prime}_{22}+m^{\prime}_{24}+m^{\prime}_{33}+m^{\prime}_{34}}(p_3-h_2+1-d/2)_{m^{\prime}_2}}\\
&\qquad\times\frac{(-h_2)_{m^{\prime}_1+m^{\prime}_2-m^{\prime}_3+m^{\prime}_{11}+m^{\prime}_{22}+m^{\prime}_{24}+m^{\prime}_{34}+m+m_{23}}(p_2+h_2)_{m^{\prime}_1-m^{\prime}_2+m^{\prime}_3+m^{\prime}_{13}+m^{\prime}_{23}+m^{\prime}_{44}}}{(p_2)_{2m^{\prime}_{1}+m^{\prime}_{11}+m^{\prime}_{13}+m^{\prime}_{22}+m^{\prime}_{23}+m^{\prime}_{24}+m^{\prime}_{34}+m^{\prime}_{44}+m+m_{23}}(p_2+1-d/2)_{m^{\prime}_1}}\\
&\qquad\times\frac{(-h_3)_{m^{\prime}_1+m^{\prime}_4+m^{\prime}_{11}+m^{\prime}_{13}+m^{\prime}_{22}+m^{\prime}_{24}+m^{\prime}_{44}+m^{\prime}_{34}+m_{23}+m}(-h_{5})_{m^{\prime}_{3}+m^{\prime}_{13}+m^{\prime}_{23}+m^{\prime}_{33}+m^{\prime}_{44}}}{(\bar{p}_{3}+h_{2})_{2m^{\prime}_{3}+m^{\prime}_{12}+m^{\prime}_{13}+m^{\prime}_{23}+m^{\prime}_{33}+m^{\prime}_{44}-m_{23}-m}(\bar{p}_3+h_2+1-d/2)_{m^{\prime}_3}}\\
&\qquad\times\frac{(p_{3}-h_2+h_{4})_{m^{\prime}_2+m^{\prime}_{11}+m^{\prime}_{34}}(-h_{4})_{m^{\prime}_2+m^{\prime}_{12}+m^{\prime}_{22}+m^{\prime}_{24}+m^{\prime}_{33}}}{m!}\\
&\qquad\times\frac{(-1)^{m_{23}}(-m_{12}^{\prime})_{m+m_{23}}}{m_{23}!}F_7\prod_{1\leq a\leq4}\frac{(u_a^7)^{m^{\prime}_a}}{m^{\prime}_a!}\prod_{1\leq a\leq b\leq4}\frac{(1-v_{ab}^7)^{m^{\prime}_{ab}}}{m^{\prime}_{ab}!}.
}

Finally, we introduce $m_{23}=r-m$ and evaluate the sums over $m$ and $r$,\footnote{Again, the binomial identity \eqref{EqBinom} with $v=1$ implies that the sum over $m$ forces $r=0$.} such that
\eqna{
&G_{7R}=\sum\frac{(p_4-m^{\prime}_{1})_{m^{\prime}_4+m^{\prime}_{14}}(-h_{6})_{m^{\prime}_4+m^{\prime}_{14}+m^{\prime}_{24}+m^{\prime}_{34}+m^{\prime}_{44}}(p_{4}-h_3+h_{6})_{m^{\prime}_4+m^{\prime}_{11}+m^{\prime}_{13}+m^{\prime}_{22}}}{(p_2+h_3+m^{\prime}_1)_{-m^{\prime}_4}(p_{4}-h_3)_{2m^{\prime}_4+m^{\prime}_{11}+m^{\prime}_{13}+m^{\prime}_{14}+m^{\prime}_{22}+m^{\prime}_{24}+m^{\prime}_{34}+m^{\prime}_{44}}(p_4-h_3+1-d/2)_{m^{\prime}_4}}\\
&\qquad\times\frac{(p_2+h_3)_{m^{\prime}_1-m^{\prime}_4+m^{\prime}_{23}}(p_3)_{-m^{\prime}_1+m^{\prime}_2+m^{\prime}_3+m^{\prime}_{12}+m^{\prime}_{33}}(p_{3}-h_2+h_{4})_{m^{\prime}_2+m^{\prime}_{11}+m^{\prime}_{34}}(\bar{p}_3+h_2+h_5)_{m^{\prime}_{3}+m_{12}^{\prime}}}{(p_{3}-h_2)_{2m^{\prime}_2+m^{\prime}_{11}+m^{\prime}_{12}+m^{\prime}_{22}+m^{\prime}_{24}+m^{\prime}_{33}+m^{\prime}_{34}}(p_3-h_2+1-d/2)_{m^{\prime}_2}}\\
&\qquad\times\frac{(-h_2)_{m^{\prime}_1+m^{\prime}_2-m^{\prime}_3+m^{\prime}_{11}+m^{\prime}_{22}+m^{\prime}_{24}+m^{\prime}_{34}}(p_2+h_2)_{m^{\prime}_1-m^{\prime}_2+m^{\prime}_3+m^{\prime}_{13}+m^{\prime}_{23}+m^{\prime}_{44}}}{(p_2)_{2m^{\prime}_{1}+m^{\prime}_{11}+m^{\prime}_{13}+m^{\prime}_{22}+m^{\prime}_{23}+m^{\prime}_{24}+m^{\prime}_{34}+m^{\prime}_{44}}(p_2+1-d/2)_{m^{\prime}_1}}\\
&\qquad\times\frac{(-h_3)_{m^{\prime}_1+m^{\prime}_4+m^{\prime}_{11}+m^{\prime}_{13}+m^{\prime}_{22}+m^{\prime}_{24}+m^{\prime}_{34}+m^{\prime}_{44}}(-h_{4})_{m^{\prime}_2+m^{\prime}_{12}+m^{\prime}_{22}+m^{\prime}_{24}+m^{\prime}_{33}}(-h_{5})_{m^{\prime}_{3}+m^{\prime}_{13}+m^{\prime}_{23}+m^{\prime}_{33}+m^{\prime}_{44}}}{(\bar{p}_{3}+h_{2})_{2m^{\prime}_{3}+m^{\prime}_{12}+m^{\prime}_{13}+m^{\prime}_{23}+m^{\prime}_{33}+m^{\prime}_{44}}(\bar{p}_3+h_2+1-d/2)_{m^{\prime}_3}}\\
&\qquad\times F_7\prod_{1\leq a\leq4}\frac{(u_a^7)^{m^{\prime}_a}}{m^{\prime}_a!}\prod_{1\leq a\leq b\leq4}\frac{(1-v_{ab}^7)^{m^{\prime}_{ab}}}{m^{\prime}_{ab}!}\\
&\qquad=G_7.
}
This ends our proof of the invariance of the scalar seven-point correlation functions under reflections \eqref{EqSymRef}.

%%%%%%%%%%%%%%%%%%%%%%%%%%%%%%%%%%%%%%%%%%%%%%%%%%

\subsection{Dendrite Permutations of the First Kind}

Invariance under dendrite permutations of the first kind implies the identity \eqref{EqSymPerm1}, which we rewrite as $G_7=G_{7P_1}$ to simplify the notation.  Once again, $F_7$ is invariant under the generator, as can be seen from \eqref{EqCB7F}.  Expressing $G_{7P_1}$ in terms of the original conformal cross-ratios and expanding, we find
\eqna{
&G_{7P_1}=\sum \frac{(p_4-m_{1})_{m_4+m_{14}}(-h_{6})_{m_4+m_{14}+m_{24}+m_{34}+m_{44}}(p_{4}-h_3+h_{6})_{m_4+m_{11}+m_{13}+m_{22}}}{(p_{4}-h_3)_{2m_4+m_{11}+m_{13}+m_{14}+m_{22}+m_{24}+m_{34}+m_{44}}(p_4-h_3+1-d/2)_{m_4}}\\
&\qquad\times\frac{(p_2+h_3)_{m_{1}-m_4+m_{23}}(-h_4)_{m_{2}+m_{11}+m_{34}}(\bar{p}_{3}+h_2+h_{5})_{m_3+m_{12}}}{(p_2+h_3+m_1)_{-m_4}(\bar{p}_{3}+h_2)_{2m_3+m_{13}+m_{44}+m_{12}+m_{23}+m_{33}}(\bar{p}_3+h_2+1-d/2)_{m_3}}\\
&\qquad\times\frac{(p_3)_{-m_1+m_{2}+m_3+m_{12}+m_{33}}(p_2+h_2)_{m_1-m_2+m_3+m_{13}+m_{44}+m_{23}}(-h_2)_{m_1+m_2-m_3+m_{11}+m_{22}+m_{24}+m_{34}}}{(p_2)_{2m_1+m_{11}+m_{13}+m_{22}+m_{23}+m_{24}+m_{34}+m_{44}}(p_2+1-d/2)_{m_1}}\\
&\qquad\times\frac{(-h_3)_{m_1+m_4+m_{11}+m_{13}+m_{22}+m_{24}+m_{34}+m_{44}}(p_3-h_2+h_{4})_{m_{2}+m_{12}+m_{22}+m_{24}+m_{33}}(-h_{5})_{m_3+m_{13}+m_{23}+m_{33}+m_{44}}}{(p_{3}-h_{2})_{2m_{2}+m_{11}+m_{12}+m_{22}+m_{24}+m_{33}+m_{34}}(p_3-h_2+1-d/2)_{m_2}}\\
&\qquad\times\binom{m_{12}}{k_{12}}\binom{m_{13}}{k_{13}}\binom{m_{23}}{k_{23}}\binom{m_{33}}{k_{33}}\binom{m_{44}}{k_{44}}\binom{-p_3-h_5+m_1-m_2-k_{12}+k_{13}+k_{23}+k_{44}}{m^{\prime}_{12}}\\
&\qquad\times\binom{k_{13}}{m^{\prime}_{13}}\binom{k_{23}}{m^{\prime}_{23}}\binom{h_5-m_3-k_{13}-k_{23}-k_{33}-k_{44}}{m^{\prime}_{33}}\binom{k_{44}}{m^{\prime}_{44}}\\
&\qquad\times(-1)^{k_{12}+k_{13}+k_{23}+k_{33}+k_{44}+m^{\prime}_{12}+m^{\prime}_{13}+m^{\prime}_{23}+m^{\prime}_{33}+m^{\prime}_{44}}F_7\prod_{1\leq a\leq4}\frac{(u_a^7)^{m_a}}{m_a!}\prod_{1\leq a\leq b\leq4}\frac{(1-v_{ab}^7)^{m_{ab}}}{m_{ab}!},
}
and we must now re-sum the extra sums to rewrite the result as $G_7$.\footnote{For simplicity, we wrote the product over the $v_{ab}$ cross-ratios although some of the cross-ratios are mixed as per \eqref{EqSymPerm1}.  They will be unmixed at the end of the proof.}

We first evaluate the sums over $k_{12}$ and $k_{33}$, leading to
\eqna{
&G_{7P_1}=\sum \frac{(p_4-m_{1})_{m_4+m_{14}}(-h_{6})_{m_4+m_{14}+m_{24}+m_{34}+m_{44}}(p_{4}-h_3+h_{6})_{m_4+m_{11}+m_{13}+m_{22}}}{(p_{4}-h_3)_{2m_4+m_{11}+m_{13}+m_{14}+m_{22}+m_{24}+m_{34}+m_{44}}(p_4-h_3+1-d/2)_{m_4}}\\
&\qquad\times\frac{(p_2+h_3)_{m_{1}-m_4+m_{23}}(-h_4)_{m_{2}+m_{11}+m_{34}}(\bar{p}_{3}+h_2+h_{5})_{m_3+m_{12}}}{(p_2+h_3+m_1)_{-m_4}(\bar{p}_{3}+h_2)_{2m_3+m_{13}+m_{44}+m_{12}+m_{23}+m_{33}}(\bar{p}_3+h_2+1-d/2)_{m_3}}\\
&\qquad\times\frac{(p_3)_{-m_1+m_{2}+m_3+m_{12}+m_{33}}(p_2+h_2)_{m_1-m_2+m_3+m_{13}+m_{44}+m_{23}}(-h_2)_{m_1+m_2-m_3+m_{11}+m_{22}+m_{24}+m_{34}}}{(p_2)_{2m_1+m_{11}+m_{13}+m_{22}+m_{23}+m_{24}+m_{34}+m_{44}}(p_2+1-d/2)_{m_1}}\\
&\qquad\times\frac{(-h_3)_{m_1+m_4+m_{11}+m_{13}+m_{22}+m_{24}+m_{34}+m_{44}}(p_3-h_2+h_{4})_{m_{2}+m_{12}+m_{22}+m_{24}+m_{33}}(-h_{5})_{m_3+m_{13}+m_{23}+m_{33}+m_{44}}}{(p_{3}-h_{2})_{2m_{2}+m_{11}+m_{12}+m_{22}+m_{24}+m_{33}+m_{34}}(p_3-h_2+1-d/2)_{m_2}}\\
&\qquad\times\binom{m_{13}}{k_{13}}\binom{m_{23}}{k_{23}}\binom{m_{44}}{k_{44}}\frac{(-m_{12}^{\prime})_{m_{12}}(p_3+h_5-m_1+m_2+m_{12}-k_{13}-k_{23}-k_{44})_{m^{\prime}_{12}-m_{12}}}{m^{\prime}_{12}!}\\
&\qquad\times\binom{k_{13}}{m^{\prime}_{13}}\binom{k_{23}}{m^{\prime}_{23}}\binom{k_{44}}{m^{\prime}_{44}}\frac{(-m^{\prime}_{33})_{m_{33}}(-h_5+m_3+m_{33}+k_{13}+k_{23}+k_{44})_{m_{33}^{\prime}-m_{33}}}{m^{\prime}_{33}!}\\
&\qquad\times(-1)^{k_{13}+k_{23}+k_{44}+m^{\prime}_{13}+m^{\prime}_{23}+m^{\prime}_{44}}F_7\prod_{1\leq a\leq4}\frac{(u_a^7)^{m_a}}{m_a!}\prod_{1\leq a\leq b\leq4}\frac{(1-v_{ab}^7)^{m_{ab}}}{m_{ab}!}.
}

Using the following identity
\eqna{
&(p_3+h_5-m_1+m_2+m_{12}-k_{13}-k_{23}-k_{44})_{m^{\prime}_{12}-m_{12}}\\
&\qquad=\sum_{j_1\geq 0}\binom{m^{\prime}_{12}-m_{12}}{j_1}(p_3+h_5-m_1+m_2+m_{12}-m^{\prime}_{13}-k_{23}-k_{44})_{m^{\prime}_{12}-m_{12}-j_1}(-k_{13}+m^{\prime}_{13})_{j_1},
}
and then changing the variable such that $k_{13}\to k_{13}+m^{\prime}_{13}+j_1$, we can then evaluate the sum over $k_{13}$ to obtain
\eqna{
&G_{7P_1}=\sum \frac{(p_4-m_{1})_{m_4+m_{14}}(-h_{6})_{m_4+m_{14}+m_{24}+m_{34}+m_{44}}(p_{4}-h_3+h_{6})_{m_4+m_{11}+m_{13}+m_{22}}}{(p_{4}-h_3)_{2m_4+m_{11}+m_{13}+m_{14}+m_{22}+m_{24}+m_{34}+m_{44}}(p_4-h_3+1-d/2)_{m_4}}\\
&\qquad\times\frac{(p_2+h_3)_{m_{1}-m_4+m_{23}}(-h_4)_{m_{2}+m_{11}+m_{34}}(\bar{p}_{3}+h_2+h_{5})_{m_3+m_{12}}}{(p_2+h_3+m_1)_{-m_4}(\bar{p}_{3}+h_2)_{2m_3+m_{13}+m_{44}+m_{12}+m_{23}+m_{33}}(\bar{p}_3+h_2+1-d/2)_{m_3}}\\
&\qquad\times\frac{(p_3)_{-m_1+m_{2}+m_3+m_{12}+m_{33}}(p_2+h_2)_{m_1-m_2+m_3+m_{13}+m_{44}+m_{23}}(-h_2)_{m_1+m_2-m_3+m_{11}+m_{22}+m_{24}+m_{34}}}{(p_2)_{2m_1+m_{11}+m_{13}+m_{22}+m_{23}+m_{24}+m_{34}+m_{44}}(p_2+1-d/2)_{m_1}}\\
&\qquad\times\frac{(-h_3)_{m_1+m_4+m_{11}+m_{13}+m_{22}+m_{24}+m_{34}+m_{44}}(p_3-h_2+h_{4})_{m_{2}+m_{12}+m_{22}+m_{24}+m_{33}}(-h_{5})_{m_3+m_{13}+m_{23}+m_{33}+m_{44}}}{(p_{3}-h_{2})_{2m_{2}+m_{11}+m_{12}+m_{22}+m_{24}+m_{33}+m_{34}}(p_3-h_2+1-d/2)_{m_2}}\\
&\qquad\times\binom{m_{23}}{k_{23}}\binom{m_{44}}{k_{44}}\frac{m_{13}!(-m_{12}^{\prime})_{j_1+m_{12}}(p_3+h_5-m_1+m_2+m_{12}-m^{\prime}_{13}-k_{23}-k_{44})_{m^{\prime}_{12}-m_{12}-j_1}}{m^{\prime}_{12}!m^{\prime}_{13}!(m_{13}-m^{\prime}_{13}-j_1)!}\\
&\qquad\times\binom{k_{23}}{m^{\prime}_{23}}\binom{k_{44}}{m^{\prime}_{44}}\frac{(-m^{\prime}_{33})_{m_{13}+m_{33}-m^{\prime}_{13}-j_1}(-h_5+m_3+m_{13}+m_{33}+k_{23}+k_{44})_{j_1+m_{13}^{\prime}+m_{33}^{\prime}-m_{13}-m_{33}}}{j_1!m^{\prime}_{33}!}\\
&\qquad\times(-1)^{j_1+k_{23}+k_{44}+m^{\prime}_{23}+m^{\prime}_{44}}F_7\prod_{1\leq a\leq4}\frac{(u_a^7)^{m_a}}{m_a!}\prod_{1\leq a\leq b\leq4}\frac{(1-v_{ab}^7)^{m_{ab}}}{m_{ab}!}.
}

In a similar way, we compute the sums over $k_{23}$ and $k_{44}$, giving us
\eqna{
&G_{7P_1}=\sum \frac{(p_4-m_{1})_{m_4+m_{14}}(-h_{6})_{m_4+m_{14}+m_{24}+m_{34}+m_{44}}(p_{4}-h_3+h_{6})_{m_4+m_{11}+m_{13}+m_{22}}}{(p_{4}-h_3)_{2m_4+m_{11}+m_{13}+m_{14}+m_{22}+m_{24}+m_{34}+m_{44}}(p_4-h_3+1-d/2)_{m_4}}\\
&\qquad\times\frac{(p_2+h_3)_{m_{1}-m_4+m_{23}}(-h_4)_{m_{2}+m_{11}+m_{34}}(\bar{p}_{3}+h_2+h_{5})_{m_3+m_{12}}}{(p_2+h_3+m_1)_{-m_4}(\bar{p}_{3}+h_2)_{2m_3+m_{13}+m_{44}+m_{12}+m_{23}+m_{33}}(\bar{p}_3+h_2+1-d/2)_{m_3}}\\
&\qquad\times\frac{(p_3)_{-m_1+m_{2}+m_3+m_{12}+m_{33}}(p_2+h_2)_{m_1-m_2+m_3+m_{13}+m_{44}+m_{23}}(-h_2)_{m_1+m_2-m_3+m_{11}+m_{22}+m_{24}+m_{34}}}{(p_2)_{2m_1+m_{11}+m_{13}+m_{22}+m_{23}+m_{24}+m_{34}+m_{44}}(p_2+1-d/2)_{m_1}}\\
&\qquad\times\frac{(-h_3)_{m_1+m_4+m_{11}+m_{13}+m_{22}+m_{24}+m_{34}+m_{44}}(p_3-h_2+h_{4})_{m_{2}+m_{12}+m_{22}+m_{24}+m_{33}}}{(p_{3}-h_{2})_{2m_{2}+m_{11}+m_{12}+m_{22}+m_{24}+m_{33}+m_{34}}(p_3-h_2+1-d/2)_{m_2}}\\
&\qquad\times\frac{(-m_{12}^{\prime})_{j_1+j_2+j_3+m_{12}}(p_3+h_5-m_1+m_2+m_{12}-m^{\prime}_{13}-m^{\prime}_{23}-m^{\prime}_{44})_{m^{\prime}_{12}-m_{12}-j_1-j_2-j_3}}{(m_{13}-m^{\prime}_{13}-j_1)!(m_{23}-m^{\prime}_{23}-j_2)!}\\
&\qquad\times\frac{(-h_{5})_{m_3+m^{\prime}_{13}+m^{\prime}_{23}+m^{\prime}_{33}+m^{\prime}_{44}+j_1+j_2+j_3}(-m^{\prime}_{33})_{m_{13}+m_{23}+m_{33}+m_{44}-m^{\prime}_{13}-m^{\prime}_{23}-m^{\prime}_{44}-j_1-j_2-j_3}}{j_1!j_2!j_3!m_{12}!m_{33}!(m_{44}-m^{\prime}_{44}-j_3)!}\\
&\qquad\times(-1)^{j_1+j_2+j_3}F_7\prod_{1\leq a\leq4}\frac{(u_a^7)^{m_a}}{m_a!}\prod_{1\leq a\leq b\leq4}\frac{(1-v_{ab}^7)^{m^{\prime}_{ab}}}{m^{\prime}_{ab}!}.
}

We then define $m_{13}=m-m_{44}$ and shift $m_{44}$ by $m_{44}+m^{\prime}_{44}+j_3$.  This allows us to sum over $m_{44}$, leading to
\eqna{
&G_{7P_1}=\sum \frac{(p_4-m_{1})_{m_4+m_{14}}(-h_{6})_{m_4+m^{\prime}_{44}+m_{14}+m_{24}+m_{34}+j_3}(p_{4}-h_3+h_{6})_{m_4+m^{\prime}_{13}+m_{11}+m_{22}+j_1}}{(p_{4}-h_3)_{2m_4+m^{\prime}_{13}+m^{\prime}_{44}+m_{11}+m_{14}+m_{22}+m_{24}+m_{34}+j_1+j_3}(p_4-h_3+1-d/2)_{m_4}}\\
&\qquad\times\frac{(p_2+h_3)_{m_{1}-m_4+m_{23}}(-h_4)_{m_{2}+m_{11}+m_{34}}(\bar{p}_{3}+h_2+h_{5})_{m_3+m_{12}}}{(p_2+h_3+m_1)_{-m_4}(\bar{p}_{3}+h_2)_{2m_3+m_{12}+m_{23}+m_{33}+m}(\bar{p}_3+h_2+1-d/2)_{m_3}}\\
&\qquad\times\frac{(p_3)_{-m_1+m_{2}+m_3+m_{12}+m_{33}}(p_2+h_2)_{m_1-m_2+m_3+m_{23}+m}(-h_2)_{m_1+m_2-m_3+m_{11}+m_{22}+m_{24}+m_{34}}}{(p_2)_{2m_1+m_{11}+m_{22}+m_{23}+m_{24}+m_{34}+m}(p_2+1-d/2)_{m_1}}\\
&\qquad\times\frac{(-h_3)_{m_1+m_4+m_{11}+m_{22}+m_{24}+m_{34}+m}(p_3-h_2+h_{4})_{m_{2}+m_{12}+m_{22}+m_{24}+m_{33}}}{(p_{3}-h_{2})_{2m_{2}+m_{11}+m_{12}+m_{22}+m_{24}+m_{33}+m_{34}}(p_3-h_2+1-d/2)_{m_2}}\\
&\qquad\times\frac{(-m_{12}^{\prime})_{j_1+j_2+j_3+m_{12}}(p_3+h_5-m_1+m_2+m_{12}-m^{\prime}_{13}-m^{\prime}_{23}-m^{\prime}_{44})_{m^{\prime}_{12}-m_{12}-j_1-j_2-j_3}}{(m-m^{\prime}_{13}-m^{\prime}_{44}-j_1-j_3)!(m_{23}-m^{\prime}_{23}-j_2)!}\\
&\qquad\times\frac{(-h_{5})_{m_3+m^{\prime}_{13}+m^{\prime}_{23}+m^{\prime}_{33}+m^{\prime}_{44}+j_1+j_2+j_3}(-m^{\prime}_{33})_{m_{23}+m_{33}+m-m^{\prime}_{13}-m^{\prime}_{23}-m^{\prime}_{44}-j_1-j_2-j_3}}{j_1!j_2!j_3!m_{12}!m_{33}!}\\
&\qquad\times(-1)^{j_1+j_2+j_3}F_7\prod_{1\leq a\leq4}\frac{(u_a^7)^{m_a}}{m_a!}\prod_{1\leq a\leq b\leq4}\frac{(1-v_{ab}^7)^{m^{\prime}_{ab}}}{m^{\prime}_{ab}!}.
}

Defining $j_3=j-j_1$, the sum over $j_1$ thus gives
\eqna{
&G_{7P_1}=\sum \frac{(p_4-m_{1})_{m_4+m_{14}}(-h_{6})_{m_4+m^{\prime}_{44}+m_{14}+m_{24}+m_{34}}(p_{4}-h_3+h_{6})_{m_4+m^{\prime}_{13}+m_{11}+m_{22}}}{(p_{4}-h_3)_{2m_4+m^{\prime}_{13}+m^{\prime}_{44}+m_{11}+m_{14}+m_{22}+m_{24}+m_{34}}(p_4-h_3+1-d/2)_{m_4}}\\
&\qquad\times\frac{(p_2+h_3)_{m_{1}-m_4+m_{23}}(-h_4)_{m_{2}+m_{11}+m_{34}}(\bar{p}_{3}+h_2+h_{5})_{m_3+m_{12}}}{(p_2+h_3+m_1)_{-m_4}(\bar{p}_{3}+h_2)_{2m_3+m_{12}+m_{23}+m_{33}+m}(\bar{p}_3+h_2+1-d/2)_{m_3}}\\
&\qquad\times\frac{(p_3)_{-m_1+m_{2}+m_3+m_{12}+m_{33}}(p_2+h_2)_{m_1-m_2+m_3+m_{23}+m}(-h_2)_{m_1+m_2-m_3+m_{11}+m_{22}+m_{24}+m_{34}}}{(p_2)_{2m_1+m_{11}+m_{22}+m_{23}+m_{24}+m_{34}+m}(p_2+1-d/2)_{m_1}}\\
&\qquad\times\frac{(-h_3)_{m_1+m_4+m_{11}+m_{22}+m_{24}+m_{34}+m}(p_3-h_2+h_{4})_{m_{2}+m_{12}+m_{22}+m_{24}+m_{33}}}{(p_{3}-h_{2})_{2m_{2}+m_{11}+m_{12}+m_{22}+m_{24}+m_{33}+m_{34}}(p_3-h_2+1-d/2)_{m_2}}\\
&\qquad\times\frac{(-m_{12}^{\prime})_{j+j_2+m_{12}}(p_3+h_5-m_1+m_2+m_{12}-m^{\prime}_{13}-m^{\prime}_{23}-m^{\prime}_{44})_{m^{\prime}_{12}-m_{12}-j-j_2}}{(m-m^{\prime}_{13}-m^{\prime}_{44}-j)!(m_{23}-m^{\prime}_{23}-j_2)!}\\
&\qquad\times\frac{(-h_{5})_{m_3+m^{\prime}_{13}+m^{\prime}_{23}+m^{\prime}_{33}+m^{\prime}_{44}+j+j_2}(-m^{\prime}_{33})_{m_{23}+m_{33}+m-m^{\prime}_{13}-m^{\prime}_{23}-m^{\prime}_{44}-j-j_2}}{j!j_2!m_{12}!m_{33}!}\\
&\qquad\times(-1)^{j+j_2}F_7\prod_{1\leq a\leq4}\frac{(u_a^7)^{m_a}}{m_a!}\prod_{1\leq a\leq b\leq4}\frac{(1-v_{ab}^7)^{m^{\prime}_{ab}}}{m^{\prime}_{ab}!}.
}

We now define $m_{23}=k-m$ and then change $m$ by $m\to m+m^{\prime}_{13}+m^{\prime}_{44}+j$.  The sums over $m$ and $k$ (with the extra change $k\to k+m^{\prime}_{13}+m^{\prime}_{23}+m^{\prime}_{44}+j+j_2$) can be performed and lead to
\eqna{
&G_{7P_1}=\sum \frac{(p_4-m_{1})_{m_4+m_{14}}(-h_{6})_{m_4+m^{\prime}_{44}+m_{14}+m_{24}+m_{34}}(p_{4}-h_3+h_{6})_{m_4+m^{\prime}_{13}+m_{11}+m_{22}}}{(p_{4}-h_3)_{2m_4+m^{\prime}_{13}+m^{\prime}_{44}+m_{11}+m_{14}+m_{22}+m_{24}+m_{34}}(p_4-h_3+1-d/2)_{m_4}}\\
&\qquad\times\frac{(p_2+h_3)_{m_{1}-m_4+m^{\prime}_{23}+j_2}(-h_4)_{m_{2}+m_{11}+m_{34}}(\bar{p}_{3}+h_2+h_{5})_{m_3+m_{12}}}{(p_2+h_3+m_1)_{-m_4}(\bar{p}_{3}+h_2)_{2m_3+m_{12}+m^{\prime}_{13}+m^{\prime}_{23}+m^{\prime}_{33}+m^{\prime}_{44}+j+j_2}(\bar{p}_3+h_2+1-d/2)_{m_3}}\\
&\qquad\times\frac{(p_3)_{-m_1+m_{2}+m_3+m_{12}+m^{\prime}_{33}}(p_2+h_2)_{m_1-m_2+m_3+m^{\prime}_{13}+m^{\prime}_{23}+m^{\prime}_{44}+j+j_2}(-h_2)_{m_1+m_2-m_3+m_{11}+m_{22}+m_{24}+m_{34}}}{(p_2)_{2m_1+m^{\prime}_{13}+m^{\prime}_{23}+m^{\prime}_{44}+m_{11}+m_{22}+m_{24}+m_{34}+j+j_2}(p_2+1-d/2)_{m_1}}\\
&\qquad\times\frac{(-h_3)_{m_1+m_4+m^{\prime}_{13}+m^{\prime}_{44}+m_{11}+m_{22}+m_{24}+m_{34}+j}(p_3-h_2+h_{4})_{m_{2}+m_{12}+m_{22}+m_{24}+m_{33}}}{(p_{3}-h_{2})_{2m_{2}+m_{11}+m_{12}+m_{22}+m_{24}+m_{33}+m_{34}}(p_3-h_2+1-d/2)_{m_2}}\\
&\qquad\times\frac{(-m^{\prime}_{33})_{m_{33}}(-m_{12}^{\prime})_{j+j_2+m_{12}}(p_3+h_5-m_1+m_2+m_{12}-m^{\prime}_{13}-m^{\prime}_{23}-m^{\prime}_{44})_{m^{\prime}_{12}-m_{12}-j-j_2}}{j!j_2!m_{12}!m_{33}!}\\
&\qquad\times(-h_{5})_{m_3+m^{\prime}_{13}+m^{\prime}_{23}+m^{\prime}_{33}+m^{\prime}_{44}+j+j_2}(-1)^{j+j_2}F_7\prod_{1\leq a\leq4}\frac{(u_a^7)^{m_a}}{m_a!}\prod_{1\leq a\leq b\leq4}\frac{(1-v_{ab}^7)^{m^{\prime}_{ab}}}{m^{\prime}_{ab}!}.
}

It is now possible to re-sum over $m_{33}$ to get
\eqna{
&G_{7P_1}=\sum \frac{(p_4-m_{1})_{m_4+m_{14}}(-h_{6})_{m_4+m^{\prime}_{44}+m_{14}+m_{24}+m_{34}}(p_{4}-h_3+h_{6})_{m_4+m^{\prime}_{13}+m_{11}+m_{22}}}{(p_{4}-h_3)_{2m_4+m^{\prime}_{13}+m^{\prime}_{44}+m_{11}+m_{14}+m_{22}+m_{24}+m_{34}}(p_4-h_3+1-d/2)_{m_4}}\\
&\qquad\times\frac{(p_2+h_3)_{m_{1}-m_4+m^{\prime}_{23}+j_2}(-h_4)_{m_{2}+m_{11}+m_{34}+m^{\prime}_{33}}(\bar{p}_{3}+h_2+h_{5})_{m_3+m_{12}}}{(p_2+h_3+m_1)_{-m_4}(\bar{p}_{3}+h_2)_{2m_3+m_{12}+m^{\prime}_{13}+m^{\prime}_{23}+m^{\prime}_{33}+m^{\prime}_{44}+j+j_2}(\bar{p}_3+h_2+1-d/2)_{m_3}}\\
&\qquad\times\frac{(p_3)_{-m_1+m_{2}+m_3+m_{12}+m^{\prime}_{33}}(p_2+h_2)_{m_1-m_2+m_3+m^{\prime}_{13}+m^{\prime}_{23}+m^{\prime}_{44}+j+j_2}(-h_2)_{m_1+m_2-m_3+m_{11}+m_{22}+m_{24}+m_{34}}}{(p_2)_{2m_1+m^{\prime}_{13}+m^{\prime}_{23}+m^{\prime}_{44}+m_{11}+m_{22}+m_{24}+m_{34}+j+j_2}(p_2+1-d/2)_{m_1}}\\
&\qquad\times\frac{(-h_3)_{m_1+m_4+m^{\prime}_{13}+m^{\prime}_{44}+m_{11}+m_{22}+m_{24}+m_{34}+j}(p_3-h_2+h_{4})_{m_{2}+m_{12}+m_{22}+m_{24}}}{(p_{3}-h_{2})_{2m_{2}+m_{11}+m_{12}+m_{22}+m_{24}+m^{\prime}_{33}+m_{34}}(p_3-h_2+1-d/2)_{m_2}}\\
&\qquad\times\frac{(-m_{12}^{\prime})_{j+j_2+m_{12}}(p_3+h_5-m_1+m_2+m_{12}-m^{\prime}_{13}-m^{\prime}_{23}-m^{\prime}_{44})_{m^{\prime}_{12}-m_{12}-j-j_2}}{j!j_2!m_{12}!}\\
&\qquad\times(-h_{5})_{m_3+m^{\prime}_{13}+m^{\prime}_{23}+m^{\prime}_{33}+m^{\prime}_{44}+j+j_2}(-1)^{j+j_2}F_7\prod_{1\leq a\leq4}\frac{(u_a^7)^{m_a}}{m_a!}\prod_{1\leq a\leq b\leq4}\frac{(1-v_{ab}^7)^{m^{\prime}_{ab}}}{m^{\prime}_{ab}!}.
}

We finally shift $j$ by $j\to j-j_2$ and evaluate the sums over $j_2$, $j$, and $m_{12}$,\footnote{Before proceeding with the sum over $j$, we use the first identity in \eqref{Eq3F2}.} to obtain
\eqna{
&G_{7P_1}=\sum \frac{(p_4-m_{1})_{m_4+m_{14}}(-h_{6})_{m_4+m^{\prime}_{44}+m_{14}+m_{24}+m_{34}}(p_{4}-h_3+h_{6})_{m_4+m^{\prime}_{13}+m_{11}+m_{22}}}{(p_{4}-h_3)_{2m_4+m^{\prime}_{13}+m^{\prime}_{44}+m_{11}+m_{14}+m_{22}+m_{24}+m_{34}}(p_4-h_3+1-d/2)_{m_4}}\\
&\qquad\times\frac{(p_2+h_3)_{m_{1}-m_4+m^{\prime}_{23}}(-h_4)_{m_{2}+m_{11}+m_{34}+m^{\prime}_{12}+m^{\prime}_{33}}(\bar{p}_{3}+h_2+h_{5})_{m_3+m^{\prime}_{12}}}{(p_2+h_3+m_1)_{-m_4}(\bar{p}_{3}+h_2)_{2m_3+m^{\prime}_{12}+m^{\prime}_{13}+m^{\prime}_{23}+m^{\prime}_{33}+m^{\prime}_{44}}(\bar{p}_3+h_2+1-d/2)_{m_3}}\\
&\qquad\times\frac{(p_3)_{-m_1+m_{2}+m_3+m^{\prime}_{12}+m^{\prime}_{33}}(p_2+h_2)_{m_1-m_2+m_3+m^{\prime}_{13}+m^{\prime}_{23}+m^{\prime}_{44}}(-h_2)_{m_1+m_2-m_3+m_{11}+m_{22}+m_{24}+m_{34}}}{(p_2)_{2m_1+m^{\prime}_{13}+m^{\prime}_{23}+m^{\prime}_{44}+m_{11}+m_{22}+m_{24}+m_{34}}(p_2+1-d/2)_{m_1}}\\
&\qquad\times\frac{(-h_3)_{m_1+m_4+m^{\prime}_{13}+m^{\prime}_{44}+m_{11}+m_{22}+m_{24}+m_{34}}(p_3-h_2+h_{4})_{m_{2}+m_{22}+m_{24}}(-h_{5})_{m_3+m^{\prime}_{13}+m^{\prime}_{23}+m^{\prime}_{33}+m^{\prime}_{44}}}{(p_{3}-h_{2})_{2m_{2}+m_{11}+m^{\prime}_{12}+m_{22}+m_{24}+m^{\prime}_{33}+m_{34}}(p_3-h_2+1-d/2)_{m_2}}\\
&\qquad\times F_7\prod_{1\leq a\leq4}\frac{(u_a^7)^{m_a}}{m_a!}\prod_{1\leq a\leq b\leq4}\frac{(1-v_{ab}^7)^{m^{\prime}_{ab}}}{m^{\prime}_{ab}!}.
}
Using the fact that $m_{11}=m^{\prime}_{22}$, $m_{22}=m^{\prime}_{11}$, $m_{14}=m^{\prime}_{14}$, $m_{24}=m^{\prime}_{34}$ and $m_{34}=m^{\prime}_{24}$, we find that $G_{7P}=G_7$ as expected from \eqref{EqSymPerm1}, proving invariance of the scalar seven-point conformal blocks in the extended snowflake channel under dendrite permutations of the first kind.

%%%%%%%%%%%%%%%%%%%%%%%%%%%%%%%%%%%%%%%%%%%%%%%%%%

\subsection{Dendrite Permutations of the Second Kind}

We finally focus on the proof of the invariance of the scalar seven-point correlation functions under dendrite permutations of the second kind \eqref{EqSymPerm2}, which we write as $G_7=G_{7P_2}$ for notational simplicity.

Taking into account the fact that $F_7$ is invariant under dendrite permutations of the second kind, we find that
\eqna{
&G_{7P_2}=\sum\frac{(p_4-m_{1})_{m_4+m_{14}}(p_4-h_3+h_{6})_{m_4+m_{14}+m_{24}+m_{34}+m_{44}}(-h_{6})_{m_4+m_{11}+m_{13}+m_{22}}}{(p_2+h_3+m_1)_{-m_4}(p_{4}-h_3)_{2m_4+m_{11}+m_{13}+m_{14}+m_{22}+m_{24}+m_{34}+m_{44}}(p_4-h_3+1-d/2)_{m_4}}\\
&\qquad\times\frac{(p_2+h_3)_{m_1-m_4+m_{23}}(p_3)_{-m_1+m_2+m_3+m_{12}+m_{33}}(p_3-h_2+h_4)_{m_{2}+m_{11}+m_{34}}(\bar{p}_{3}+h_2+h_{5})_{m_3+m_{12}}}{(\bar{p}_{3}+h_2)_{2m_3+m_{12}+m_{13}+m_{23}+m_{33}+m_{44}}(\bar{p}_3+h_2+1-d/2)_{m_3}}\\
&\qquad\times\frac{(p_2+h_2)_{m_1-m_2+m_3+m_{13}+m_{23}+m_{44}}(-h_2)_{m_1+m_2-m_3+m_{11}+m_{22}+m_{24}+m_{34}}}{(p_2)_{2m_{1}+m_{11}+m_{13}+m_{22}+m_{23}+m_{24}+m_{34}+m_{44}}(p_2+1-d/2)_{m_1}}\\
&\qquad\times\frac{(-h_3)_{m_1+m_4+m_{11}+m_{13}+m_{22}+m_{24}+m_{34}+m_{44}}(-h_{4})_{m_{2}+m_{12}+m_{22}+m_{24}+m_{33}}(-h_{5})_{m_3+m_{13}+m_{23}+m_{33}+m_{44}}}{(p_{3}-h_{2})_{2m_{2}+m_{11}+m_{12}+m_{22}+m_{24}+m_{33}+m_{34}}(p_3-h_2+1-d/2)_{m_2}}\\
&\qquad\times(-1)^{k_{14}+m^{\prime}_{14}}\frac{m^{\prime}_{14}!}{m_{14}!} \binom{m_{14}}{k_{14}}\binom{-p_4+m_1-m_4-k_{14}}{m^{\prime}_{14}}F_7\prod_{1\leq a\leq4}\frac{(u_a^7)^{m_a}}{m_a!}\prod_{1\leq a\leq b\leq4}\frac{(1-v_{ab}^7)^{m^{\prime}_{ab}}}{m^{\prime}_{ab}!},
}
with
\eqn{
\begin{gathered}
m_{11}=m^{\prime}_{34},\qquad m_{12}=m^{\prime}_{12},\qquad m_{22}=m^{\prime}_{24},\\
m_{13}=m^{\prime}_{44},\qquad m_{23}=m^{\prime}_{23},\qquad m_{33}=m^{\prime}_{33},\\
m_{24}=m^{\prime}_{22},\qquad m_{34}=m^{\prime}_{11},\qquad m_{44}=m^{\prime}_{13}.
\end{gathered}
}

Computing the sum over $k_{14}$ leads to
\eqna{
&G_{7P_2}=\sum\frac{(p_4-m_{1})_{m_4+m^{\prime}_{14}}(p_4-h_3+h_{6})_{m_4+m_{14}+m_{24}+m_{34}+m_{44}}(-h_{6})_{m_4+m_{11}+m_{13}+m_{22}}}{(p_2+h_3+m_1)_{-m_4}(p_{4}-h_3)_{2m_4+m_{11}+m_{13}+m_{14}+m_{22}+m_{24}+m_{34}+m_{44}}(p_4-h_3+1-d/2)_{m_4}}\\
&\qquad\times\frac{(p_2+h_3)_{m_1-m_4+m_{23}}(p_3)_{-m_1+m_2+m_3+m_{12}+m_{33}}(p_3-h_2+h_4)_{m_{2}+m_{11}+m_{34}}(\bar{p}_{3}+h_2+h_{5})_{m_3+m_{12}}}{(\bar{p}_{3}+h_2)_{2m_3+m_{12}+m_{13}+m_{23}+m_{33}+m_{44}}(\bar{p}_3+h_2+1-d/2)_{m_3}}\\
&\qquad\times\frac{(p_2+h_2)_{m_1-m_2+m_3+m_{13}+m_{23}+m_{44}}(-h_2)_{m_1+m_2-m_3+m_{11}+m_{22}+m_{24}+m_{34}}}{(p_2)_{2m_{1}+m_{11}+m_{13}+m_{22}+m_{23}+m_{24}+m_{34}+m_{44}}(p_2+1-d/2)_{m_1}}\\
&\qquad\times\frac{(-h_3)_{m_1+m_4+m_{11}+m_{13}+m_{22}+m_{24}+m_{34}+m_{44}}(-h_{4})_{m_{2}+m_{12}+m_{22}+m_{24}+m_{33}}(-h_{5})_{m_3+m_{13}+m_{23}+m_{33}+m_{44}}}{(p_{3}-h_{2})_{2m_{2}+m_{11}+m_{12}+m_{22}+m_{24}+m_{33}+m_{34}}(p_3-h_2+1-d/2)_{m_2}}\\
&\qquad\times\frac{(-m^{\prime}_{14})_{m_{14}}}{m_{14}!}F_7\prod_{1\leq a\leq4}\frac{(u_a^7)^{m_a}}{m_a!}\prod_{1\leq a\leq b\leq4}\frac{(1-v_{ab}^7)^{m^{\prime}_{ab}}}{m^{\prime}_{ab}!}.
}

We then evaluate the sum over $m_{14}$, giving
\eqna{
&G_{7P_2}=\sum\frac{(p_4-m_{1})_{m_4+m^{\prime}_{14}}(p_4-h_3+h_{6})_{m_4+m_{24}+m_{34}+m_{44}}(-h_{6})_{m_4+m_{11}+m_{13}+m^{\prime}_{14}+m_{22}}}{(p_2+h_3+m_1)_{-m_4}(p_{4}-h_3)_{2m_4+m_{11}+m_{13}+m^{\prime}_{14}+m_{22}+m_{24}+m_{34}+m_{44}}(p_4-h_3+1-d/2)_{m_4}}\\
&\qquad\times\frac{(p_2+h_3)_{m_1-m_4+m_{23}}(p_3)_{-m_1+m_2+m_3+m_{12}+m_{33}}(p_3-h_2+h_4)_{m_{2}+m_{11}+m_{34}}(\bar{p}_{3}+h_2+h_{5})_{m_3+m_{12}}}{(\bar{p}_{3}+h_2)_{2m_3+m_{12}+m_{13}+m_{23}+m_{33}+m_{44}}(\bar{p}_3+h_2+1-d/2)_{m_3}}\\
&\qquad\times\frac{(p_2+h_2)_{m_1-m_2+m_3+m_{13}+m_{23}+m_{44}}(-h_2)_{m_1+m_2-m_3+m_{11}+m_{22}+m_{24}+m_{34}}}{(p_2)_{2m_{1}+m_{11}+m_{13}+m_{22}+m_{23}+m_{24}+m_{34}+m_{44}}(p_2+1-d/2)_{m_1}}\\
&\qquad\times\frac{(-h_3)_{m_1+m_4+m_{11}+m_{13}+m_{22}+m_{24}+m_{34}+m_{44}}(-h_{4})_{m_{2}+m_{12}+m_{22}+m_{24}+m_{33}}(-h_{5})_{m_3+m_{13}+m_{23}+m_{33}+m_{44}}}{(p_{3}-h_{2})_{2m_{2}+m_{11}+m_{12}+m_{22}+m_{24}+m_{33}+m_{34}}(p_3-h_2+1-d/2)_{m_2}}\\
&\qquad\times F_7\prod_{1\leq a\leq4}\frac{(u_a^7)^{m_a}}{m_a!}\prod_{1\leq a\leq b\leq4}\frac{(1-v_{ab}^7)^{m^{\prime}_{ab}}}{m^{\prime}_{ab}!}\\
&\qquad=G_7,
}
which ends the proof of the invariance of the scalar seven-point correlation functions under dendrite permutations of the second kind \eqref{EqSymPerm2}.

%%%%%%%%%%%%%%%%%%%%%%%%%%%%%%%%%%%%%%%%%%%%%%%%%%
%%%%%%%%%%%%%%%%%%%%%%%%%%%%%%%%%%%%%%%%%%%%%%%%%%

\section{OPE Limit and Limit of Unit Operator}\label{SAppLim}

This appendix presents the remaining proofs for the OPE limit as well as the limit of unit operator.  As usual, all re-summations are performed with the help of \eqref{EqBinom}, \eqref{Eq2F1} and \eqref{Eq3F2}.

%%%%%%%%%%%%%%%%%%%%%%%%%%%%%%%%%%%%%%%%%%%%%%%%%%

\subsection{OPE Limit}

In the OPE limit $\eta_3\to\eta_4$, we expect from the topologies that
\eqn{\left.I_{7(\Delta_{k_1},\Delta_{k_2},\Delta_{k_3},\Delta_{k_4})}^{(\Delta_{i_2},\Delta_{i_3},\Delta_{i_4},\Delta_{i_5},\Delta_{i_6},\Delta_{i_7},\Delta_{i_1})}\right|_{\substack{\text{extended}\\\text{snowflake}}}\underset{\eta_3\to\eta_4}{\to}(\eta_{34})^{-p_5}\left.I_{6(\Delta_{k_3},\Delta_{k_1},\Delta_{k_4})}^{(\Delta_{i_5},\Delta_{i_6},\Delta_{i_4},\Delta_{i_2},\Delta_{i_7},\Delta_{i_1})}\right|_{\text{comb}}.}[EqOPELim34]
Defining all quantities in the vectors $\boldsymbol{h}$ and $\boldsymbol{p}$ on the RHS of \eqref{EqOPELim34} with primes, this leads to
\eqn{
\begin{gathered}
L_7\prod_{1\leq a\leq4}(u_a^7)^{\frac{\Delta_{k_a}}{2}}\to(v_{23}^6)^{-\bar{p}^{\prime}_4-\bar{h}^{\prime}_5}(v_{12}^6)^{h^{\prime}_5}L_6\prod_{1\leq a\leq3}(u_a^6)^{\frac{\Delta^{\prime}_{k_a}}{2}},\\
u_1^7\to\frac{u_2^6}{v_{23}^6},\qquad u_2^7\to0,\qquad u_3^7\to u_1^6,\qquad u_4^7\to\frac{u_3^6}{v_{12}^6},\\
v_{11}^7\to\frac{v_{13}^6}{v_{23}^6},\qquad v_{12}^7\to1,\qquad v_{13}^7\to\frac{v_{33}^6}{v_{23}^6},\qquad v_{14}^7\to\frac{v_{23}^6}{v_{12}^6},\\
v_{22}^7\to\frac{v_{13}^6}{v_{23}^6},\qquad v_{23}^7\to v_{11}^6,\qquad v_{24}^7\to\frac{1}{v_{12}^6},\qquad v_{33}^7\to1,\\
v_{34}^7\to\frac{1}{v_{12}^6},\qquad v_{44}^7\to\frac{v_{22}^6}{v_{12}^6},
\end{gathered}
}
which implies the identity
\eqna{
&G_{7|\substack{\text{extended}\\\text{snowflake}}}^{(d,h_2,h_3,h_4,h_5,h_6;p_2,p_3,p_4,p_5,p_6,p_7)}\left(\frac{u_2^6}{v_{23}^6},0,u_1^6,\frac{u_3^6}{v_{12}^6};\frac{v_{13}^6}{v_{23}^6},1,\frac{v_{33}^6}{v_{23}^6},\frac{v_{23}^6}{v_{12}^6},\frac{v_{13}^6}{v_{23}^6},v_{11}^6,\frac{1}{v_{12}^6},1,\frac{1}{v_{12}^6},\frac{v_{22}^6}{v_{12}^6}\right)\\
&\qquad=(v_{23}^6)^{\bar{p}^{\prime}_4+\bar{h}^{\prime}_5}(v_{12}^6)^{-h^{\prime}_5}G_{6|\text{comb}}^{(d,h'_2,h'_3,h'_4,h'_5;p'_2,p'_3,p'_4,p'_5,p'_6)}(u_1^6,u_2^6,u_3^6;v_{11}^6,v_{12}^6,v_{13}^6,v_{22}^6,v_{23}^6,v_{33}^6).
}

Here the scalar six-point conformal blocks in the comb channel is given by
\eqna{
&G_{6|\text{comb}}^{(d,h^{\prime}_2,h^{\prime}_3,h^{\prime}_4,h^{\prime}_5;p^{\prime}_2,p^{\prime}_3,p^{\prime}_4,p^{\prime}_5,p^{\prime}_6)}\\
&\qquad=\sum\frac{(-h^{\prime}_4+m_2-m_3)_{m_{11}}(p^{\prime}_5-m_2)_{m_3}(\bar{p}^{\prime}_5+\bar{h}^{\prime}_5)_{m_3+m_{13}+m_{23}+m_{33}}(-h_5^{\prime})_{m_3+m_{12}+m_{22}}}{(\bar{p}^{\prime}_5+\bar{h}^{\prime}_4)_{2m_3+m_{12}+m_{22}+m_{13}+m_{23}+m_{33}}(\bar{p}^{\prime}_5+\bar{h}^{\prime}_4+1-d/2)_{m_3}}\\
&\qquad\phantom{=}\times\frac{(p_4^{\prime}-m_1)_{m_2+m_{13}}(p_3^{\prime})_{m_1+m_{11}+m_{22}+m_{33}}(-h^{\prime}_3)_{m_1}(p^{\prime}_2+h^{\prime}_2)_{m_1+m_{23}+m_{12}}(-h^{\prime}_4)_{m_2}}{(\bar{p}^{\prime}_3+h^{\prime}_2)_{2m_1+m_{11}+m_{23}+m_{12}+m_{22}+m_{33}}(\bar{p}^{\prime}_3+h^{\prime}_2+1-d/2)_{m_1}(\bar{p}^{\prime}_4+\bar{h}^{\prime}_3+1-d/2)_{m_2}}\\
&\qquad\phantom{=}\times\frac{(\bar{p}^{\prime}_3+\bar{h}^{\prime}_3)_{m_1+m_2+m_{11}+m_{23}+m_{12}+m_{22}+m_{33}}(\bar{p}^{\prime}_4+\bar{h}^{\prime}_4)_{m_2+m_3+m_{12}+m_{22}+m_{13}+m_{23}+m_{33}}}{(\bar{p}^{\prime}_4+\bar{h}^{\prime}_3)_{2m_2+m_{11}+m_{12}+m_{22}+m_{13}+m_{23}+m_{33}}}\\
&\qquad\phantom{=}\times F_6\prod_{1\leq a\leq3}\frac{(u_a^6)^{m_a}}{(m_a)!}\prod_{1\leq a\leq b\leq3}\frac{(1-v_{ab}^6)^{m_{ab}}}{(m_{ab})!},
}
with
\eqn{F_6={}_3F_2\left[\begin{array}{c}-m_1,-m_2,-\bar{p}^{\prime}_3-\bar{h}^{\prime}_2+d/2-m_1;\\p^{\prime}_4-m_1,h^{\prime}_3+1-m_1\end{array};1\right]{}_3F_2\left[\begin{array}{c}-m_2,-m_3,-\bar{p}^{\prime}_4-\bar{h}^{\prime}_3+d/2-m_2;\\p^{\prime}_5-m_2,h^{\prime}_4+1-m_2\end{array};1\right],}
and
\eqn{
\begin{gathered}
h_2\to-p^{\prime}_4,\qquad h_3\to-\bar{p}^{\prime}_4-\bar{h}^{\prime}_4,\qquad h_5\to -p^{\prime}_3,\qquad h_6\to h^{\prime}_5,\\
p_2\to\bar{p}^{\prime}_4+\bar{h}^{\prime}_3,\qquad p_3\to-h^{\prime}_3,\qquad p_4\to p^{\prime}_5,\qquad \\
p_2+h_3\to -h^{\prime}_4,\qquad \bar{p}_3+h_2+h_5\to p^{\prime}_2+h^{\prime}_2,\qquad p_4-h_3+h_6\to \bar{p}^{\prime}_5+\bar{h}^{\prime}_5,\\
p_2+h_2\to \bar{p}^{\prime}_3+\bar{h}_3^{\prime},\qquad \bar{p}_3+\bar{h}_2\to \bar{p}^{\prime}_3+\bar{h}_2^{\prime},\qquad p_4-h_3\to \bar{p}^{\prime}_5+\bar{h}_4^{\prime}.
\end{gathered}
}

To prove \eqref{EqOPELim34}, which we rewrite as $G_{7|\eta_3\to\eta_4}=G_6$ for simplicity, we first note that
\eqna{
F_7&=\frac{(-\bar{p}_3-h_2+d/2-m^{\prime}_1)_{m^{\prime}_1}}{(p_3)_{-m^{\prime}_2}}\sum\frac{(-m^{\prime}_1)_{t_2}(\bar{p}_3-d/2)_{t_2}(p_3)_{t_2}}{t_2!(p_3-m^{\prime}_2)_{t_2}(\bar{p}_3+h_2+1-d/2)_{t_2}}\\
&\qquad\times\frac{(-m^{\prime}_2)_{t_3}(-m^{\prime}_3)_{t_3}(-p_2+d/2-m^{\prime}_2)_{t_{3}}}{t_3!(1-p_2-h_3-m^{\prime}_2)_{t_3}(p_4-m^{\prime}_2)_{t_3}}\\
&=\frac{(-h_2-m^{\prime}_1)_{m^{\prime}_1}}{(p_3)_{-m^{\prime}_2}}\sum\frac{(-m^{\prime}_1)_{t_2}(-m^{\prime}_2)_{t_2}(\bar{p}_3-d/2)_{t_2}}{t_2!(p_3-m^{\prime}_2)_{t_2}(-h_2-m^{\prime}_1)_{t_2}}\\
&\qquad\times\frac{(-m^{\prime}_2)_{t_3}(-m^{\prime}_3)_{t_3}(-p_2+d/2-m^{\prime}_2)_{t_{3}}}{t_3!(1-p_2-h_3-m^{\prime}_2)_{t_3}(p_4-m^{\prime}_2)_{t_3}}\\
&=\frac{(-h_2-m^{\prime}_1)_{m^{\prime}_1}(p_3)_{m^{\prime}_1}}{(p_3)_{m^{\prime}_1-m^{\prime}_2}}\sum\frac{(-m^{\prime}_1)_{t_2}(-m^{\prime}_2)_{t_2}(-\bar{p}_3-h_2+d/2-m^{\prime}_1)_{t_2}}{t_2!(1-p_3-m^{\prime}_1)_{t_2}(-h_2-m^{\prime}_1)_{t_2}}\\
&\qquad\times\frac{(-m^{\prime}_2)_{t_3}(-m^{\prime}_3)_{t_3}(-p_2+d/2-m^{\prime}_2)_{t_{3}}}{t_3!(1-p_2-h_3-m^{\prime}_2)_{t_3}(p_4-m^{\prime}_2)_{t_3}}\\
&=\frac{(-h_2-m^{\prime}_1)_{m^{\prime}_1}(p_3)_{m^{\prime}_1}}{(p_3)_{m^{\prime}_1-m^{\prime}_2}}F_6.
}
Thus, multiplying $G_7$ by $(v^6_{23})^{h_3-h_6}(v^6_{12})^{h_6}$ and taking the OPE limit $\eta_3\to\eta_4$, we need to recover $G_6$ from
\eqna{
&G_{7|\eta_3\to\eta_4}=\sum\frac{(p_4-m^{\prime}_{2})_{m^{\prime}_3+m_{14}}(-h_{6})_{m^{\prime}_3+m_{14}+m_{24}+m_{34}+m_{44}}(p_{4}-h_3+h_{6})_{m^{\prime}_3+m_{11}+m_{13}+m_{22}}}{(p_{4}-h_3)_{2m^{\prime}_3+m_{11}+m_{13}+m_{14}+m_{22}+m_{24}+m_{34}+m_{44}}(p_4-h_3+1-d/2)_{m^{\prime}_3}}\\
&\qquad\times\frac{(p_2+h_3)_{m^{\prime}_{2}-m^{\prime}_3+m^{\prime}_{11}}(p_3-h_2+h_4)_{m_{11}+m_{34}}(\bar{p}_{3}+h_2+h_{5})_{m^{\prime}_1}}{(p_2+h_3+m^{\prime}_2)_{-m^{\prime}_3}(\bar{p}_{3}+h_2)_{2m^{\prime}_1+m_{13}+m_{44}+m^{\prime}_{11}}(\bar{p}_3+h_2+1-d/2)_{m^{\prime}_1}}\\
&\qquad\times\frac{(p_3)_{m^{\prime}_1}(p_2+h_2)_{m^{\prime}_2+m^{\prime}_1+m_{13}+m_{44}+m^{\prime}_{11}}(-h_2-m^{\prime}_1)_{m_1^{\prime}}(-h_2)_{m^{\prime}_2-m^{\prime}_1+m_{11}+m_{22}+m_{24}+m_{34}}}{(p_2)_{2m^{\prime}_2+m_{11}+m_{13}+m_{22}+m^{\prime}_{11}+m_{24}+m_{34}+m_{44}}(p_2+1-d/2)_{m^{\prime}_2}}\\
&\qquad\times\frac{(-h_3)_{m^{\prime}_2+m^{\prime}_3+m_{11}+m_{13}+m_{22}+m_{24}+m_{34}+m_{44}}(-h_{4})_{m_{22}+m_{24}}(-h_{5})_{m^{\prime}_1+m_{13}+m^{\prime}_{11}+m_{44}}}{(p_{3}-h_{2})_{m_{11}+m_{22}+m_{24}+m_{34}}}\\
&\qquad\times\binom{m_{11}}{k_{11}}\binom{m_{13}}{k_{13}}\binom{m_{14}}{k_{14}}\binom{m_{22}}{k_{22}}\binom{m_{24}}{k_{24}}\binom{m_{34}}{k_{34}}\binom{m_{44}}{k_{44}}\binom{h_6-m^{\prime}_3-k_{14}-k_{24}-k_{34}-k_{44}}{m^{\prime}_{12}}\\
&\qquad\times\binom{k_{11}}{l_{13}}\binom{k_{22}}{m^{\prime}_{13}-l_{13}}\binom{k_{44}}{m^{\prime}_{22}}\binom{h_3-h_6-m^{\prime}_2-k_{11}-k_{13}-k_{22}+k_{14}}{m^{\prime}_{23}}\binom{k_{13}}{m^{\prime}_{33}}\\
&\qquad\times(-1)^{k_{11}+k_{13}+k_{14}+k_{22}+k_{24}+k_{34}+k_{44}+m^{\prime}_{12}+m^{\prime}_{13}+m^{\prime}_{22}+m^{\prime}_{23}+m^{\prime}_{33}}\\
&\qquad\times F_6\prod_{1\leq a\leq3}\frac{(u_a^6)^{m^{\prime}_a}}{m^{\prime}_a!}\frac{m_{12}!m_{33}!\prod_{1\leq a\leq b\leq3}(1-v_{ab}^6)^{m^{\prime}_{ab}}}{\prod_{1\leq a\leq b\leq4}m_{ab}!}.
}

To proceed, we first evaluate the sums over $k_{13}$, $k_{22}$, $k_{24}$, $k_{34}$, and finally $k_{44}$,\footnote{We first change variables such that $k_{13}\to k_{13}+m^{\prime}_{33}$ and $k_{44}\to k_{44}+m^{\prime}_{22}$.} which gives
\eqna{
&G_{7|\eta_3\to\eta_4}=\sum\frac{(p_4-m^{\prime}_{2})_{m^{\prime}_3+m_{14}}(-h_{6})_{m^{\prime}_3+m_{14}+m_{24}+m_{34}+m_{44}}(p_{4}-h_3+h_{6})_{m^{\prime}_3+m_{11}+m_{13}+m_{22}}}{(p_{4}-h_3)_{2m^{\prime}_3+m_{11}+m_{13}+m_{14}+m_{22}+m_{24}+m_{34}+m_{44}}(p_4-h_3+1-d/2)_{m^{\prime}_3}}\\
&\qquad\times\frac{(p_2+h_3)_{m^{\prime}_{2}-m^{\prime}_3+m^{\prime}_{11}}(p_3-h_2+h_4)_{m_{11}+m_{34}}(\bar{p}_{3}+h_2+h_{5})_{m^{\prime}_1}}{(p_2+h_3+m^{\prime}_2)_{-m^{\prime}_3}(\bar{p}_{3}+h_2)_{2m^{\prime}_1+m_{13}+m_{44}+m^{\prime}_{11}}(\bar{p}_3+h_2+1-d/2)_{m^{\prime}_1}}\\
&\qquad\times\frac{(p_3)_{m^{\prime}_1}(p_2+h_2)_{m^{\prime}_2+m^{\prime}_1+m_{13}+m_{44}+m^{\prime}_{11}}(-h_2-m^{\prime}_1)_{m_1^{\prime}}(-h_2)_{m^{\prime}_2-m^{\prime}_1+m_{11}+m_{22}+m_{24}+m_{34}}}{(p_2)_{2m^{\prime}_2+m_{11}+m_{13}+m_{22}+m^{\prime}_{11}+m_{24}+m_{34}+m_{44}}(p_2+1-d/2)_{m^{\prime}_2}}\\
&\qquad\times\frac{(-h_3)_{m^{\prime}_2+m^{\prime}_3+m_{11}+m_{13}+m_{22}+m_{24}+m_{34}+m_{44}}(-h_{4})_{m_{22}+m_{24}}(-h_{5})_{m^{\prime}_1+m_{13}+m^{\prime}_{11}+m_{44}}}{(p_{3}-h_{2})_{m_{11}+m_{22}+m_{24}+m_{34}}}\\
&\qquad\times\binom{m_{14}}{k_{14}}(-1)^{k_{14}}\frac{m^{\prime}_{13}!(-m^{\prime}_{23})_{m_{11}+m_{13}+m_{22}-m^{\prime}_{13}-m^{\prime}_{33}}(-m^{\prime}_{12})_{m_{24}+m_{34}+m_{44}-m^{\prime}_{22}}}{(m_{11}-l_{13})!(m_{22}-m^{\prime}_{13}+l_{13})!}\\
&\qquad\times\frac{(-h_3+h_6+m^{\prime}_2+m_{11}+m_{13}+m_{22}-k_{14})_{m^{\prime}_{13}+m^{\prime}_{23}+m^{\prime}_{33}-m_{11}-m_{13}-m_{22}}}{(m_{13}-m^{\prime}_{33})!}\\
&\qquad\times\frac{(-h_6+m^{\prime}_3+m_{24}+m_{34}+m_{44}+k_{14})_{m^{\prime}_{12}+m^{\prime}_{22}-m_{24}-m_{34}-m_{44}}}{l_{13}!(m^{\prime}_{13}-l_{13})!(m_{44}-m^{\prime}_{22})!m_{14}!m_{24}!m_{34}!}\\
&\qquad\times F_6\prod_{1\leq a\leq3}\frac{(u_a^6)^{m^{\prime}_a}}{m^{\prime}_a!}\prod_{1\leq a\leq b\leq3}\frac{(1-v_{ab}^6)^{m^{\prime}_{ab}}}{m^{\prime}_{ab}!}.
}

To eliminate the sum over $k_{14}$, we now use the following identity
\eqna{
&(-h_3+h_6+m^{\prime}_2+m_{11}+m_{13}+m_{22}-k_{14})_{m^{\prime}_{13}+m^{\prime}_{23}+m^{\prime}_{33}-m_{11}-m_{13}-m_{22}}\\
&\qquad=\sum_j\binom{m^{\prime}_{13}+m^{\prime}_{23}+m^{\prime}_{33}-m_{11}-m_{13}-m_{22}}{j}\\
&\qquad\phantom{=}\times(-h_3+h_6+m^{\prime}_2+m_{11}+m_{13}+m_{22})_{m^{\prime}_{13}+m^{\prime}_{23}+m^{\prime}_{33}-m_{11}-m_{13}-m_{22}-j}(-k_{14})_j,
}
and then change the variable by $k_{14}\to k_{14}+j$.  The sum over $k_{14}$ can be performed, leading to
\eqna{
&G_{7|\eta_3\to\eta_4}=\sum\frac{(p_4-m^{\prime}_{2})_{m^{\prime}_3+m_{14}}(-h_{6})_{m^{\prime}_3+m^{\prime}_{12}+m^{\prime}_{22}+j}(p_{4}-h_3+h_{6})_{m^{\prime}_3+m_{11}+m_{13}+m_{22}}}{(p_{4}-h_3)_{2m^{\prime}_3+m_{11}+m_{13}+m_{14}+m_{22}+m_{24}+m_{34}+m_{44}}(p_4-h_3+1-d/2)_{m^{\prime}_3}}\\
&\qquad\times\frac{(p_2+h_3)_{m^{\prime}_{2}-m^{\prime}_3+m^{\prime}_{11}}(p_3-h_2+h_4)_{m_{11}+m_{34}}(\bar{p}_{3}+h_2+h_{5})_{m^{\prime}_1}}{(p_2+h_3+m^{\prime}_2)_{-m^{\prime}_3}(\bar{p}_{3}+h_2)_{2m^{\prime}_1+m_{13}+m_{44}+m^{\prime}_{11}}(\bar{p}_3+h_2+1-d/2)_{m^{\prime}_1}}\\
&\qquad\times\frac{(p_3)_{m^{\prime}_1}(p_2+h_2)_{m^{\prime}_2+m^{\prime}_1+m_{13}+m_{44}+m^{\prime}_{11}}(-h_2-m^{\prime}_1)_{m_1^{\prime}}(-h_2)_{m^{\prime}_2-m^{\prime}_1+m_{11}+m_{22}+m_{24}+m_{34}}}{(p_2)_{2m^{\prime}_2+m_{11}+m_{13}+m_{22}+m^{\prime}_{11}+m_{24}+m_{34}+m_{44}}(p_2+1-d/2)_{m^{\prime}_2}}\\
&\qquad\times\frac{(-h_3)_{m^{\prime}_2+m^{\prime}_3+m_{11}+m_{13}+m_{22}+m_{24}+m_{34}+m_{44}}(-h_{4})_{m_{22}+m_{24}}(-h_{5})_{m^{\prime}_1+m_{13}+m^{\prime}_{11}+m_{44}}}{(p_{3}-h_{2})_{m_{11}+m_{22}+m_{24}+m_{34}}}\\
&\qquad\times\frac{m^{\prime}_{13}!(-m^{\prime}_{23})_{j+m_{11}+m_{13}+m_{22}-m^{\prime}_{13}-m^{\prime}_{33}}(-m^{\prime}_{12})_{m_{14}+m_{24}+m_{34}+m_{44}-m^{\prime}_{22}-j}}{(m_{11}-l_{13})!(m_{22}-m^{\prime}_{13}+l_{13})!}\\
&\qquad\times\frac{(-h_3+h_6+m^{\prime}_2+m_{11}+m_{13}+m_{22})_{m^{\prime}_{13}+m^{\prime}_{23}+m^{\prime}_{33}-m_{11}-m_{13}-m_{22}-j}}{l_{13}!(m^{\prime}_{13}-l_{13})!(m_{13}-m^{\prime}_{33})!}\\
&\qquad\times\frac{(-1)^j}{(m_{44}-m^{\prime}_{22})!(m_{14}-j)!j!m_{24}!m_{34}!}F_6\prod_{1\leq a\leq3}\frac{(u_a^6)^{m^{\prime}_a}}{m^{\prime}_a!}\prod_{1\leq a\leq b\leq3}\frac{(1-v_{ab}^6)^{m^{\prime}_{ab}}}{m^{\prime}_{ab}!}.
}

We now define $m_{34}=m-m_{24}$ to evaluate the sums over $m_{24}$, $m_{14}$, $m$, and finally $m_{44}$,\footnote{We first shift $m_{14}\to m_{14}+j$ and $m_{44}\to m_{44}+m^{\prime}_{22}$.} leading to
\eqna{
&G_{7|\eta_3\to\eta_4}=\sum\frac{(p_4-m^{\prime}_{2})_{m^{\prime}_3+j}(-h_{6})_{m^{\prime}_3+m^{\prime}_{12}+m^{\prime}_{22}+j}(p_{4}-h_3+h_{6})_{m^{\prime}_3+m_{11}+m_{13}+m_{22}}}{(p_{4}-h_3)_{2m^{\prime}_3+m^{\prime}_{12}+m^{\prime}_{22}+m_{11}+m_{13}+m_{22}+j}(p_4-h_3+1-d/2)_{m^{\prime}_3}}\\
&\qquad\times\frac{(p_2+h_3)_{m^{\prime}_{2}-m^{\prime}_3+m^{\prime}_{11}}(p_3-h_2+h_4)_{m_{11}}(\bar{p}_{3}+h_2+h_{5})_{m^{\prime}_1+m^{\prime}_{12}}}{(p_2+h_3+m^{\prime}_2)_{-m^{\prime}_3}(\bar{p}_{3}+h_2)_{2m^{\prime}_1+m^{\prime}_{11}+m^{\prime}_{12}+m^{\prime}_{22}+m_{13}}(\bar{p}_3+h_2+1-d/2)_{m^{\prime}_1}}\\
&\qquad\times\frac{(p_3)_{m^{\prime}_1}(p_2+h_2)_{m^{\prime}_1+m^{\prime}_2+m^{\prime}_{11}+m^{\prime}_{12}+m^{\prime}_{22}+m_{13}}(-h_2-m^{\prime}_1)_{m_1^{\prime}}(-h_2)_{m^{\prime}_2-m^{\prime}_1+m_{11}+m_{22}}}{(p_2)_{2m^{\prime}_2+m^{\prime}_{11}+m^{\prime}_{12}+m^{\prime}_{22}+m_{11}+m_{13}+m_{22}}(p_2+1-d/2)_{m^{\prime}_2}}\\
&\qquad\times\frac{(-h_3)_{m^{\prime}_2+m^{\prime}_3+m^{\prime}_{12}+m^{\prime}_{22}+m_{11}+m_{13}+m_{22}}(-h_{4})_{m_{22}}(-h_{5})_{m^{\prime}_1+m^{\prime}_{11}+m^{\prime}_{22}+m_{13}}}{(p_{3}-h_{2})_{m_{11}+m_{22}}}\\
&\qquad\times\frac{(-h_3+h_6+m^{\prime}_2+m_{11}+m_{13}+m_{22})_{m^{\prime}_{13}+m^{\prime}_{23}+m^{\prime}_{33}-m_{11}-m_{13}-m_{22}-j}}{(m^{\prime}_{13}-l_{13})!(m_{13}-m^{\prime}_{33})!}\\
&\qquad\times\frac{m^{\prime}_{13}!(-m^{\prime}_{23})_{j+m_{11}+m_{13}+m_{22}-m^{\prime}_{13}-m^{\prime}_{33}}}{l_{13}!(m_{11}-l_{13})!(m_{22}-m^{\prime}_{13}+l_{13})!}\frac{(-1)^j}{j!}F_6\prod_{1\leq a\leq3}\frac{(u_a^6)^{m^{\prime}_a}}{m^{\prime}_a!}\prod_{1\leq a\leq b\leq3}\frac{(1-v_{ab}^6)^{m^{\prime}_{ab}}}{m^{\prime}_{ab}!}.
}

With the help of the first identity involving  ${}_3F_2$ shown in \eqref{Eq3F2}, we evaluate the sum over $j$ and get
\eqna{
&G_{7|\eta_3\to\eta_4}=\sum\frac{(p_4-m^{\prime}_{2})_{m^{\prime}_3}(-h_{6})_{m^{\prime}_3+m^{\prime}_{12}+m^{\prime}_{22}}(p_{4}-h_3+h_{6})_{m^{\prime}_3+m^{\prime}_{13}+m^{\prime}_{23}+m^{\prime}_{33}}}{(p_{4}-h_3)_{2m^{\prime}_3+m^{\prime}_{12}+m^{\prime}_{13}+m^{\prime}_{22}+m^{\prime}_{23}+m^{\prime}_{33}}(p_4-h_3+1-d/2)_{m^{\prime}_3}}\\
&\qquad\times\frac{(p_2+h_3)_{m^{\prime}_{2}-m^{\prime}_3+m^{\prime}_{11}}(p_3-h_2+h_4)_{m_{11}}(\bar{p}_{3}+h_2+h_{5})_{m^{\prime}_1+m^{\prime}_{12}}}{(p_2+h_3+m^{\prime}_2)_{-m^{\prime}_3}(\bar{p}_{3}+h_2)_{2m^{\prime}_1+m^{\prime}_{11}+m^{\prime}_{12}+m^{\prime}_{22}+m_{13}}(\bar{p}_3+h_2+1-d/2)_{m^{\prime}_1}}\\
&\qquad\times\frac{(p_3)_{m^{\prime}_1}(p_2+h_2)_{m^{\prime}_1+m^{\prime}_2+m^{\prime}_{11}+m^{\prime}_{12}+m^{\prime}_{22}+m_{13}}(-h_2-m^{\prime}_1)_{m_1^{\prime}}(-h_2)_{m^{\prime}_2-m^{\prime}_1+m_{11}+m_{22}}}{(p_2)_{2m^{\prime}_2+m^{\prime}_{11}+m^{\prime}_{12}+m^{\prime}_{22}+m_{11}+m_{13}+m_{22}}(p_2+1-d/2)_{m^{\prime}_2}}\\
&\qquad\times\frac{(-h_3)_{m^{\prime}_2+m^{\prime}_3+m^{\prime}_{12}+m^{\prime}_{13}+m^{\prime}_{22}+m^{\prime}_{23}+m^{\prime}_{33}}(-h_{4})_{m_{22}}(-h_{5})_{m^{\prime}_1+m^{\prime}_{11}+m^{\prime}_{22}+m_{13}}}{(p_{3}-h_{2})_{m_{11}+m_{22}}}\\
&\qquad\times\frac{m^{\prime}_{13}!(-m^{\prime}_{23})_{m_{11}+m_{13}+m_{22}-m^{\prime}_{13}-m^{\prime}_{33}}}{l_{13}!(m^{\prime}_{13}-l_{13})!(m_{13}-m^{\prime}_{33})!(m_{11}-l_{13})!(m_{22}-m^{\prime}_{13}+l_{13})!}\\
&\qquad\times F_6\prod_{1\leq a\leq3}\frac{(u_a^6)^{m^{\prime}_a}}{m^{\prime}_a!}\prod_{1\leq a\leq b\leq3}\frac{(1-v_{ab}^6)^{m^{\prime}_{ab}}}{m^{\prime}_{ab}!}.
}

We finally change the variables by
\eqn{m_{11}\to m_{11}+l_{13},\qquad m_{22}\to m_{22}+m_{13}^{\prime}-l_{13},}
and then define $m_{22}=n-m_{11}$.  We can thus evaluate the sums over $m_{11}$, $n$, $l_{13}$, and $m_{13}$ after shifting $m_{13}\to m_{13}+m^{\prime}_{33}$, resulting in
\eqna{
&G_{7|\eta_3\to\eta_4}=\sum\frac{(p_4-m^{\prime}_{2})_{m^{\prime}_3}(-h_{6})_{m^{\prime}_3+m^{\prime}_{12}+m^{\prime}_{22}}(p_{4}-h_3+h_{6})_{m^{\prime}_3+m^{\prime}_{13}+m^{\prime}_{23}+m^{\prime}_{33}}}{(p_{4}-h_3)_{2m^{\prime}_3+m^{\prime}_{12}+m^{\prime}_{13}+m^{\prime}_{22}+m^{\prime}_{23}+m^{\prime}_{33}}(p_4-h_3+1-d/2)_{m^{\prime}_3}}\\
&\qquad\times\frac{(p_2+h_3)_{m^{\prime}_{2}-m^{\prime}_3+m^{\prime}_{11}}(\bar{p}_{3}+h_2+h_{5})_{m^{\prime}_1+m^{\prime}_{12}+m^{\prime}_{23}}}{(p_2+h_3+m^{\prime}_2)_{-m^{\prime}_3}(\bar{p}_{3}+h_2)_{2m^{\prime}_1+m^{\prime}_{11}+m^{\prime}_{12}+m^{\prime}_{22}+m^{\prime}_{23}+m^{\prime}_{33}}(\bar{p}_3+h_2+1-d/2)_{m^{\prime}_1}}\\
&\qquad\times\frac{(p_3)_{m^{\prime}_1}(p_2+h_2)_{m^{\prime}_1+m^{\prime}_2+m^{\prime}_{11}+m^{\prime}_{12}+m^{\prime}_{22}+m^{\prime}_{23}+m^{\prime}_{33}}(-h_2-m^{\prime}_1)_{m^{\prime}_2+m^{\prime}_{13}}}{(p_2)_{2m^{\prime}_2+m^{\prime}_{11}+m^{\prime}_{12}+m^{\prime}_{13}+m^{\prime}_{22}+m^{\prime}_{23}+m^{\prime}_{33}}}\\
&\qquad\times\frac{(-h_3)_{m^{\prime}_2+m^{\prime}_3+m^{\prime}_{12}+m^{\prime}_{13}+m^{\prime}_{22}+m^{\prime}_{23}+m^{\prime}_{33}}(-h_{5})_{m^{\prime}_1+m^{\prime}_{11}+m^{\prime}_{22}+m^{\prime}_{33}}}{(p_2+1-d/2)_{m^{\prime}_2}}\\
&\qquad\times F_6\prod_{1\leq a\leq3}\frac{(u_a^6)^{m^{\prime}_a}}{m^{\prime}_a!}\prod_{1\leq a\leq b\leq3}\frac{(1-v_{ab}^6)^{m^{\prime}_{ab}}}{m^{\prime}_{ab}!}.
}
Re-expressing the unprimed variables in terms of the primed variables leads to $G_6$ which completes our proof of the OPE limit $\eta_3\to\eta_4$ \eqref{EqOPELim34}.

%%%%%%%%%%%%%%%%%%%%%%%%%%%%%%%%%%%%%%%%%%%%%%%%%%

\subsection{Limit of Unit Operator}

We focus first on the limit of unit operator given by $\mathcal{O}_{i_2}(\eta_2)\to\1$, for which we have $\Delta_{i_2}=0$ as well as $\Delta_{k_4}=\Delta_{k_1}$.  It implies
\eqn{\left.I_{7(\Delta_{k_1},\Delta_{k_2},\Delta_{k_3},\Delta_{k_4})}^{(\Delta_{i_2},\Delta_{i_3},\Delta_{i_4},\Delta_{i_5},\Delta_{i_6},\Delta_{i_7},\Delta_{i_1})}\right|_{\substack{\text{extended}\\\text{snowflake}}}\underset{\mathcal{O}_{i_2}(\eta_2)\to\1}{\to}\left.I_{6(\Delta_{k_1},\Delta_{k_2},\Delta_{k_3})}^{(\Delta_{i_7},\Delta_{i_1},\Delta_{i_3},\Delta_{i_4},\Delta_{i_5},\Delta_{i_6})}\right|_{\text{snowflake}},}[EqLimUnit2Id]
with $p_4=p_2+h_3=0$.  Moreover, in this limit the legs and conformal cross-ratios of the six- and seven-point correlation functions are related by
\eqn{
\begin{gathered}
L_7\prod_{1\leq a\leq4}(u_a^7)^{\frac{\Delta_{\Delta_{k_a}}}{2}}=(v_{24}^7)^{-h_4}(v_{34}^7)^{h_4-h_2}L_6\prod_{1\leq a\leq3}(u_a^6)^{\frac{\Delta_{k_a}}{2}},\\
u_1^6=\frac{u_1^7u_4^7}{v_{34}^7},\qquad u_2^6=\frac{u_2^7}{v_{24}^7},\qquad u_3^6=u_3^7v_{34}^7,\\
v_{11}^6=\frac{v_{11}^7}{v_{34}^7},\qquad v_{12}^6=\frac{v_{12}^7v_{34}^7}{v_{24}^7},\qquad v_{13}^6=v_{13}^7,\\
v_{22}^6=\frac{v_{22}^7}{v_{24}^7},\qquad v_{23}^6=v_{44}^7,\qquad v_{33}^6=\frac{v_{33}^7v_{34}^7}{v_{24}^7}.
\end{gathered}
}
These observations lead to the identity $G_{7|\mathcal{O}_{i_2}\to\1}=(v_{24}^7)^{h_4}(v_{34}^7)^{h_2-h_4}G_6$ that we now prove.

From the vanishing of $p_4$ and $p_2+h_3$, we have
\eqn{m_1=m_4=r,\qquad m_{14}=m_{23}=0.}
As a result, we find that
\eqna{
&G_{7|\mathcal{O}_{i_2}\to\1}=\sum\frac{(-h_6)_{m_1+m_{24}+m_{34}+m^{\prime}_{23}}(p_2+h_6)_{m_1+m_{11}+m_{13}+m_{22}}(p_2+h_2)_{m_1-m_2+m_3+m_{13}+m^{\prime}_{23}}}{(p_2)_{2m_1+m_{11}+m_{13}+m_{22}+m_{24}+m_{34}+m^{\prime}_{23}}(p_2+1-d/2)_{m_1}}\\
&\qquad\times\frac{(p_3)_{-m_1+m_2+m_3+m_{12}+m_{33}}(p_3-h_2+h_4)_{m_{2}+m_{11}+m_{34}}(\bar{p}_3+h_2+h_5)_{m_3+m_{12}}}{(\bar{p}_3+h_2)_{2m_3+m_{12}+m_{13}+m_{33}+m^{\prime}_{23}}(\bar{p}_3+h_2+1-d/2)_{m_3}}\\
&\qquad\times\frac{(-h_2)_{m_1+m_2-m_3+m_{11}+m_{22}+m_{24}+m_{34}}(-h_{4})_{m_{2}+m_{12}+m_{22}+m_{24}+m_{33}}(-h_{5})_{m_3+m_{13}+m_{33}+m^{\prime}_{23}}}{(p_{3}-h_{2})_{2m_{2}+m_{11}+m_{12}+m_{22}+m_{24}+m_{33}+m_{34}}(p_3-h_2+1-d/2)_{m_2}}\\
&\qquad\times\frac{(-\bar{p}_3-h_2+d/2-m_3)_{m_3}(-p_3+h_2+d/2-m_2)_{m_2}}{(p_3)_{-m_1}}\frac{(-m_2)_{t_1}(-m_3)_{t_2}}{t_1!t_2!}\\
&\qquad\times\frac{(\bar{p}_3-d/2)_{t_1+t_2}}{(p_3-m_1)_{t_1+t_2}}\frac{(p_3)_{t_1+t_2}}{(\bar{p}_3+h_2+1-d/2)_{t_2}(p_3-h_2+1-d/2)_{t_1}}\\
&\qquad\times\binom{m_{11}}{k_{11}}\binom{m_{12}}{k_{12}}\binom{m_{22}}{k_{22}}\binom{m_{33}}{k_{33}}\binom{k_{11}}{m^{\prime}_{11}}\binom{k_{12}}{m^{\prime}_{12}}\binom{k_{22}}{m^{\prime}_{22}}\binom{k_{33}}{m^{\prime}_{33}}\\
&\qquad\times(v^7_{24})^{m_2+k_{12}+k_{22}+k_{33}}(v^7_{34})^{m_1-m_3+k_{11}-k_{12}-k_{33}}\frac{(1-v_{24}^{7})^{m_{24}}}{m_{24}!}\frac{(1-v_{34}^{7})^{m_{34}}}{m_{34}!}\\
&\qquad\times(-1)^{k_{11}+k_{12}+k_{22}+k_{33}+m^{\prime}_{11}+m^{\prime}_{12}+m^{\prime}_{22}+m^{\prime}_{33}}\frac{m_{23}!}{m^{\prime}_{23}!}\prod_{1\leq a\leq3}\frac{(u_a^6)^{m_a}}{m_a!}\prod_{1\leq a\leq b\leq3}\frac{(1-v_{ab}^6)^{m^{\prime}_{ab}}}{m_{ab}!},
}
where we defined $m_{23}^{\prime}=m_{44}$.

We then shift all of $k_{ab}$ by $k_{ab}\to k_{ab}+m^{\prime}_{ab}$ and use the fact that
\eqn{\frac{1}{m_{ab}!}\binom{m_{ab}}{k_{ab}+m^{\prime}_{ab}}\binom{k_{ab}+m^{\prime}_{ab}}{m^{\prime}_{ab}}=\frac{1}{m^{\prime}_{ab}!(m_{ab}-m^{\prime}_{ab})!}\binom{m_{ab}-m^{\prime}_{ab}}{k_{ab}},}
to write
\eqna{
&G_{7|\mathcal{O}_{i_2}\to\1}=\sum\frac{(-h_{6})_{m_1+m_{24}+m_{34}+m^{\prime}_{23}}(p_2+h_{6})_{m_1+m_{11}+m^{\prime}_{13}+m_{22}}(p_2+h_2)_{m_1-m_2+m_3+m^{\prime}_{13}+m^{\prime}_{23}}}{(p_2)_{2m_1+m_{11}+m^{\prime}_{13}+m_{22}+m_{24}+m_{34}+m^{\prime}_{23}}(p_2+1-d/2)_{m_1}}\\
&\qquad\times\frac{(p_3)_{-m_1+m_2+m_3+m_{12}+m_{33}}(p_3-h_2+h_4)_{m_{2}+m_{11}+m_{34}}(\bar{p}_{3}+h_2+h_{5})_{m_3+m_{12}}}{(\bar{p}_{3}+h_2)_{2m_3+m_{12}+m^{\prime}_{13}+m_{33}+m^{\prime}_{23}}(\bar{p}_3+h_2+1-d/2)_{m_3}}\\
&\qquad\times\frac{(-h_2)_{m_1+m_2-m_3+m_{11}+m_{22}+m_{24}+m_{34}}(-h_{4})_{m_{2}+m_{12}+m_{22}+m_{24}+m_{33}}(-h_{5})_{m_3+m^{\prime}_{13}+m_{33}+m^{\prime}_{23}}}{(p_{3}-h_{2})_{2m_{2}+m_{11}+m_{12}+m_{22}+m_{24}+m_{33}+m_{34}}(p_3-h_2+1-d/2)_{m_2}}\\
&\qquad\times\frac{(-\bar{p}_3-h_2+d/2-m_3)_{m_3}(-p_3+h_2+d/2-m_2)_{m_2}}{(p_3)_{-m_1}}\frac{(-m_2)_{t_1}(-m_3)_{t_2}}{t_1!t_2!}\\
&\qquad\times\frac{(\bar{p}_3-d/2)_{t_1+t_2}}{(p_3-m_1)_{t_1+t_2}}\frac{(p_3)_{t_1+t_2}}{(\bar{p}_3+h_2+1-d/2)_{t_2}(p_3-h_2+1-d/2)_{t_1}}\\
&\qquad\times\binom{m_{11}-m^{\prime}_{11}}{k_{11}}\binom{m_{12}-m^{\prime}_{12}}{k_{12}}\binom{m_{22}-m^{\prime}_{22}}{k_{22}}\binom{m_{33}-m^{\prime}_{33}}{k_{33}}\\
&\qquad\times\frac{(v^7_{24})^{m_2+k_{12}+k_{22}+k_{33}+m^{\prime}_{12}+m^{\prime}_{22}+m^{\prime}_{33}}(v^7_{34})^{m_1-m_3+k_{11}-k_{12}-k_{33}+m^{\prime}_{11}-m^{\prime}_{12}-m^{\prime}_{33}}}{(m_{11}-m^{\prime}_{11})!(m_{12}-m^{\prime}_{12})!(m_{22}-m^{\prime}_{22})!(m_{33}-m^{\prime}_{33})!}\\
&\qquad\times\frac{(1-v_{24}^{7})^{m_{24}}}{m_{24}!}\frac{(1-v_{34}^{7})^{m_{34}}}{m_{34}!}(-1)^{k_{11}+k_{12}+k_{22}+k_{33}}\prod_{1\leq a\leq3}\frac{(u_a^6)^{m_a}}{m_a!}\prod_{1\leq a\leq b\leq3}\frac{(1-v_{ab}^6)^{m^{\prime}_{ab}}}{m^{\prime}_{ab}!},
}
where we defined $m_{13}^{\prime}=m_{13}$.

Using the identity \eqref{EqBinom}, we can compute the sums over $k_{11}$ and $k_{22}$, leading to
\eqna{
&G_{7|\mathcal{O}_{i_2}\to\1}=\sum\frac{(-h_{6})_{m_1+m_{24}+m_{34}+m^{\prime}_{23}}(p_2+h_{6})_{m_1+m_{11}+m^{\prime}_{13}+m_{22}}(p_2+h_2)_{m_1-m_2+m_3+m^{\prime}_{13}+m^{\prime}_{23}}}{(p_2)_{2m_1+m_{11}+m^{\prime}_{13}+m_{22}+m_{24}+m_{34}+m^{\prime}_{23}}(p_2+1-d/2)_{m_1}}\\
&\qquad\times\frac{(p_3)_{-m_1+m_2+m_3+m_{12}+m_{33}}(p_3-h_2+h_4)_{m_{2}+m_{11}+m_{34}}(\bar{p}_{3}+h_2+h_{5})_{m_3+m_{12}}}{(\bar{p}_{3}+h_2)_{2m_3+m_{12}+m^{\prime}_{13}+m_{33}+m^{\prime}_{23}}(\bar{p}_3+h_2+1-d/2)_{m_3}}\\
&\qquad\times\frac{(-h_2)_{m_1+m_2-m_3+m_{11}+m_{22}+m_{24}+m_{34}}(-h_{4})_{m_{2}+m_{12}+m_{22}+m_{24}+m_{33}}(-h_{5})_{m_3+m^{\prime}_{13}+m_{33}+m^{\prime}_{23}}}{(p_{3}-h_{2})_{2m_{2}+m_{11}+m_{12}+m_{22}+m_{24}+m_{33}+m_{34}}(p_3-h_2+1-d/2)_{m_2}}\\
&\qquad\times\frac{(-\bar{p}_3-h_2+d/2-m_3)_{m_3}(-p_3+h_2+d/2-m_2)_{m_2}}{(p_3)_{-m_1}}\frac{(-m_2)_{t_1}(-m_3)_{t_2}}{t_1!t_2!}\\
&\qquad\times\frac{(\bar{p}_3-d/2)_{t_1+t_2}}{(p_3-m_1)_{t_1+t_2}}\frac{(p_3)_{t_1+t_2}}{(\bar{p}_3+h_2+1-d/2)_{t_2}(p_3-h_2+1-d/2)_{t_1}}\\
&\qquad\times\binom{m_{12}-m^{\prime}_{12}}{k_{12}}\binom{m_{33}-m^{\prime}_{33}}{k_{33}}\binom{k_{12}}{r_{12}}\binom{k_{33}}{r_{33}}\\
&\qquad\times\frac{(v^7_{24})^{m_2+m^{\prime}_{12}+m^{\prime}_{22}+m^{\prime}_{33}}(v^7_{34})^{m_1-m_3-k_{12}-k_{33}+m^{\prime}_{11}-m^{\prime}_{12}-m^{\prime}_{33}}}{(m_{11}-m^{\prime}_{11})!(m_{12}-m^{\prime}_{12})!(m_{22}-m^{\prime}_{22})!(m_{33}-m^{\prime}_{33})!}\frac{(1-v_{24}^{7})^{r_{12}+r_{33}+m_{22}+m_{24}-m^{\prime}_{22}}}{m_{24}!}\\
&\qquad\times\frac{(1-v_{34}^{7})^{m_{11}+m_{34}-m^{\prime}_{11}}}{m_{34}!}(-1)^{r_{12}+r_{33}+k_{12}+k_{33}}\prod_{1\leq a\leq3}\frac{(u_a^6)^{m_a}}{m_a!}\prod_{1\leq a\leq b\leq3}\frac{(1-v_{ab}^6)^{m^{\prime}_{ab}}}{m^{\prime}_{ab}!}.
}

We now change $k_{12}$ and $k_{33}$ by $k_{12}\to k_{12}+r_{12}$ and $k_{33}\to k_{33}+r_{33}$, respectively, giving us
\eqna{
&G_{7|\mathcal{O}_{i_2}\to\1}=\sum\frac{(-h_{6})_{m_1+m_{24}+m_{34}+m^{\prime}_{23}}(p_2+h_{6})_{m_1+m_{11}+m^{\prime}_{13}+m_{22}}(p_2+h_2)_{m_1-m_2+m_3+m^{\prime}_{13}+m^{\prime}_{23}}}{(p_2)_{2m_1+m_{11}+m^{\prime}_{13}+m_{22}+m_{24}+m_{34}+m^{\prime}_{23}}(p_2+1-d/2)_{m_1}}\\
&\qquad\times\frac{(p_3)_{-m_1+m_2+m_3+m_{12}+m_{33}}(p_3-h_2+h_4)_{m_{2}+m_{11}+m_{34}}(\bar{p}_{3}+h_2+h_{5})_{m_3+m_{12}}}{(\bar{p}_{3}+h_2)_{2m_3+m_{12}+m^{\prime}_{13}+m_{33}+m^{\prime}_{23}}(\bar{p}_3+h_2+1-d/2)_{m_3}}\\
&\qquad\times\frac{(-h_2)_{m_1+m_2-m_3+m_{11}+m_{22}+m_{24}+m_{34}}(-h_{4})_{m_{2}+m_{12}+m_{22}+m_{24}+m_{33}}(-h_{5})_{m_3+m^{\prime}_{13}+m_{33}+m^{\prime}_{23}}}{(p_{3}-h_{2})_{2m_{2}+m_{11}+m_{12}+m_{22}+m_{24}+m_{33}+m_{34}}(p_3-h_2+1-d/2)_{m_2}}\\
&\qquad\times\frac{(-\bar{p}_3-h_2+d/2-m_3)_{m_3}(-p_3+h_2+d/2-m_2)_{m_2}}{(p_3)_{-m_1}}\frac{(-m_2)_{t_1}(-m_3)_{t_2}}{t_1!t_2!}\\
&\qquad\times\frac{(\bar{p}_3-d/2)_{t_1+t_2}}{(p_3-m_1)_{t_1+t_2}}\frac{(p_3)_{t_1+t_2}}{(\bar{p}_3+h_2+1-d/2)_{t_2}(p_3-h_2+1-d/2)_{t_1}}\\
&\qquad\times\binom{m_{12}-m^{\prime}_{12}-r_{12}}{k_{12}}\binom{m_{33}-m^{\prime}_{33}-r_{33}}{k_{33}}\\
&\qquad\times\frac{(v^7_{24})^{m_2+m^{\prime}_{12}+m^{\prime}_{22}+m^{\prime}_{33}}(v^7_{34})^{m_1-m_3-r_{12}-r_{33}-k_{12}-k_{33}+m^{\prime}_{11}-m^{\prime}_{12}-m^{\prime}_{33}}}{r_{12}!r_{33}!(m_{11}-m^{\prime}_{11})!(m_{12}-m^{\prime}_{12}-r_{12})!(m_{22}-m^{\prime}_{22})!(m_{33}-m^{\prime}_{33}-r_{33})!}\\
&\qquad\times\frac{(1-v_{24}^{7})^{r_{12}+r_{33}+m_{22}+m_{24}-m^{\prime}_{22}}}{m_{24}!}\frac{(1-v_{34}^{7})^{m_{11}+m_{34}-m^{\prime}_{11}}}{m_{34}!}\\
&\qquad\times(-1)^{k_{12}+k_{33}}\prod_{1\leq a\leq3}\frac{(u_a^6)^{m_a}}{m_a!}\prod_{1\leq a\leq b\leq3}\frac{(1-v_{ab}^6)^{m^{\prime}_{ab}}}{m^{\prime}_{ab}!}.
}

The sums over $k_{12}$ and $k_{33}$ then lead to
\eqna{
&G_{7|\mathcal{O}_{i_2}\to\1}=\sum\frac{(-h_{6})_{m_1+m_{24}+m_{34}+m^{\prime}_{23}}(p_2+h_{6})_{m_1+m_{11}+m^{\prime}_{13}+m_{22}}(p_2+h_2)_{m_1-m_2+m_3+m^{\prime}_{13}+m^{\prime}_{23}}}{(p_2)_{2m_1+m_{11}+m^{\prime}_{13}+m_{22}+m_{24}+m_{34}+m^{\prime}_{23}}(p_2+1-d/2)_{m_1}}\\
&\qquad\times\frac{(p_3)_{-m_1+m_2+m_3+m_{12}+m_{33}}(p_3-h_2+h_4)_{m_{2}+m_{11}+m_{34}}(\bar{p}_{3}+h_2+h_{5})_{m_3+m_{12}}}{(\bar{p}_{3}+h_2)_{2m_3+m_{12}+m^{\prime}_{13}+m_{33}+m^{\prime}_{23}}(\bar{p}_3+h_2+1-d/2)_{m_3}}\\
&\qquad\times\frac{(-h_2)_{m_1+m_2-m_3+m_{11}+m_{22}+m_{24}+m_{34}}(-h_{4})_{m_{2}+m_{12}+m_{22}+m_{24}+m_{33}}(-h_{5})_{m_3+m^{\prime}_{13}+m_{33}+m^{\prime}_{23}}}{(p_{3}-h_{2})_{2m_{2}+m_{11}+m_{12}+m_{22}+m_{24}+m_{33}+m_{34}}(p_3-h_2+1-d/2)_{m_2}}\\
&\qquad\times\frac{(-\bar{p}_3-h_2+d/2-m_3)_{m_3}(-p_3+h_2+d/2-m_2)_{m_2}}{(p_3)_{-m_1}}\frac{(-m_2)_{t_1}(-m_3)_{t_2}}{t_1!t_2!}\\
&\qquad\times\frac{(\bar{p}_3-d/2)_{t_1+t_2}}{(p_3-m_1)_{t_1+t_2}}\frac{(p_3)_{t_1+t_2}}{(\bar{p}_3+h_2+1-d/2)_{t_2}(p_3-h_2+1-d/2)_{t_1}}\\
&\qquad\times\frac{(v^7_{24})^{m_2+m^{\prime}_{12}+m^{\prime}_{22}+m^{\prime}_{33}}(v^7_{34})^{m_1-m_3-r_{12}-r_{33}+m^{\prime}_{11}-m^{\prime}_{12}-m^{\prime}_{33}}}{r_{12}!r_{33}!(m_{11}-m^{\prime}_{11})!(m_{12}-m^{\prime}_{12}-r_{12})!(m_{22}-m^{\prime}_{22})!(m_{33}-m^{\prime}_{33}-r_{33})!}\\
&\qquad\times\frac{(1-v_{24}^{7})^{r_{12}+r_{33}+m_{22}+m_{24}-m^{\prime}_{22}}}{m_{24}!}\frac{(1-\frac{1}{v^7_{34}})^{m_{12}+m_{33}-m^{\prime}_{12}-m^{\prime}_{33}-r_{12}-r_{33}}(1-v_{34}^{7})^{m_{11}+m_{34}-m^{\prime}_{11}}}{m_{34}!}\\
&\qquad\times\prod_{1\leq a\leq3}\frac{(u_a^6)^{m_a}}{m_a!}\prod_{1\leq a\leq b\leq3}\frac{(1-v_{ab}^6)^{m^{\prime}_{ab}}}{m^{\prime}_{ab}!}.
}

Changing variables such that
\eqn{m_{11}\to m_{11}+m^{\prime}_{11},\qquad m_{12}\to m_{12}+m^{\prime}_{11}+r_{12},\qquad m_{22}\to m_{22}+m^{\prime}_{22},\qquad m_{33}\to m_{33}+m^{\prime}_{33}+r_{33},}
we obtain
\eqna{
&G_{7|\mathcal{O}_{i_2}\to\1}=\sum\frac{(-h_{6})_{m_1+m_{24}+m_{34}+m^{\prime}_{23}}(p_2+h_{6})_{m_1+m_{11}+m^{\prime}_{11}+m^{\prime}_{13}+m_{22}+m^{\prime}_{22}}(p_2+h_2)_{m_1-m_2+m_3+m^{\prime}_{13}+m^{\prime}_{23}}}{(p_2)_{2m_1+m_{11}+m^{\prime}_{11}+m^{\prime}_{13}+m_{22}+m^{\prime}_{22}+m_{24}+m_{34}+m^{\prime}_{23}}(p_2+1-d/2)_{m_1}}\\
&\qquad\times\frac{(p_3)_{-m_1+m_2+m_3+m_{12}+m^{\prime}_{12}+r_{12}+m_{33}+m^{\prime}_{33}+r_{33}}(p_3-h_2+h_4)_{m_{2}+m_{11}+m^{\prime}_{11}+m_{34}}}{(\bar{p}_{3}+h_2)_{2m_3+m_{12}+m^{\prime}_{12}+r_{12}+m^{\prime}_{13}+m_{33}+m^{\prime}_{33}+r_{33}+m^{\prime}_{23}}(\bar{p}_3+h_2+1-d/2)_{m_3}}\\
&\qquad\times\frac{(-h_2)_{m_1+m_2-m_3+m_{11}+m^{\prime}_{11}+m_{22}+m^{\prime}_{22}+m_{24}+m_{34}}(-h_{4})_{m_{2}+m_{12}+m^{\prime}_{12}+r_{12}+m_{22}+m^{\prime}_{22}+m_{24}+m_{33}+m^{\prime}_{33}+r_{33}}}{(p_{3}-h_{2})_{2m_{2}+m_{11}+m^{\prime}_{11}+m_{12}+m^{\prime}_{12}+r_{12}+m_{22}+m^{\prime}_{22}+m_{24}+m_{33}+m^{\prime}_{33}+r_{33}+m_{34}}(p_3-h_2+1-d/2)_{m_2}}\\
&\qquad\times\frac{(-h_{5})_{m_3+m^{\prime}_{13}+m_{33}+m^{\prime}_{33}+r_{33}+m^{\prime}_{23}}(-\bar{p}_3-h_2+d/2-m_3)_{m_3}(-p_3+h_2+d/2-m_2)_{m_2}}{(p_3)_{-m_1}}\\
&\qquad\times\frac{(-m_2)_{t_1}(-m_3)_{t_2}}{t_1!t_2!}\frac{(\bar{p}_3-d/2)_{t_1+t_2}}{(p_3-m_1)_{t_1+t_2}}\frac{(p_3)_{t_1+t_2}(\bar{p}_{3}+h_2+h_{5})_{m_3+m_{12}+m^{\prime}_{12}+r_{12}}}{(\bar{p}_3+h_2+1-d/2)_{t_2}(p_3-h_2+1-d/2)_{t_1}}\\
&\qquad\times\frac{(v^7_{24})^{m_2+m^{\prime}_{12}+m^{\prime}_{22}+m^{\prime}_{33}}(v^7_{34})^{m_1-m_3-r_{12}-r_{33}+m^{\prime}_{11}-m^{\prime}_{12}-m^{\prime}_{33}}}{r_{12}!r_{33}!m_{11}!m_{12}!m_{22}!m_{33}!}\\
&\qquad\times\frac{(1-v_{24}^{7})^{r_{12}+r_{33}+m_{22}+m_{24}}}{m_{24}!}\frac{(1-\frac{1}{v^7_{34}})^{m_{12}+m_{33}}(1-v_{34}^{7})^{m_{11}+m_{34}}}{m_{34}!}\prod_{1\leq a\leq3}\frac{(u_a^6)^{m_a}}{m_a!}\prod_{1\leq a\leq b\leq3}\frac{(1-v_{ab}^6)^{m^{\prime}_{ab}}}{m^{\prime}_{ab}!}.
}

We now define
\eqn{m_{24}=n_{24}-m_{22}-r_{12}-r_{33},\qquad m_{34}=n_{34}-m_{11},\qquad m_{33}=s_{34}-m_{12},}
such that
\eqna{
&G_{7|\mathcal{O}_{i_2}\to\1}=\sum\frac{(-h_{6})_{m_1+n_{24}+n_{34}+m^{\prime}_{23}-m_{11}-m_{22}-r_{12}-r_{33}}(p_2+h_{6})_{m_1+m_{11}+m^{\prime}_{11}+m^{\prime}_{13}+m_{22}+m^{\prime}_{22}}}{(p_2)_{2m_1+m^{\prime}_{11}+m^{\prime}_{13}+m^{\prime}_{22}+n_{24}+n_{34}-r_{12}-r_{33}+m^{\prime}_{23}}(p_2+1-d/2)_{m_1}}\\
&\qquad\times\frac{(p_3)_{-m_1+m_2+m_3+m^{\prime}_{12}+r_{12}+s_{34}+m^{\prime}_{33}+r_{33}}(p_3-h_2+h_4)_{m_{2}+m^{\prime}_{11}+n_{34}}(\bar{p}_{3}+h_2+h_{5})_{m_3+m_{12}+m^{\prime}_{12}+r_{12}}}{(\bar{p}_{3}+h_2)_{2m_3+m^{\prime}_{12}+r_{12}+m^{\prime}_{13}+s_{34}+m^{\prime}_{33}+r_{33}+m^{\prime}_{23}}(\bar{p}_3+h_2+1-d/2)_{m_3}}\\
&\qquad\times\frac{(-h_2)_{m_1+m_2-m_3+m^{\prime}_{11}+m^{\prime}_{22}+n_{24}+n_{34}-r_{12}-r_{33}}(-h_{4})_{m_{2}+m^{\prime}_{12}+m^{\prime}_{22}+n_{24}+s_{34}+m^{\prime}_{33}}}{(p_{3}-h_{2})_{2m_{2}+m^{\prime}_{11}+m^{\prime}_{12}+m^{\prime}_{22}+m^{\prime}_{33}+n_{24}+n_{34}+s_{34}}(p_3-h_2+1-d/2)_{m_2}}\\
&\qquad\times\frac{(-h_{5})_{m_3+m^{\prime}_{13}+s_{34}-m_{12}+m^{\prime}_{33}+r_{33}+m^{\prime}_{23}}(-\bar{p}_3-h_2+d/2-m_3)_{m_3}(-p_3+h_2+d/2-m_2)_{m_2}}{(p_3)_{-m_1}}\\
&\qquad\times\frac{(-m_2)_{t_1}(-m_3)_{t_2}}{t_1!t_2!}\frac{(\bar{p}_3-d/2)_{t_1+t_2}}{(p_3-m_1)_{t_1+t_2}}\frac{(p_2+h_2)_{m_1-m_2+m_3+m^{\prime}_{13}+m^{\prime}_{23}}(p_3)_{t_1+t_2}}{(\bar{p}_3+h_2+1-d/2)_{t_2}(p_3-h_2+1-d/2)_{t_1}}\\
&\qquad\times\frac{(v^7_{24})^{m_2+m^{\prime}_{12}+m^{\prime}_{22}+m^{\prime}_{33}}(v^7_{34})^{m_1-m_3-r_{12}-r_{33}+m^{\prime}_{11}-m^{\prime}_{12}-m^{\prime}_{33}}}{r_{12}!r_{33}!m_{11}!m_{12}!m_{22}!(s_{34}-m_{12})!}\\
&\qquad\times\frac{(1-v_{24}^{7})^{n_{24}}}{(n_{24}-m_{22}-r_{12}-r_{33})!}\frac{(1-\frac{1}{v^7_{34}})^{s_{34}}(1-v_{34}^{7})^{n_{34}}}{(n_{34}-m_{11})!}\prod_{1\leq a\leq3}\frac{(u_a^6)^{m_a}}{m_a!}\prod_{1\leq a\leq b\leq3}\frac{(1-v_{ab}^6)^{m^{\prime}_{ab}}}{m^{\prime}_{ab}!}.
}

The sums over $m_{11}$, $m_{12}$, and $m_{22}$ can be performed, leading to
\eqna{
&G_{7|\mathcal{O}_{i_2}\to\1}=\sum\frac{(-h_{6})_{m_1+m^{\prime}_{23}}(p_2+h_{6})_{m_1+m^{\prime}_{11}+m^{\prime}_{13}+m^{\prime}_{22}}(p_2+h_2)_{m_1-m_2+m_3+m^{\prime}_{13}+m^{\prime}_{23}}}{(p_2)_{2m_1+m^{\prime}_{11}+m^{\prime}_{13}+m^{\prime}_{22}+m^{\prime}_{23}}(p_2+1-d/2)_{m_1}}\\
&\qquad\times\frac{(p_3)_{-m_1+m_2+m_3+m^{\prime}_{12}+r_{12}+s_{34}+m^{\prime}_{33}+r_{33}}(p_3-h_2+h_4)_{m_{2}+m^{\prime}_{11}+n_{34}}(\bar{p}_{3}+h_2+h_{5})_{m_3+m^{\prime}_{12}+r_{12}}}{(\bar{p}_{3}+h_2)_{2m_3+m^{\prime}_{12}+r_{12}+m^{\prime}_{13}+m^{\prime}_{33}+r_{33}+m^{\prime}_{23}}(\bar{p}_3+h_2+1-d/2)_{m_3}}\\
&\qquad\times\frac{(-h_2)_{m_1+m_2-m_3+m^{\prime}_{11}+m^{\prime}_{22}+n_{24}+n_{34}-r_{12}-r_{33}}(-h_{4})_{m_{2}+m^{\prime}_{12}+m^{\prime}_{22}+n_{24}+s_{34}+m^{\prime}_{33}}}{(p_{3}-h_{2})_{2m_{2}+m^{\prime}_{11}+m^{\prime}_{12}+m^{\prime}_{22}+m^{\prime}_{33}+n_{24}+n_{34}+s_{34}}(p_3-h_2+1-d/2)_{m_2}}\\
&\qquad\times\frac{(-h_{5})_{m_3+m^{\prime}_{13}+m^{\prime}_{33}+r_{33}+m^{\prime}_{23}}(-\bar{p}_3-h_2+d/2-m_3)_{m_3}(-p_3+h_2+d/2-m_2)_{m_2}}{(p_3)_{-m_1}}\\
&\qquad\times\frac{(-m_2)_{t_1}(-m_3)_{t_2}}{t_1!t_2!}\frac{(\bar{p}_3-d/2)_{t_1+t_2}}{(p_3-m_1)_{t_1+t_2}}\frac{(p_3)_{t_1+t_2}}{(\bar{p}_3+h_2+1-d/2)_{t_2}(p_3-h_2+1-d/2)_{t_1}}\\
&\qquad\times\frac{(v^7_{24})^{m_2+m^{\prime}_{12}+m^{\prime}_{22}+m^{\prime}_{33}}(v^7_{34})^{m_1-m_3-r_{12}-r_{33}+m^{\prime}_{11}-m^{\prime}_{12}-m^{\prime}_{33}}}{r_{12}!r_{33}!s_{34}!}\\
&\qquad\times\frac{(1-v_{24}^{7})^{n_{24}}}{(n_{24}-r_{12}-r_{33})!}\frac{(1-\frac{1}{v^7_{34}})^{s_{34}}(1-v_{34}^{7})^{n_{34}}}{n_{34}!}\prod_{1\leq a3}\frac{(u_a^6)^{m_a}}{m_a!}\prod_{1\leq a\leq b\leq3}\frac{(1-v_{ab}^6)^{m^{\prime}_{ab}}}{m^{\prime}_{ab}!}.
}

We now redefine $r_{33}=r-r_{12}$ to evaluate the sum over $r_{12}$ and get
\eqna{
&G_{7|\mathcal{O}_{i_2}\to\1}=\sum\frac{(-h_{6})_{m_1+m^{\prime}_{23}}(p_2+h_{6})_{m_1+m^{\prime}_{11}+m^{\prime}_{13}+m^{\prime}_{22}}(p_2+h_2)_{m_1-m_2+m_3+m^{\prime}_{13}+m^{\prime}_{23}}}{(p_2)_{2m_1+m^{\prime}_{11}+m^{\prime}_{13}+m^{\prime}_{22}+m^{\prime}_{23}}(p_2+1-d/2)_{m_1}}\\
&\qquad\times\frac{(p_3)_{-m_1+m_2+m_3+m^{\prime}_{12}+s_{34}+m^{\prime}_{33}+r}(p_3-h_2+h_4)_{m_{2}+m^{\prime}_{11}+n_{34}}(\bar{p}_{3}+h_2+h_{5})_{m_3+m^{\prime}_{12}}}{(\bar{p}_{3}+h_2)_{2m_3+m^{\prime}_{12}+m^{\prime}_{13}+m^{\prime}_{23}+m^{\prime}_{33}}(\bar{p}_3+h_2+1-d/2)_{m_3}}\\
&\qquad\times\frac{(-h_2)_{m_1+m_2-m_3+m^{\prime}_{11}+m^{\prime}_{22}+n_{24}+n_{34}-r}(-h_{4})_{m_{2}+m^{\prime}_{12}+m^{\prime}_{22}+n_{24}+s_{34}+m^{\prime}_{33}}}{(p_{3}-h_{2})_{2m_{2}+m^{\prime}_{11}+m^{\prime}_{12}+m^{\prime}_{22}+m^{\prime}_{33}+n_{24}+n_{34}+s_{34}}(p_3-h_2+1-d/2)_{m_2}}\\
&\qquad\times\frac{(-h_{5})_{m_3+m^{\prime}_{13}+m^{\prime}_{23}+m^{\prime}_{33}}(-\bar{p}_3-h_2+d/2-m_3)_{m_3}(-p_3+h_2+d/2-m_2)_{m_2}}{(p_3)_{-m_1}}\\
&\qquad\times\frac{(-m_2)_{t_1}(-m_3)_{t_2}}{t_1!t_2!}\frac{(\bar{p}_3-d/2)_{t_1+t_2}}{(p_3-m_1)_{t_1+t_2}}\frac{(p_3)_{t_1+t_2}}{(\bar{p}_3+h_2+1-d/2)_{t_2}(p_3-h_2+1-d/2)_{t_1}}\\
&\qquad\times\frac{(v^7_{24})^{m_2+m^{\prime}_{12}+m^{\prime}_{22}+m^{\prime}_{33}}(v^7_{34})^{m_1-m_3+m^{\prime}_{11}-m^{\prime}_{12}-m^{\prime}_{33}}}{r!s_{34}!}\\
&\qquad\times\binom{r}{j}\frac{(1-v_{24}^{7})^{n_{24}}}{(n_{24}-r)!}\frac{(1-\frac{1}{v^7_{34}})^{j+s_{34}}(1-v_{34}^{7})^{n_{34}}}{n_{34}!}(-1)^j\prod_{1\leq a\leq3}\frac{(u_a^6)^{m_a}}{m_a!}\prod_{1\leq a\leq b\leq3}\frac{(1-v_{ab}^6)^{m^{\prime}_{ab}}}{m^{\prime}_{ab}!},
}
where we also expanded $(v_{34}^7)^{-r}$ in a power series in $1-v_{34}^7$.

After shifting $r$ by $r\to r+j$, we evaluate the sum over $r$, leading to
\eqna{
&G_{7|\mathcal{O}_{i_2}\to\1}=\sum\frac{(-h_{6})_{m_1+m^{\prime}_{23}}(p_2+h_{6})_{m_1+m^{\prime}_{11}+m^{\prime}_{13}+m^{\prime}_{22}}(p_2+h_2)_{m_1-m_2+m_3+m^{\prime}_{13}+m^{\prime}_{23}}}{(p_2)_{2m_1+m^{\prime}_{11}+m^{\prime}_{13}+m^{\prime}_{22}+m^{\prime}_{23}}(p_2+1-d/2)_{m_1}}\\
&\qquad\times\frac{(p_3)_{-m_1+m_2+m_3+m^{\prime}_{12}+s_{34}+m^{\prime}_{33}+j}(p_3-h_2+h_4)_{m_{2}+m^{\prime}_{11}+n_{34}}(\bar{p}_{3}+h_2+h_{5})_{m_3+m^{\prime}_{12}}}{(\bar{p}_{3}+h_2)_{2m_3+m^{\prime}_{12}+m^{\prime}_{13}+m^{\prime}_{23}+m^{\prime}_{33}}(\bar{p}_3+h_2+1-d/2)_{m_3}}\\
&\qquad\times\frac{(-h_2)_{m_1+m_2-m_3+m^{\prime}_{11}+m^{\prime}_{22}+n_{34}}(-h_{4})_{m_{2}+m^{\prime}_{12}+m^{\prime}_{22}+n_{24}+s_{34}+m^{\prime}_{33}}}{(p_{3}-h_{2})_{2m_{2}+m^{\prime}_{11}+m^{\prime}_{12}+m^{\prime}_{22}+m^{\prime}_{33}+j+n_{34}+s_{34}}(p_3-h_2+1-d/2)_{m_2}}\\
&\qquad\times\frac{(-h_{5})_{m_3+m^{\prime}_{13}+m^{\prime}_{23}+m^{\prime}_{33}}(-\bar{p}_3-h_2+d/2-m_3)_{m_3}(-p_3+h_2+d/2-m_2)_{m_2}}{(p_3)_{-m_1}}\\
&\qquad\times\frac{(-m_2)_{t_1}(-m_3)_{t_2}}{t_1!t_2!}\frac{(\bar{p}_3-d/2)_{t_1+t_2}}{(p_3-m_1)_{t_1+t_2}}\frac{(p_3)_{t_1+t_2}}{(\bar{p}_3+h_2+1-d/2)_{t_2}(p_3-h_2+1-d/2)_{t_1}}\\
&\qquad\times\frac{(v^7_{24})^{m_2+m^{\prime}_{12}+m^{\prime}_{22}+m^{\prime}_{33}}(v^7_{34})^{m_1-m_3+m^{\prime}_{11}-m^{\prime}_{12}-m^{\prime}_{33}}}{j!s_{34}!}\\
&\qquad\times\frac{(1-v_{24}^{7})^{n_{24}}}{(n_{24}-j)!}\frac{(1-\frac{1}{v^7_{34}})^{j+s_{34}}(1-v_{34}^{7})^{n_{34}}}{n_{34}!}(-1)^j\prod_{1\leq a\leq3}\frac{(u_a^6)^{m_a}}{m_a!}\prod_{1\leq a\leq b\leq3}\frac{(1-v_{ab}^6)^{m^{\prime}_{ab}}}{m^{\prime}_{ab}!}.
}

Changing $n_{24}$ by $n_{24}+j$, the sum over $n_{24}$ gives 
\eqna{
&G_{7|\mathcal{O}_{i_2}\to\1}=\sum\frac{(-h_{6})_{m_1+m^{\prime}_{23}}(p_2+h_{6})_{m_1+m^{\prime}_{11}+m^{\prime}_{13}+m^{\prime}_{22}}(p_2+h_2)_{m_1-m_2+m_3+m^{\prime}_{13}+m^{\prime}_{23}}}{(p_2)_{2m_1+m^{\prime}_{11}+m^{\prime}_{13}+m^{\prime}_{22}+m^{\prime}_{23}}(p_2+1-d/2)_{m_1}}\\
&\qquad\times\frac{(p_3)_{-m_1+m_2+m_3+m^{\prime}_{12}+s_{34}+m^{\prime}_{33}+j}(p_3-h_2+h_4)_{m_{2}+m^{\prime}_{11}+n_{34}}(\bar{p}_{3}+h_2+h_{5})_{m_3+m^{\prime}_{12}}}{(\bar{p}_{3}+h_2)_{2m_3+m^{\prime}_{12}+m^{\prime}_{13}+m^{\prime}_{23}+m^{\prime}_{33}}(\bar{p}_3+h_2+1-d/2)_{m_3}}\\
&\qquad\times\frac{(-h_2)_{m_1+m_2-m_3+m^{\prime}_{11}+m^{\prime}_{22}+n_{34}}(-h_{4})_{m_{2}+m^{\prime}_{12}+m^{\prime}_{22}+j+s_{34}+m^{\prime}_{33}}}{(p_{3}-h_{2})_{2m_{2}+m^{\prime}_{11}+m^{\prime}_{12}+m^{\prime}_{22}+m^{\prime}_{33}+j+n_{34}+s_{34}}(p_3-h_2+1-d/2)_{m_2}}\\
&\qquad\times\frac{(-h_{5})_{m_3+m^{\prime}_{13}+m^{\prime}_{23}+m^{\prime}_{33}}(-\bar{p}_3-h_2+d/2-m_3)_{m_3}(-p_3+h_2+d/2-m_2)_{m_2}}{(p_3)_{-m_1}}\\
&\qquad\times\frac{(-m_2)_{t_1}(-m_3)_{t_2}}{t_1!t_2!}\frac{(\bar{p}_3-d/2)_{t_1+t_2}}{(p_3-m_1)_{t_1+t_2}}\frac{(p_3)_{t_1+t_2}}{(\bar{p}_3+h_2+1-d/2)_{t_2}(p_3-h_2+1-d/2)_{t_1}}\\
&\qquad\times\frac{(v^7_{24})^{h_4-j-s_{34}}(v^7_{34})^{m_1-m_3+m^{\prime}_{11}-m^{\prime}_{12}-m^{\prime}_{33}}}{s_{34}!}\\
&\qquad\times\frac{(1-v_{24}^{7})^{j}}{j!}\frac{(1-\frac{1}{v^7_{34}})^{j+s_{34}}(1-v_{34}^{7})^{n_{34}}}{n_{34}!}(-1)^j\prod_{1\leq a3}\frac{(u_a^6)^{m_a}}{m_a!}\prod_{1\leq a\leq b\leq3}\frac{(1-v_{ab}^6)^{m^{\prime}_{ab}}}{m^{\prime}_{ab}!}.
}

We now define $s_{34}=s-j$ and evaluate the sum over $j$ to reach
\eqna{
&G_{7|\mathcal{O}_{i_2}\to\1}=\sum\frac{(-h_{6})_{m_1+m^{\prime}_{23}}(p_2+h_{6})_{m_1+m^{\prime}_{11}+m^{\prime}_{13}+m^{\prime}_{22}}(p_2+h_2)_{m_1-m_2+m_3+m^{\prime}_{13}+m^{\prime}_{23}}}{(p_2)_{2m_1+m^{\prime}_{11}+m^{\prime}_{13}+m^{\prime}_{22}+m^{\prime}_{23}}(p_2+1-d/2)_{m_1}}\\
&\qquad\times\frac{(p_3)_{-m_1+m_2+m_3+m^{\prime}_{12}+s+m^{\prime}_{33}}(p_3-h_2+h_4)_{m_{2}+m^{\prime}_{11}+n_{34}}(\bar{p}_{3}+h_2+h_{5})_{m_3+m^{\prime}_{12}}}{(\bar{p}_{3}+h_2)_{2m_3+m^{\prime}_{12}+m^{\prime}_{13}+m^{\prime}_{23}+m^{\prime}_{33}}(\bar{p}_3+h_2+1-d/2)_{m_3}}\\
&\qquad\times\frac{(-h_2)_{m_1+m_2-m_3+m^{\prime}_{11}+m^{\prime}_{22}+n_{34}}(-h_{4})_{m_{2}+m^{\prime}_{12}+m^{\prime}_{22}+s+m^{\prime}_{33}}}{(p_{3}-h_{2})_{2m_{2}+m^{\prime}_{11}+m^{\prime}_{12}+m^{\prime}_{22}+m^{\prime}_{33}+n_{34}+s}(p_3-h_2+1-d/2)_{m_2}}\\
&\qquad\times\frac{(-h_{5})_{m_3+m^{\prime}_{13}+m^{\prime}_{23}+m^{\prime}_{33}}(-\bar{p}_3-h_2+d/2-m_3)_{m_3}(-p_3+h_2+d/2-m_2)_{m_2}}{(p_3)_{-m_1}}\frac{(-m_2)_{t_1}(-m_3)_{t_2}}{t_1!t_2!}\\
&\qquad\times\frac{(\bar{p}_3-d/2)_{t_1+t_2}}{(p_3-m_1)_{t_1+t_2}}\frac{(p_3)_{t_1+t_2}}{(\bar{p}_3+h_2+1-d/2)_{t_2}(p_3-h_2+1-d/2)_{t_1}}\\
&\qquad\times\frac{(v^7_{24})^{h_4}(v^7_{34})^{m_1-m_3+m^{\prime}_{11}-m^{\prime}_{12}-m^{\prime}_{33}}}{s!}\frac{(1-\frac{1}{v^7_{34}})^{s}(1-v_{34}^{7})^{n_{34}}}{n_{34}!}\prod_{1\leq a\leq3}\frac{(u_a^6)^{m_a}}{m_a!}\prod_{1\leq a\leq b\leq3}\frac{(1-v_{ab}^6)^{m^{\prime}_{ab}}}{m^{\prime}_{ab}!}.
}

We then express the sum over $s$ in terms of a hypergeometric function and use the first identity in \eqref{Eq2F1} to rewrite the summation over $s$, leading to
\eqna{
&G_{7|\mathcal{O}_{i_2}\to\1}=\sum\frac{(-h_{6})_{m_1+m^{\prime}_{23}}(p_2+h_{6})_{m_1+m^{\prime}_{11}+m^{\prime}_{13}+m^{\prime}_{22}}(p_2+h_2)_{m_1-m_2+m_3+m^{\prime}_{13}+m^{\prime}_{23}}}{(p_2)_{2m_1+m^{\prime}_{11}+m^{\prime}_{13}+m^{\prime}_{22}+m^{\prime}_{23}}(p_2+1-d/2)_{m_1}}\\
&\qquad\times\frac{(p_3)_{-m_1+m_2+m_3+m^{\prime}_{12}+m^{\prime}_{33}}(p_3-h_2+h_4)_{m_{2}+m^{\prime}_{11}+s+n_{34}}(\bar{p}_{3}+h_2+h_{5})_{m_3+m^{\prime}_{12}}}{(\bar{p}_{3}+h_2)_{2m_3+m^{\prime}_{12}+m^{\prime}_{13}+m^{\prime}_{23}+m^{\prime}_{33}}(\bar{p}_3+h_2+1-d/2)_{m_3}}\\
&\qquad\times\frac{(-h_2)_{m_1+m_2-m_3+m^{\prime}_{11}+m^{\prime}_{22}+s+n_{34}}(-h_{4})_{m_{2}+m^{\prime}_{12}+m^{\prime}_{22}+m^{\prime}_{33}}}{(p_{3}-h_{2})_{2m_{2}+m^{\prime}_{11}+m^{\prime}_{12}+m^{\prime}_{22}+m^{\prime}_{33}+n_{34}+s}(p_3-h_2+1-d/2)_{m_2}}\\
&\qquad\times\frac{(-h_{5})_{m_3+m^{\prime}_{13}+m^{\prime}_{23}+m^{\prime}_{33}}(-\bar{p}_3-h_2+d/2-m_3)_{m_3}(-p_3+h_2+d/2-m_2)_{m_2}}{(p_3)_{-m_1}}\frac{(-m_2)_{t_1}(-m_3)_{t_2}}{t_1!t_2!}\\
&\qquad\times\frac{(\bar{p}_3-d/2)_{t_1+t_2}}{(p_3-m_1)_{t_1+t_2}}\frac{(p_3)_{t_1+t_2}}{(\bar{p}_3+h_2+1-d/2)_{t_2}(p_3-h_2+1-d/2)_{t_1}}\\
&\qquad\times(-1)^s\frac{(v^7_{24})^{h_4}(v^7_{34})^{h_2-h_4-s-n_{34}}}{s!}\frac{(1-v_{34}^{7})^{s+n_{34}}}{n_{34}!}\prod_{1\leq a\leq3}\frac{(u_a^6)^{m_a}}{m_a!}\prod_{1\leq a\leq b\leq3}\frac{(1-v_{ab}^6)^{m^{\prime}_{ab}}}{m^{\prime}_{ab}!},
}
where we used $(1-1/v_{34}^7)^s=(-1)^s(v_{34}^7)^{-s}(1-v_{34}^7)^s$.

After defining $n_{34}=m-s$, we evaluate the sum over $s$ to get
\eqna{
&G_{7|\mathcal{O}_{i_2}\to\1}=(v^7_{24})^{h_4}(v^7_{34})^{h_2-h_4}\\
&\qquad\times\sum\frac{(-h_{6})_{m_1+m^{\prime}_{23}}(p_2+h_{6})_{m_1+m^{\prime}_{11}+m^{\prime}_{13}+m^{\prime}_{22}}(p_2+h_2)_{m_1-m_2+m_3+m^{\prime}_{13}+m^{\prime}_{23}}}{(p_2)_{2m_1+m^{\prime}_{11}+m^{\prime}_{13}+m^{\prime}_{22}+m^{\prime}_{23}}(p_2+1-d/2)_{m_1}}\\
&\qquad\times\frac{(p_3)_{-m_1+m_2+m_3+m^{\prime}_{12}+m^{\prime}_{33}}(p_3-h_2+h_4)_{m_{2}+m^{\prime}_{11}}(\bar{p}_{3}+h_2+h_{5})_{m_3+m^{\prime}_{12}}}{(\bar{p}_{3}+h_2)_{2m_3+m^{\prime}_{12}+m^{\prime}_{13}+m^{\prime}_{23}+m^{\prime}_{33}}(\bar{p}_3+h_2+1-d/2)_{m_3}}\\
&\qquad\times\frac{(-h_2)_{m_1+m_2-m_3+m^{\prime}_{11}+m^{\prime}_{22}}(-h_{4})_{m_{2}+m^{\prime}_{12}+m^{\prime}_{22}+m^{\prime}_{33}}}{(p_{3}-h_{2})_{2m_{2}+m^{\prime}_{11}+m^{\prime}_{12}+m^{\prime}_{22}+m^{\prime}_{33}}(p_3-h_2+1-d/2)_{m_2}}\\
&\qquad\times\frac{(-h_{5})_{m_3+m^{\prime}_{13}+m^{\prime}_{23}+m^{\prime}_{33}}(-\bar{p}_3-h_2+d/2-m_3)_{m_3}(-p_3+h_2+d/2-m_2)_{m_2}}{(p_3)_{-m_1}}\frac{(-m_2)_{t_1}(-m_3)_{t_2}}{t_1!t_2!}\\
&\qquad\times\frac{(\bar{p}_3-d/2)_{t_1+t_2}}{(p_3-m_1)_{t_1+t_2}}\frac{(p_3)_{t_1+t_2}}{(\bar{p}_3+h_2+1-d/2)_{t_2}(p_3-h_2+1-d/2)_{t_1}}\prod_{1\leq a\leq3}\frac{(u_a^6)^{m_a}}{m_a!}\prod_{1\leq a\leq b\leq3}\frac{(1-v_{ab}^6)^{m^{\prime}_{ab}}}{m^{\prime}_{ab}!}\\
&\qquad=(v^7_{24})^{h_4}(v^7_{34})^{h_2-h_4}G_6,
}
which complete our proof for the limit of unit operator $\mathcal{O}_2(\eta_2)\to\1$ \eqref{EqLimUnit2Id}.

The last limit of unit operator to verify is $\mathcal{O}_{i_3}(\eta_3)=\1$ which leads to
\eqn{\left.I_{7(\Delta_{k_1},\Delta_{k_2},\Delta_{k_3},\Delta_{k_4})}^{(\Delta_{i_2},\Delta_{i_3},\Delta_{i_4},\Delta_{i_5},\Delta_{i_6},\Delta_{i_7},\Delta_{i_1})}\right|_{\substack{\text{extended}\\\text{snowflake}}}\underset{\mathcal{O}_{i_2}(\eta_2)\to\1}{\to}\left.I_{6(\Delta_{k_3},\Delta_{k_1},\Delta_{k_4})}^{(\Delta_{i_5},\Delta_{i_6},\Delta_{i_4},\Delta_{i_2},\Delta_{i_7},\Delta_{i_1})}\right|_{\text{comb}}.}[EqLimUnit3Id]
Moreover, in this limit we have $\Delta_{i_3}=0$, $\Delta_{k_2}=\Delta_{i_4}$ as well as
\eqn{
\begin{gathered}
L_7\prod_{1\leq a\leq4}(u_a^7)^{\frac{\Delta_{\Delta_{k_a}}}{2}}=(v_{23}^6)^{h_3-h_6}(v_{12}^6)^{h_6}L_6\prod_{1\leq a\leq3}(u_i^6)^{\frac{\Delta_{k^{\prime}_a}}{2}},\\
u_1^6=u_3^7,\qquad u_2^6=\frac{u_1^7v_{14}^7}{v_{34}^7},\qquad u_3^6=\frac{u_4^7}{v_{34}^7},\\
v_{11}^6=v_{23}^7,\qquad v_{12}^6=\frac{1}{v_{34}^7},\qquad v_{13}^6=\frac{v_{11}^7v_{14}^7}{v_{34}^7},\\
v_{22}^6=\frac{v_{44}^7}{v_{34}^7},\qquad v_{23}^6=\frac{v_{14}^7}{v_{34}^7},\qquad v_{33}^6=\frac{v_{13}^7v_{14}^7}{v_{34}^7},
\end{gathered}
}
for the legs and conformal cross-ratios.  Here, the primed parameters correspond to the parameters relevant for the RHS of \eqref{EqLimUnit3Id}.  Considering that $h_4=p_5=p_2+h_3=0$, the limit of unit operator \eqref{EqLimUnit3Id} implies the identity $G_{7|\mathcal{O}_{i_3}\to\1}\equiv(v^6_{23})^{h_3-h_6}(v^6_{12})^{h_6}G_7=G_6$.

From the vanishing components of $\boldsymbol{h}$ and $\boldsymbol{p}$, we have
\eqn{m_2=m_{12}=m_{22}=m_{24}=m_{33}=0.}
As a result, we find that
\eqna{
F_7&=\frac{(-\bar{p}_3-h_2+d/2-m_3)_{m_3}}{(p_3)_{-m_1}}\sum\frac{(-m_3)_{t_2}(\bar{p}_3-d/2)_{t_2}}{t_2!(p_3-m_1)_{t_2}(p_3)_{t_2}(\bar{p}_3+h_2+1-d/2)_{t_2}}\\
&\phantom{=}\qquad\times\frac{(-m_1)_{r}(-m_4)_r(-p_2+d/2-m_1)_r}{r!(1-p_2-h_3-m_1)_r(p_4-m_1)_r}\\
&=\frac{(-h_2-m^{\prime}_{1})_{m^{\prime}_1}(p_3)_{m^{\prime}_1}}{(p_3)_{m^{\prime}_1-m^{\prime}_2}}F_{6|\text{comb}}.
}
Thus, $G_{7|\mathcal{O}_{i_3}\to\1}$ becomes
\eqna{
G_{7|\mathcal{O}_{i_3}\to\1}&=\sum\frac{(p_4-m^{\prime}_{2})_{m^{\prime}_3+m_{14}}(-h_{6})_{m^{\prime}_3+m_{14}+m_{34}+m_{44}}(p_{4}-h_3+h_{6})_{m^{\prime}_3+m_{11}+m_{13}}}{(p_2+h_3+m^{\prime}_2)_{-m^{\prime}_3}(p_{4}-h_3)_{2m^{\prime}_3+m_{11}+m_{13}+m_{14}+m_{34}+m_{44}}(p_4-h_3+1-d/2)_{m^{\prime}_3}}\\
&\qquad\times\frac{(p_2+h_3)_{m^{\prime}_2-m^{\prime}_3+m^{\prime}_{11}}(p_3)_{m^{\prime}_1}(p_3-h_2)_{m_{11}+m_{34}}(\bar{p}_{3}+h_2+h_{5})_{m^{\prime}_1}}{(\bar{p}_{3}+h_2)_{2m^{\prime}_1+m_{13}+m^{\prime}_{11}+m_{44}}(\bar{p}_3+h_2+1-d/2)_{m^{\prime}_1}}\\
&\qquad\times\frac{(p_2+h_2)_{m^{\prime}_1+m^{\prime}_2+m_{13}+m^{\prime}_{11}+m_{44}}(-h_2-m^{\prime}_1)_{m^{\prime}_1}(-h_2)_{m^{\prime}_1-m^{\prime}_2+m_{11}+m_{34}}}{(p_2)_{2m^{\prime}_{2}+m_{11}+m_{13}+m^{\prime}_{11}+m_{34}+m_{44}}(p_2+1-d/2)_{m^{\prime}_2}}\\
&\qquad\times\frac{(-h_3)_{m^{\prime}_2+m^{\prime}_3+m_{11}+m_{13}+m_{34}+m_{44}}(-h_{5})_{m^{\prime}_2+m_{13}+m^{\prime}_{11}+m_{44}}}{(p_{3}-h_{2})_{m_{11}+m_{34}}}\\
&\qquad\times\binom{m_{11}}{k_{11}}\binom{m_{13}}{k_{13}}\binom{m_{14}}{k_{14}}\binom{m_{34}}{k_{34}}\binom{m_{44}}{k_{44}}\binom{h_6-m^{\prime}_3-k_{14}-k_{34}-k_{44}}{m^{\prime}_{12}}\binom{k_{11}}{m^{\prime}_{13}}\binom{k_{44}}{m^{\prime}_{22}}\\
&\qquad\times\binom{h_3-h_6-m^{\prime}_2-k_{11}-k_{13}+k_{14}}{m^{\prime}_{23}}\binom{k_{13}}{m^{\prime}_{33}}(-1)^{k_{11}+k_{13}+k_{14}+k_{34}+k_{44}+m^{\prime}_{12}+m^{\prime}_{13}+m^{\prime}_{22}+m^{\prime}_{23}+m^{\prime}_{33}}\\
&\qquad\times\frac{(1-v_{13}^6)^{m^{\prime}_{13}}}{m_{11}!}\frac{(1-v_{33}^6)^{m^{\prime}_{33}}}{m_{13}!}\frac{(1-v_{11}^6)^{m^{\prime}_{11}}}{m^{\prime}_{11}!}\frac{(1-v_{22}^6)^{m^{\prime}_{22}}}{m_{44}!}\frac{(1-v_{23}^6)^{m^{\prime}_{23}}}{m_{14}!}\frac{(1-v_{12}^6)^{m^{\prime}_{12}}}{m_{34}!}\\
&\qquad\times F_{6|\text{comb}}\prod_{1\leq a\leq3}\frac{(u^6_a)^{m^{\prime}_a}}{m^{\prime}_a!}.
}
At this point, the re-summations mirror the ones of the previous limit of unit operator.  As such, they are left for the interested reader.

%%%%%%%%%%%%%%%%%%%%%%%%%%%%%%%%%%%%%%%%%%%%%%%%%%
%%%%%%%%%%%%%%%%%%%%%%%%%%%%%%%%%%%%%%%%%%%%%%%%%%

\bibliography{ExtendedSnowflake}

%bibliography generated by nb.bst v1.01 (C) 2003-2010 Niklas Beisert
\begin{thebibliography}{10}
\ifx\href\asklfhas\newcommand{\href}[2]{#2}\fi
\ifx\arxivref\asklfhas\newcommand{\arxivref}[2]{\href{http://arxiv.org/abs/#1}{#2}}\fi
\ifx\doiref\asklfhas\newcommand{\doiref}[2]{\href{http://dx.doi.org/#1}{#2}}\fi
\parskip 0pt
\normalsize

\bibitem{Ferrara:1973yt}
S.~Ferrara, A.~F. Grillo \& R.~Gatto,
\textit{``{Tensor representations of conformal algebra and conformally
  covariant operator product expansion}''},
\doiref{10.1016/0003-4916(73)90446-6}{Annals~Phys. \textbf{76}, 161
  (1973)\ignorespaces}\ignorespaces
\bibitem{Polyakov:1974gs}
A.~M. Polyakov,
\textit{``{Nonhamiltonian approach to conformal quantum field theory}''},
Zh.~Eksp.~Teor.~Fiz. \textbf{66}, 23 (1974)\ignorespaces\ignorespaces,
[Sov. Phys. JETP39,9(1974)]\ignorespaces
\bibitem{Dolan:2003hv}
F.~A. Dolan \& H.~Osborn,
\textit{``{Conformal partial waves and the operator product expansion}''},
\doiref{10.1016/j.nuclphysb.2003.11.016}{Nucl.~Phys. \textbf{B678}, 491
  (2004)\ignorespaces}\ignorespaces,
\normalsize{\texttt{\arxivref{hep-th/0309180}{hep-th/0309180}}}\ignorespaces
\bibitem{Dolan:2011dv}
F.~A. Dolan \& H.~Osborn,
\textit{``{Conformal Partial Waves: Further Mathematical Results}''},
\normalsize{\texttt{\arxivref{1108.6194}{arXiv:1108.6194}}}\ignorespaces
\bibitem{Kravchuk:2017dzd}
P.~Kravchuk,
\textit{``{Casimir recursion relations for general conformal blocks}''},
\doiref{10.1007/JHEP02(2018)011}{JHEP \textbf{1802}, 011
  (2018)\ignorespaces}\ignorespaces,
\normalsize{\texttt{\arxivref{1709.05347}{arXiv:1709.05347}}}\ignorespaces,
[,164(2017)]\ignorespaces
\bibitem{Ferrara:1972xe}
S.~Ferrara \& G.~Parisi,
\textit{``{Conformal covariant correlation functions}''},
\doiref{10.1016/0550-3213(72)90480-4}{Nucl.~Phys. \textbf{B42}, 281
  (1972)\ignorespaces}\ignorespaces
\bibitem{Ferrara:1972uq}
S.~Ferrara, A.~F. Grillo, G.~Parisi \& R.~Gatto,
\textit{``{The shadow operator formalism for conformal algebra. Vacuum
  expectation values and operator products}''},
\doiref{10.1007/BF02907130}{Lett.~Nuovo~Cim. \textbf{4S2}, 115
  (1972)\ignorespaces}\ignorespaces,
[Lett. Nuovo Cim.4,115(1972)]\ignorespaces
\bibitem{SimmonsDuffin:2012uy}
D.~Simmons-Duffin,
\textit{``{Projectors, Shadows, and Conformal Blocks}''},
\doiref{10.1007/JHEP04(2014)146}{JHEP \textbf{1404}, 146
  (2014)\ignorespaces}\ignorespaces,
\normalsize{\texttt{\arxivref{1204.3894}{arXiv:1204.3894}}}\ignorespaces
\bibitem{Karateev:2017jgd}
D.~Karateev, P.~Kravchuk \& D.~Simmons-Duffin,
\textit{``{Weight Shifting Operators and Conformal Blocks}''},
\doiref{10.1007/JHEP02(2018)081}{JHEP \textbf{1802}, 081
  (2018)\ignorespaces}\ignorespaces,
\normalsize{\texttt{\arxivref{1706.07813}{arXiv:1706.07813}}}\ignorespaces,
[,91(2017)]\ignorespaces
\bibitem{Costa:2018mcg}
M.~S. Costa \& T.~Hansen,
\textit{``{AdS Weight Shifting Operators}''},
\doiref{10.1007/JHEP09(2018)040}{JHEP \textbf{1809}, 040
  (2018)\ignorespaces}\ignorespaces,
\normalsize{\texttt{\arxivref{1805.01492}{arXiv:1805.01492}}}\ignorespaces
\bibitem{Isachenkov:2016gim}
M.~Isachenkov \& V.~Schomerus,
\textit{``{Superintegrability of $d$-dimensional Conformal Blocks}''},
\doiref{10.1103/PhysRevLett.117.071602}{Phys.~Rev.~Lett. \textbf{117}, 071602
  (2016)\ignorespaces}\ignorespaces,
\normalsize{\texttt{\arxivref{1602.01858}{arXiv:1602.01858}}}\ignorespaces
\bibitem{Schomerus:2016epl}
V.~Schomerus, E.~Sobko \& M.~Isachenkov,
\textit{``{Harmony of Spinning Conformal Blocks}''},
\doiref{10.1007/JHEP03(2017)085}{JHEP \textbf{1703}, 085
  (2017)\ignorespaces}\ignorespaces,
\normalsize{\texttt{\arxivref{1612.02479}{arXiv:1612.02479}}}\ignorespaces
\bibitem{Schomerus:2017eny}
V.~Schomerus \& E.~Sobko,
\textit{``{From Spinning Conformal Blocks to Matrix Calogero-Sutherland
  Models}''},
\doiref{10.1007/JHEP04(2018)052}{JHEP \textbf{1804}, 052
  (2018)\ignorespaces}\ignorespaces,
\normalsize{\texttt{\arxivref{1711.02022}{arXiv:1711.02022}}}\ignorespaces
\bibitem{Isachenkov:2017qgn}
M.~Isachenkov \& V.~Schomerus,
\textit{``{Integrability of conformal blocks. Part I. Calogero-Sutherland
  scattering theory}''},
\doiref{10.1007/JHEP07(2018)180}{JHEP \textbf{1807}, 180
  (2018)\ignorespaces}\ignorespaces,
\normalsize{\texttt{\arxivref{1711.06609}{arXiv:1711.06609}}}\ignorespaces
\bibitem{Buric:2019dfk}
I.~Burić, V.~Schomerus \& M.~Isachenkov,
\textit{``{Conformal Group Theory of Tensor Structures}''},
\normalsize{\texttt{\arxivref{1910.08099}{arXiv:1910.08099}}}\ignorespaces
\bibitem{Hijano:2015zsa}
E.~Hijano, P.~Kraus, E.~Perlmutter \& R.~Snively,
\textit{``{Witten Diagrams Revisited: The AdS Geometry of Conformal Blocks}''},
\doiref{10.1007/JHEP01(2016)146}{JHEP \textbf{1601}, 146
  (2016)\ignorespaces}\ignorespaces,
\normalsize{\texttt{\arxivref{1508.00501}{arXiv:1508.00501}}}\ignorespaces
\bibitem{Nishida:2016vds}
M.~Nishida \& K.~Tamaoka,
\textit{``{Geodesic Witten diagrams with an external spinning field}''},
\doiref{10.1093/ptep/ptx055}{PTEP \textbf{2017}, 053B06
  (2017)\ignorespaces}\ignorespaces,
\normalsize{\texttt{\arxivref{1609.04563}{arXiv:1609.04563}}}\ignorespaces
\bibitem{Castro:2017hpx}
A.~Castro, E.~Llabrés \& F.~Rejon-Barrera,
\textit{``{Geodesic Diagrams, Gravitational Interactions \& OPE Structures}''},
\doiref{10.1007/JHEP06(2017)099}{JHEP \textbf{1706}, 099
  (2017)\ignorespaces}\ignorespaces,
\normalsize{\texttt{\arxivref{1702.06128}{arXiv:1702.06128}}}\ignorespaces
\bibitem{Dyer:2017zef}
E.~Dyer, D.~Z. Freedman \& J.~Sully,
\textit{``{Spinning Geodesic Witten Diagrams}''},
\doiref{10.1007/JHEP11(2017)060}{JHEP \textbf{1711}, 060
  (2017)\ignorespaces}\ignorespaces,
\normalsize{\texttt{\arxivref{1702.06139}{arXiv:1702.06139}}}\ignorespaces
\bibitem{Chen:2017yia}
H.-Y. Chen, E.-J. Kuo \& H.~Kyono,
\textit{``{Anatomy of Geodesic Witten Diagrams}''},
\doiref{10.1007/JHEP05(2017)070}{JHEP \textbf{1705}, 070
  (2017)\ignorespaces}\ignorespaces,
\normalsize{\texttt{\arxivref{1702.08818}{arXiv:1702.08818}}}\ignorespaces
\bibitem{Sleight:2017fpc}
C.~Sleight \& M.~Taronna,
\textit{``{Spinning Witten Diagrams}''},
\doiref{10.1007/JHEP06(2017)100}{JHEP \textbf{1706}, 100
  (2017)\ignorespaces}\ignorespaces,
\normalsize{\texttt{\arxivref{1702.08619}{arXiv:1702.08619}}}\ignorespaces
\bibitem{Ferrara:1971vh}
S.~Ferrara, A.~F. Grillo \& R.~Gatto,
\textit{``{Manifestly conformal covariant operator-product expansion}''},
\doiref{10.1007/BF02770435}{Lett.~Nuovo~Cim. \textbf{2S2}, 1363
  (1971)\ignorespaces}\ignorespaces,
[Lett. Nuovo Cim.2,1363(1971)]\ignorespaces
\bibitem{Ferrara:1971zy}
S.~Ferrara, R.~Gatto \& A.~F. Grillo,
\textit{``{Conformal invariance on the light cone and canonical dimensions}''},
\doiref{10.1016/0550-3213(71)90333-6}{Nucl.~Phys. \textbf{B34}, 349
  (1971)\ignorespaces}\ignorespaces
\bibitem{Ferrara:1972cq}
S.~Ferrara, A.~F. Grillo \& R.~Gatto,
\textit{``{Manifestly conformal-covariant expansion on the light cone}''},
\doiref{10.1103/PhysRevD.5.3102}{Phys.~Rev. \textbf{D5}, 3102
  (1972)\ignorespaces}\ignorespaces
\bibitem{Ferrara:1973eg}
S.~Ferrara, P.~Gatto \& A.~F. Grilla,
\textit{``{Conformal algebra in spacetime and operator product expansion}''},
\doiref{10.1007/BFb0111104}{Springer~Tracts~Mod.~Phys. \textbf{67}, 1
  (1973)\ignorespaces}\ignorespaces
\bibitem{Ferrara:1973vz}
S.~Ferrara, A.~F. Grillo, G.~Parisi \& R.~Gatto,
\textit{``{Covariant expansion of the conformal four-point function}''},
\doiref{10.1016/0550-3213(72)90587-1, 10.1016/0550-3213(73)90467-7}{Nucl.~Phys.
  \textbf{B49}, 77 (1972)\ignorespaces}\ignorespaces,
[Erratum: Nucl. Phys.B53,643(1973)]\ignorespaces
\bibitem{Ferrara:1974nf}
S.~Ferrara, A.~F. Grillo, R.~Gatto \& G.~Parisi,
\textit{``{Analyticity properties and asymptotic expansions of conformal
  covariant green's functions}''},
\doiref{10.1007/BF02813413}{Nuovo~Cim. \textbf{A19}, 667
  (1974)\ignorespaces}\ignorespaces
\bibitem{Dolan:2000ut}
F.~A. Dolan \& H.~Osborn,
\textit{``{Conformal four point functions and the operator product
  expansion}''},
\doiref{10.1016/S0550-3213(01)00013-X}{Nucl.~Phys. \textbf{B599}, 459
  (2001)\ignorespaces}\ignorespaces,
\normalsize{\texttt{\arxivref{hep-th/0011040}{hep-th/0011040}}}\ignorespaces
\bibitem{Fortin:2016lmf}
J.-F. Fortin \& W.~Skiba,
\textit{``{Conformal Bootstrap in Embedding Space}''},
\doiref{10.1103/PhysRevD.93.105047}{Phys.~Rev. \textbf{D93}, 105047
  (2016)\ignorespaces}\ignorespaces,
\normalsize{\texttt{\arxivref{1602.05794}{arXiv:1602.05794}}}\ignorespaces
\bibitem{Fortin:2016dlj}
J.-F. Fortin \& W.~Skiba,
\textit{``{Conformal Differential Operator in Embedding Space and its
  Applications}''},
\doiref{10.1007/JHEP07(2019)093}{JHEP \textbf{1907}, 093
  (2019)\ignorespaces}\ignorespaces,
\normalsize{\texttt{\arxivref{1612.08672}{arXiv:1612.08672}}}\ignorespaces
\bibitem{Comeau:2019xco}
V.~Comeau, J.-F. Fortin \& W.~Skiba,
\textit{``{Further Results on a Function Relevant for Conformal Blocks}''},
\normalsize{\texttt{\arxivref{1902.08598}{arXiv:1902.08598}}}\ignorespaces
\bibitem{Fortin:2019fvx}
J.-F. Fortin \& W.~Skiba,
\textit{``{A recipe for conformal blocks}''},
\normalsize{\texttt{\arxivref{1905.00036}{arXiv:1905.00036}}}\ignorespaces
\bibitem{Fortin:2019dnq}
J.-F. Fortin \& W.~Skiba,
\textit{``{New methods for conformal correlation functions}''},
\doiref{10.1007/JHEP06(2020)028}{JHEP \textbf{2006}, 028
  (2020)\ignorespaces}\ignorespaces,
\normalsize{\texttt{\arxivref{1905.00434}{arXiv:1905.00434}}}\ignorespaces
\bibitem{Fortin:2019xyr}
J.-F. Fortin, V.~Prilepina \& W.~Skiba,
\textit{``{Conformal two-point correlation functions from the operator product
  expansion}''},
\doiref{10.1007/JHEP04(2020)114}{JHEP \textbf{2004}, 114
  (2020)\ignorespaces}\ignorespaces,
\normalsize{\texttt{\arxivref{1906.12349}{arXiv:1906.12349}}}\ignorespaces
\bibitem{Fortin:2019pep}
J.-F. Fortin, V.~Prilepina \& W.~Skiba,
\textit{``{Conformal Three-Point Correlation Functions from the Operator
  Product Expansion}''},
\normalsize{\texttt{\arxivref{1907.08599}{arXiv:1907.08599}}}\ignorespaces
\bibitem{Fortin:2019gck}
J.-F. Fortin, V.~Prilepina \& W.~Skiba,
\textit{``{Conformal Four-Point Correlation Functions from the Operator Product
  Expansion}''},
\normalsize{\texttt{\arxivref{1907.10506}{arXiv:1907.10506}}}\ignorespaces
\bibitem{Fortin:2020ncr}
J.-F. Fortin, W.-J. Ma, V.~Prilepina \& W.~Skiba,
\textit{``{Efficient Rules for All Conformal Blocks}''},
\normalsize{\texttt{\arxivref{2002.09007}{arXiv:2002.09007}}}\ignorespaces
\bibitem{Alkalaev:2015fbw}
K.~B. Alkalaev \& V.~A. Belavin,
\textit{``{From global to heavy-light: 5-point conformal blocks}''},
\doiref{10.1007/JHEP03(2016)184}{JHEP \textbf{1603}, 184
  (2016)\ignorespaces}\ignorespaces,
\normalsize{\texttt{\arxivref{1512.07627}{arXiv:1512.07627}}}\ignorespaces
\bibitem{Rosenhaus:2018zqn}
V.~Rosenhaus,
\textit{``{Multipoint Conformal Blocks in the Comb Channel}''},
\doiref{10.1007/JHEP02(2019)142}{JHEP \textbf{1902}, 142
  (2019)\ignorespaces}\ignorespaces,
\normalsize{\texttt{\arxivref{1810.03244}{arXiv:1810.03244}}}\ignorespaces
\bibitem{Goncalves:2019znr}
V.~Gonçalves, R.~Pereira \& X.~Zhou,
\textit{``{$20'$ Five-Point Function from $AdS_5\times S^5$ Supergravity}''},
\doiref{10.1007/JHEP10(2019)247}{JHEP \textbf{1910}, 247
  (2019)\ignorespaces}\ignorespaces,
\normalsize{\texttt{\arxivref{1906.05305}{arXiv:1906.05305}}}\ignorespaces
\bibitem{Parikh:2019ygo}
S.~Parikh,
\textit{``{Holographic dual of the five-point conformal block}''},
\doiref{10.1007/JHEP05(2019)051}{JHEP \textbf{1905}, 051
  (2019)\ignorespaces}\ignorespaces,
\normalsize{\texttt{\arxivref{1901.01267}{arXiv:1901.01267}}}\ignorespaces
\bibitem{Jepsen:2019svc}
C.~B. Jepsen \& S.~Parikh,
\textit{``{Propagator identities, holographic conformal blocks, and
  higher-point AdS diagrams}''},
\doiref{10.1007/JHEP10(2019)268}{JHEP \textbf{1910}, 268
  (2019)\ignorespaces}\ignorespaces,
\normalsize{\texttt{\arxivref{1906.08405}{arXiv:1906.08405}}}\ignorespaces
\bibitem{Parikh:2019dvm}
S.~Parikh,
\textit{``{A multipoint conformal block chain in $d$ dimensions}''},
\doiref{10.1007/JHEP05(2020)120}{JHEP \textbf{2005}, 120
  (2020)\ignorespaces}\ignorespaces,
\normalsize{\texttt{\arxivref{1911.09190}{arXiv:1911.09190}}}\ignorespaces
\bibitem{Fortin:2019zkm}
J.-F. Fortin, W.~Ma \& W.~Skiba,
\textit{``{Higher-Point Conformal Blocks in the Comb Channel}''},
\normalsize{\texttt{\arxivref{1911.11046}{arXiv:1911.11046}}}\ignorespaces
\bibitem{Fortin:2020yjz}
J.-F. Fortin, W.-J. Ma \& W.~Skiba,
\textit{``{Six-Point Conformal Blocks in the Snowflake Channel}''},
\normalsize{\texttt{\arxivref{2004.02824}{arXiv:2004.02824}}}\ignorespaces
\bibitem{Anous:2020vtw}
T.~Anous \& F.~M. Haehl,
\textit{``{On the Virasoro six-point identity block and chaos}''},
\normalsize{\texttt{\arxivref{2005.06440}{arXiv:2005.06440}}}\ignorespaces
\bibitem{exton1976multiple}
H.~Exton,
\textit{``Multiple hypergeometric functions and applications''}
\bibitem{srivastava1985multiple}
H.~M. Srivastava \& P.~W. Karlsson,
\textit{``Multiple Gaussian Hypergeometric Series, Halsted Press (Ellis Horwood
  Limited, Chichester)''}
\end{thebibliography}

\end{document}